\documentclass[twocolumn]{aastex63}

\usepackage{graphicx}
\usepackage{color}
\usepackage{amsmath}
\usepackage{natbib}
\usepackage{enumitem}
\usepackage{booktabs}
\usepackage{hyperref}
\hypersetup{
    colorlinks=true,
    linkcolor=blue,
    filecolor=magenta,      
    urlcolor=cyan,
}
\newcommand{\tess}{{\it TESS}}
\newcommand{\kepler}{{\it Kepler}}

\begin{document} 

\title{Vetting of 384 TESS Objects of Interest with TRICERATOPS\\and Statistical Validation of 12 Planet Candidates}

\correspondingauthor{Steven Giacalone}
\email{steven$\_$giacalone@berkeley.edu}

\author[0000-0002-8965-3969]{Steven Giacalone}
\affil{Department of Astronomy, University of California Berkeley, Berkeley, CA 94720-3411, USA}

\author[0000-0001-8189-0233]{Courtney D. Dressing}
\affiliation{Department of Astronomy, University of California Berkeley, Berkeley, CA 94720-3411, USA}

%ejensen1@swarthmore.edu
\author[0000-0002-4625-7333]{Eric L. N. Jensen}
\affiliation{Dept.\ of Physics \& Astronomy, Swarthmore College, Swarthmore PA 19081, USA}

%karen.collins@cfa.harvard.edu
\author[0000-0001-6588-9574]{Karen A.\ Collins}
\affiliation{Center for Astrophysics \textbar \ Harvard \& Smithsonian, 60 Garden Street, Cambridge, MA 02138, USA}

\author{George R. Ricker}
\affiliation{Department of Physics and Kavli Institute for Astrophysics and Space Research, Massachusetts Institute of Technology, \\Cambridge, MA 02139, USA}

\author[0000-0001-6763-6562]{Roland Vanderspek}
\affiliation{Department of Physics and Kavli Institute for Astrophysics and Space Research, Massachusetts Institute of Technology, \\Cambridge, MA 02139, USA}

% \author[0000-0001-9911-7388]{David W. Latham}
% \affiliation{Center for Astrophysics \textbar \ Harvard \& Smithsonian, 60 Garden Street, Cambridge, MA 02138, USA}

\author[0000-0002-6892-6948]{S.~Seager}
\affiliation{Department of Physics and Kavli Institute for Astrophysics and Space Research, Massachusetts Institute of Technology, \\Cambridge, MA 02139, USA}
\affiliation{Department of Earth, Atmospheric and Planetary Sciences, Massachusetts Institute of Technology, \\Cambridge, MA 02139, USA}
\affiliation{Department of Aeronautics and Astronautics, MIT, 77 Massachusetts Avenue, Cambridge, MA 02139, USA}

\author[0000-0002-4265-047X]{Joshua N. Winn}
\affiliation{Department of Astrophysical Sciences, Princeton University, Princeton, NJ 08544, USA }

%jon.jenkins@nasa.gov
\author[0000-0002-4715-9460]{Jon M. Jenkins}
\affiliation{NASA Ames Research Center, Moffett Field, CA 94035, USA}

%thomas.barclay@nasa.gov
\author[0000-0001-7139-2724]{Thomas~Barclay}
\affiliation{NASA Goddard Space Flight Center, 8800 Greenbelt Road, Greenbelt, MD 20771, USA}
\affiliation{University of Maryland, Baltimore County, 1000 Hilltop Circle, Baltimore, MD 21250, USA}

%khalid.barkaoui@doct.uliege.be
\author[0000-0003-1464-9276]{Khalid Barkaoui}
\affiliation{Astrobiology Research Unit, Universit\'e de Li\`ege, 19C All\`ee du 6 Ao\^ut, 4000 Li\`ege, Belgium}
\affiliation{Oukaimeden Observatory, High Energy Physics and Astrophysics Laboratory, Cadi Ayyad University, \\Marrakech, Morocco}

%zach.bertathompson@colorado.edu
% \author[0000-0002-3321-4924]{Zachory Berta-Thompson}
% \affil{Department of Astrophysical and Planetary Sciences, University of Colorado, Boulder, CO 80309, USA}

%carolyn.brown@usq.edu.au
% \author[0000-0001-6649-4531]{Carolyn J. Brown} 
% \affiliation{University of Southern Queensland, Centre for Astrophysics, Toowoomba QLD 4350, Australia}

%cadieux@astro.umontreal.ca
\author[0000-0001-9291-5555]{Charles Cadieux}
\affil{\href{http://www.exoplanetes.ca/}{Institute for Research on Exoplanets} (IREx), Universit\'e de Montr\'eal, D\'epartement de Physique, C.P.~6128 Succ. Centre-ville, Montr\'eal, QC H3C~3J7, Canada}

%douglas.caldwell@nasa.gov
% \author[0000-0003-1963-9616]{Douglas A. Caldwell}
% \affiliation{SETI Institute, 189 Bernardo Ave, Suite 200, Mountain View, CA 94043, USA}
% \affiliation{NASA Ames Research Center, Moffett Field, CA 94035, USA}

%dcharbonneau@cfa.harvard.edu
\author[0000-0002-9003-484X]{David Charbonneau}
\affiliation{Center for Astrophysics \textbar \ Harvard \& Smithsonian, 60 Garden Street, Cambridge, MA 02138, USA}

%kcolli3@gmu.edu
\author[0000-0003-2781-3207]{Kevin I.\ Collins}
\affiliation{George Mason University, 4400 University Drive, Fairfax, VA, 22030 USA}

%dennis@astrodennis.com
\author[0000-0003-2239-0567]{Dennis M.\ Conti}
\affiliation{American Association of Variable Star Observers, 49 Bay State Road, Cambridge, MA 02138, USA}

%doyon@astro.umontreal.ca
\author[0000-0001-5485-4675]{Ren\'e Doyon} \affil{\href{http://www.exoplanetes.ca/}{Institute for Research on Exoplanets} (IREx), Universit\'e de Montr\'eal, D\'epartement de Physique, C.P.~6128 Succ. Centre-ville, Montr\'eal, QC H3C~3J7, Canada}

%phil@astrofizz.com
\author[0000-0002-5674-2404]{Phil Evans}
\affiliation{El Sauce Observatory, Coquimbo Province, Chile}

%mr.ghachoui@gmail.com
\author{Mourad Ghachoui}
\affiliation{Oukaimeden Observatory, High Energy Physics and Astrophysics Laboratory, Cadi Ayyad University, Marrakech, Morocco}

%Michael.Gillon@uliege.be
\author[0000-0003-1462-7739]{Micha\"el Gillon}
\affiliation{Astrobiology Research Unit, Universit\'e de Li\`ege, 19C All\`ee du 6 Ao\^ut, 4000 Li\`ege, Belgium}

%nmg@mit.edu
\author[0000-0002-5169-9427]{Natalia~M.~Guerrero}
\affiliation{Department of Physics and Kavli Institute for Astrophysics and Space Research, Massachusetts Institute of Technology, \\Cambridge, MA 02139, USA}

% No email - unfortunately this is a posthumous publication as he passed away recently. 
\author{Rhodes Hart}
\affiliation{Centre for Astrophysics, University of Southern Queensland, Toowoomba, QLD, 4350, Australia}

%jhartman@astro.princeton.edu
% \author[0000-0001-8732-6166]{J. D. Hartman}
% \affil{Department of Astrophysical Sciences, Princeton University, NJ 08544, USA}

%jirwin@cfa.harvard.edu
% \author{Jonathan M. Irwin}
% \affiliation{Center for Astrophysics \textbar \ Harvard \& Smithsonian, 60 Garden Street, Cambridge, MA 02138, USA}

%ejehin@uliege.be
\author{Emmanu\"el Jehin} 
\affiliation{Space Sciences, Technologies and Astrophysics Research (STAR) Institute, Université de Liège, Allée du 6 août 19C, 4000 Liège, Belgium}

%kielkopf@louisville.edu
\author[0000-0003-0497-2651]{John F.\ Kielkopf}
\affiliation{Department of Physics and Astronomy, University of Louisville, Louisville, KY 40292, USA}

%scott@proto-logic.com
% \author{Scott~McDermott}
% \affiliation{Proto-Logic LLC, 1718 Euclid Street NW, Washington, DC 20009, USA}

%mclean@stsci.edu
\author[0000-0002-8058-643X]{Brian McLean}
\affiliation{Space Telescope Science Institute, 3700 San Martin Drive, Baltimore, MD 21218}

%fmurgas@iac.es
\author{Felipe Murgas}
\affiliation{Instituto de Astrof\'isica de Canarias (IAC), E-38205 La Laguna, Tenerife, Spain}
\affiliation{Departamento de Astrof\'isica, Universidad de La Laguna (ULL), E-38206 La Laguna, Tenerife, Spain}

%epalle@iac.es
\author{Enric Palle}
\affiliation{Instituto de Astrof\'\i sica de Canarias (IAC), 38205 La Laguna, Tenerife, Spain}
\affiliation{Departamento de Astrof\'\i sica, Universidad de La Laguna (ULL), 38206, La Laguna, Tenerife, Spain}

%hannu@iac.es
\author[0000-0001-5519-1391]{Hannu Parviainen}
\affiliation{Instituto de Astrof\'\i sica de Canarias (IAC), 38205 La Laguna, Tenerife, Spain}
\affiliation{Departamento de Astrof\'\i sica, Universidad de La Laguna (ULL), 38206, La Laguna, Tenerife, Spain}

%fjpozuelos@uliege.be
\author[0000-0003-1572-7707]{Francisco J. Pozuelos} 
\affiliation{Space Sciences, Technologies and Astrophysics Research (STAR) Institute, Université de Liège, 19C Allée du 6 Août, 4000 Liège, Belgium} 
\affiliation{Astrobiology Research Unit, Université de Liège, 19C Allée du 6 Août, 4000 Liège, Belgium}

%rellesh@yahoo.com
\author{Howard M. Relles}
\affiliation{Center for Astrophysics \textbar \ Harvard \& Smithsonian, 60 Garden Street, Cambridge, MA 02138, USA}

%shporeravi@gmail.com
\author[0000-0002-1836-3120]{Avi Shporer}
\affiliation{Department of Physics and Kavli Institute for Astrophysics and Space Research, Massachusetts Institute of Technology, \\Cambridge, MA 02139, USA}

%qjsocia@email.arizona.edu
\author[0000-0002-7434-0863]{Quentin Socia}
\affiliation{Department of Astronomy/Steward Observatory, The University of Arizona, 933 N. Cherry Avenue, Tucson, AZ 85721, USA}
\altaffiliation{Earths in Other Solar Systems Team}

%thestockdalefamily@bigpond.com
\author[0000-0003-2163-1437]{Chris Stockdale}
\affiliation{Hazelwood Observatory, Australia}

%tgtan@bigpond.net.au
\author[0000-0001-5603-6895]{Thiam-Guan Tan}
\affiliation{Perth Exoplanet Survey Telescope, Perth, Western Australia}

%gtorres@cfa.harvard.edu
\author[0000-0002-5286-0251]{Guillermo Torres}
\affiliation{Center for Astrophysics \textbar \ Harvard \& Smithsonian, 60 Garden Street, Cambridge, MA 02138, USA}

%joseph.twicken@nasa.gov
\author[0000-0002-6778-7552]{Joseph~D.~Twicken}
\affiliation{NASA Ames Research Center, Moffett Field, CA 94035, USA}
\affiliation{SETI Institute, Mountain View, CA 94043, USA}

%william.waalkes@colorado.edu
\author[0000-0002-8961-0352]{William C. Waalkes} \affil{Department of Astrophysical and Planetary Sciences, University of Colorado, Boulder, CO 80309, USA}

%ian.waite@usq.edu.au
\author[0000-0002-3249-3538]{Ian A. Waite}
\affiliation{Centre for Astrophysics, University of Southern Queensland, Toowoomba, QLD, 4350, Australia}

\shortauthors{S. Giacalone, C.D. Dressing, et al.}
\shorttitle{TRICERATOPS: A TESS Vetting and Validation Tool}

\accepted{October 30, 2020}

\begin{abstract}

We present \texttt{TRICERATOPS}, a new Bayesian tool that can be used to vet and validate \tess\ Objects of Interest (TOIs). We test the tool on 68 TOIs that have been previously confirmed as planets or rejected as astrophysical false positives. By looking in the false positive probability (FPP) -- nearby false positive probability (NFPP) plane, we define criteria that TOIs must meet to be classified as validated planets (${\rm FPP} < 0.015$ and ${\rm NFPP} < 10^{-3}$), likely planets (${\rm FPP} < 0.5$ and ${\rm NFPP} < 10^{-3}$), and likely nearby false positives (${\rm NFPP} > 10^{-1}$). We apply this procedure on 384 unclassified TOIs and statistically validate 12, classify 125 as likely planets, and classify 52 as likely nearby false positives. Of the 12 statistically validated planets, 9 are newly validated. \texttt{TRICERATOPS} is currently the only \tess\ vetting and validation tool that models transits from nearby contaminant stars in addition to the target star. We therefore encourage use of this tool to prioritize follow-up observations that confirm bona fide planets and identify false positives originating from nearby stars.

\end{abstract}

\section{Introduction}\label{sec:1}

Over the last decade, the  \kepler\  Space Telescope has revolutionized our understanding of exoplanets by facilitating the discovery of thousands of planets that transit in front of their host stars. Among other things, these planets have been useful for investigating the frequency of planets as a function of size and orbital period \citep[e.g.,][]{howard2012planet, dong2013fast, dressing2013occurrence, fressin2013false, petigura2013prevalence, foreman2014exoplanet, morton2014radius, sanchis2014study, burke2015terrestrial, dressing2015occurrence, mulders2015increase, mulders2015stellar, fulton2017california, hsu2018improving}, as well as testing theories of planet formation and evolution \citep[e.g.,][]{lopez2013role, swift2013characterizing,lee2017magnetospheric, konigl2017origin, giacalone2017test}. To ensure the veracity of their results, studies that utilized the  \kepler\  dataset required that: (1) the measured radii of these planets were accurate, and (2) that the discovered objects were actually planets. However, due to the limited $4\arcsec$/pixel resolution of the camera used by  \kepler, these two requirements could not always be assumed true. Because it was not uncommon for \kepler\ field stars of comparable brightness to reside $< 4\arcsec$ apart, the presence of multiple unresolved stars within a given set of pixels could not be discounted. This uncertainty was problematic because the existence of unresolved stars could cause an underestimation of the radius of a transiting object, sometimes to the extent that an eclipsing binary star could be mistaken for a transiting planet with a fraction of the size.

A number of methods have been used to constrain the possibility of an unresolved star residing within a given pixel. One method used is to search for offsets in the centroid of the source during transit, a signal indicative of another star residing elsewhere in the pixel \citep[e.g.,][]{bryson2013identification, coughlin2014contamination}. Multi-band time-series photometry has also been used to search for unresolved stars, as one would expect a different transit depth in different photometric bands if the transiting object is around a star of a different color than the target \citep[e.g.,][]{alonso2004tres}. Spectra of the target star can also be useful in this vetting process. High-precision radial velocities can rule out bound stellar companions by measuring the masses of transiting objects and monitoring for longer-period secondaries \citep[e.g.,][]{errmann2014investigation}, and reconnaissance spectroscopy can rule out bright unresolved stars by searching for additional lines in the spectrum of the target star \citep[e.g.,][]{santerne2012sophie, kolbl2014detection}. Finally, high-resolution imaging can rule out unresolved stars beyond a fraction of an arcsecond from the target star \citep[e.g.,][]{crossfield2016197, mayo2018275}. Unfortunately, these techniques do not cover the full allowed parameter space individually, and  \kepler\  planet candidate hosts were often too faint for precise radial velocity measurements. For this reason, it was common to turn to vetting and statistical validation to assess the genuineness of  \kepler\  planet candidates.

When speaking of vetting, we refer to the process of scrutinizing the photometry of threshold-crossing events (TCEs, periodic transit-like signals originating from target stars) and classifying them as planet candidates and false positives of instrumental or astrophysical origin. Vetting procedures typically make use of automated decision-making algorithms to determine the natures of these events. Autovetter \citep{mccauliff2015automatic, catanzarite2015autovetter} and Robovetter \citep{thompson2018planetary} are a \kepler-era vetting procedures that classify TCEs based on \kepler\ data using a random-forest and decision tree algorithms. \texttt{DAVE} \citep{kostov2019discovery} is a vetting tool that calculates metrics based on centroid position and transit shape to classify $K2$ and \tess\ TCEs. Lastly, Exonet \citep{shallue2018identifying} and Astronet \citep{ansdell2018scientific} make use of convolutional neural networks to classify TCEs based on transit shape. By distinguishing planet candidates from false positives, these tools have allowed others to focus planetary confirmation and characterization efforts on the most promising targets.

When speaking of statistical validation, we refer to the process of statistically ruling out astrophysical false positive scenarios to a degree of certainty high enough to advance the status of a planet candidate to one similar to that of a planet confirmed via mass measurement. In addition to information gleaned from the light curve of a planet candidate, validation algorithms typically incorporate constraints obtained from follow-up observations like those described previously. A number of statistical validation algorithms were used during the \kepler\ era in order to grow the dataset with which large-scale studies of planetary system properties could be conducted.

The first  \kepler-era validation framework was \texttt{BLENDER} \citep{torres2004testing, torres2005testing, torres2010modeling}. \texttt{BLENDER} begins by generating synthetic light curves using models of transiting planets and astrophysical false positives involving blended eclipsing binaries. Next, it calculates the $\chi^2$ of the best-fit planetary scenario and the $\chi^2$ values for several false positive scenarios over a grid of model parameters. For each false positive scenario, the region of parameter space where the scenario is viable (defined by where $\chi^2$ differs from the best-fit planetary $\chi^2$ with a confidence level $< 3 \sigma$) is identified. The properties of the blended stars in these viable instances are then compared to constraints obtained from supplementary follow-up, such as high-resolution imaging and spectroscopy, to determine if they are physically possible. In addition to this light curve analysis, \texttt{BLENDER} calculates the multi-color photometry one would expect to measure for each false positive scenario to compare to the actual observed colors. If the properties of all viable false positive scenarios are ruled out by the information from these external observations, the planet candidate is considered validated. 

\texttt{BLENDER} offered a robust option for the statistical validation of transiting planet candidates during the  \kepler\  era. However, the hands-on nature of the algorithm and the long computation times required to simulate the many false positive scenarios involved in its analysis made it inefficient for validating planet candidates in bulk. This led to the formulation of a different validation procedure by the name of \texttt{VESPA} \citep{morton2012efficient, 2015ascl.soft03011M}. In addition to being fully automated, \texttt{VESPA} provides a more computationally expedient option for validating planet candidates by replacing the physical transit models employed in \texttt{BLENDER} with a simpler trapezoidal model, which can capture the most important features of the transit shape with fewer free parameters. 

\texttt{VESPA} works in a Bayesian framework where the probabilities of several transit-producing scenarios are computed. For every scenario, \texttt{VESPA} uses the TRILEGAL galactic model \citep{girardi2005star} to simulate a population of stars with properties consistent with the target star in a cone around the line of sight to the target. The properties of these simulated stars are inferred using archival photometry of the target star and isochrone interpolation, which ensures agreement with observational constraints. For each instance of each population, the transit shape is characterized using a trapezoidal model, which allows for the generation of a trapezoidal parameter prior distribution for each scenario. \texttt{VESPA} then uses a Markov Chain Monte Carlo routine to fit the \kepler\ light curve to the same trapezoidal model to determine the region of parameter space the target occupies. Next, the marginal likelihood is calculated for every scenario by integrating the product of the trapezoidal likelihood and parameter prior over the predetermined region of parameter space. These marginal likelihoods are multiplied by model priors based on the geometries of simulated systems and assumptions relating to the occurrence of planets and close binaries. Lastly, the probability of the transiting planet scenario is assessed by comparing this product for the transiting planet scenario with those of all false positive scenarios, with the planet candidate being validated if the overall false positive probability is $< 1 \%$. Like \texttt{BLENDER}, \texttt{VESPA} can also incorporate follow-up observations to obtain tighter constraints on this probability.

Another procedure used to validate exoplanet candidates is \texttt{PASTIS} \citep{diaz2014pastis, santerne2015pastis}. \texttt{PASTIS} provides a rigorous option for the statistical validation of small planetary transits by calculating the Bayesian odds ratio between the transiting planet scenario and all possible false positive scenarios for a given target star. Prior probabilities are computed for each scenario by combining information about the target, including that contained within ground-based follow-up observations, with knowledge of stellar multiplicity and planet occurrence rates. In addition, for false positive scenarios that involve an unresolved foreground or background star, TRILEGAL is used to simulate a population of stars around the line of sight to target to calculate the prior probability of such a chance alignment. Like in \texttt{VESPA}, these priors are combined with marginal likelihoods, which \texttt{PASTIS} calculates using importance sampling. However, unlike \texttt{VESPA}, \texttt{PASTIS} additionally models the radial velocities of its targets and uses physical light curve models in its analysis. Like those utilized with \texttt{BLENDER}, these light curve models are more complex than the trapezoidal model, meaning \texttt{PASTIS} must sample over a wider parameter space when computing the marginal likelihood of each scenario. While ensuring that all possible parameter combinations for each scenario are considered, this method requires significantly more time to run for a given target than \texttt{VESPA} does.

Each of the aforementioned procedures was designed to work with minimal information about a given target star in order to argue for the existence of a transiting planet around it. This design mainly grew out of necessity, as information about many planet candidate hosts and the region of sky in which they were located was sparse in the absence of additional observations. For instance, the number of stars within each pixel was often unknown, and the stars that were known were not always precisely characterized. These facts imposed limitations on the functionalities of the procedures. Specifically, they restricted testable false positive scenarios to those involving the target star and a single unresolved star, even though there could have been a multitude of unknown stars in the group of pixels used to extract a given light curve. Additionally, poorly characterized target stars forced these procedures to use stellar models and isochrone interpolation to estimate host star properties, which comes at the cost of computation time.

These design features make previous validation algorithms poorly optimized for use on planet candidates identified by the Transiting Exoplanet Survey Satellite (TESS, \citealt{ricker2010transiting}). \tess\ differs from  \kepler\  by being an all-sky survey that focuses on the nearest and brightest stars in order to find planets that are well-suited for mass measurement and atmospheric characterization. However, this increased sky coverage comes at the cost of resolution. The \tess\ cameras contain pixels that span $21 \arcsec$, which means each pixel covers an area of sky roughly 25$\times$ larger than those utilized by  \kepler\ . Because of this, the assumption that there is at most one additional star contributing to the flux in a given aperture is unlikely to be true. In addition to scenarios involving a bound stellar companion or a chance alignment of a non-associated star near the target star, a \tess\ validation procedure must be capable of considering false positive scenarios involving a multitude of known stars near a given target.\footnote{It should be noted that because \tess\ focuses on brighter stars than  \kepler\  did and the field density of brighter stars is low compared to the field density of fainter stars, most of these contaminating stars will contribute only a small fraction of the total flux within the pixel. By contrast, stars blended within a  \kepler\  pixel had a higher probability of having comparable brightnesses.} While tools like \texttt{VESPA} have been used to validate planet candidates detected by \tess\ after ruling out false positives due to nearby stars with supplementary follow-up observations \citep[e.g.,][]{cloutier2019independent,gunther2019super,quinn2019near,vanderspek2019tess,cloutier2020toi, eisner2020planet,gilbert2020first,huang2020tess}, no tool exists as of yet that can perform a multi-star analysis on its own.

\begin{deluxetable*}{cll}[t]

\tabletypesize{\footnotesize}
% \tablewidth{\pagewidth}
 \tablecaption{Scenarios Tested by \texttt{TRICERATOPS} \label{tab:Scenarios}}
 \tablehead{
 \colhead{Scenario} & \colhead{Configuration} & \colhead{Parameter Vector, $\theta_j$}
 }
 \startdata 
 TP     & No unresolved companion. Transiting planet with $P_{\rm orb}$ around target star. & ($i$, $R_p$) \\
 EB     & No unresolved companion. Eclipsing binary with $P_{\rm orb}$ around target star.  & ($i$, $q_{\rm short}$) \\
 EBx2P  & No unresolved companion. Eclipsing binary with $2 \times P_{\rm orb}$ around target star. & ($i$, $q_{\rm short}$) \\
 PTP    & Unresolved bound companion. Transiting planet with $P_{\rm orb}$ around primary star. & ($i$, $R_p$, $q_{\rm long}$) \\
 PEB    & Unresolved bound companion. Eclipsing binary with $P_{\rm orb}$ around primary star. & ($i$, $q_{\rm short}$, $q_{\rm long}$) \\
 PEBx2P & Unresolved bound companion. Eclipsing binary with $2 \times P_{\rm orb}$ around primary star.      & ($i$, $q_{\rm short}$, $q_{\rm long}$) \\
 STP    & Unresolved bound companion. Transiting planet with $P_{\rm orb}$ around secondary star. & ($i$, $R_p$, $q_{\rm long}$) \\
 SEB    & Unresolved bound companion. Eclipsing binary with $P_{\rm orb}$ around secondary star. & ($i$, $q_{\rm short}$, $q_{\rm long}$) \\
 SEBx2P & Unresolved bound companion. Eclipsing binary with $2 \times P_{\rm orb}$ around secondary star. & ($i$, $q_{\rm short}$, $q_{\rm long}$) \\
 DTP    & Unresolved background star. Transiting planet with $P_{\rm orb}$ around target star. & ($i$, $R_p$, simulated star) \\
 DEB    & Unresolved background star. Eclipsing binary with $P_{\rm orb}$ around target star. & ($i$, $q_{\rm short}$, simulated star) \\
 DEBx2P & Unresolved background star. Eclipsing binary with $2 \times P_{\rm orb}$ around target star. & ($i$, $q_{\rm short}$, simulated star) \\
 BTP    & Unresolved background star. Transiting planet with $P_{\rm orb}$ around background star. & ($i$, $R_p$, simulated star) \\
 BEB    & Unresolved background star. Eclipsing binary with $P_{\rm orb}$ around background star. & ($i$, $q_{\rm short}$, simulated star) \\
 BEBx2P & Unresolved background star. Eclipsing binary with $2 \times P_{\rm orb}$ around background star. & ($i$, $q_{\rm short}$, simulated star) \\
 NTP    & No unresolved companion. Transiting planet with $P_{\rm orb}$ around nearby star. & ($i$, $R_p$) \\
 NEB    & No unresolved companion. Eclipsing binary with $P_{\rm orb}$ around nearby star. & ($i$, $q_{\rm short}$) \\
 NEBx2P & No unresolved companion. Eclipsing binary with $2 \times P_{\rm orb}$ around nearby star. & ($i$, $q_{\rm short}$)
 \enddata
%  \vspace{-0.5cm}
%  \tablecomments{Transit scenarios considered in POPTART validation procedure.}
\end{deluxetable*}

Luckily, the drawback of decreased resolution is counteracted by the wealth of information on nearby stars provided by the second \textit{Gaia} data release (DR2, \citealt{brown2018gaia}). DR2 provides optical photometry, astrometry, and positions for over one billion of the nearest stars in the Galaxy. Perhaps most importantly, it is reported that DR2 consistently resolves individual point sources that reside more than $2\farcs2$ apart, which allows for the identification of stars blended within a \tess\ pixel to levels previously only possible with supplementary follow-up. With this knowledge, one can test for false positive scenarios around known nearby stars and conduct more precise centroid analyses. In addition, the focus on nearby and bright stars means that most \tess\ planet candidate hosts can be more easily characterized using archival and follow-up data. In fact, the properties of millions of \tess\ targets have already been compiled in the \tess\ Input Catalog (TIC, \citealt{stassun2018tess}). It would benefit a validation procedure for \tess\ planet candidates to leverage these known stellar properties, rather than use stellar models to estimate them.

In this work, we present \texttt{TRICERATOPS} (\textbf{T}ool for \textbf{R}ating \textbf{I}nteresting \textbf{C}andidate \textbf{E}xoplanets and \textbf{R}eliability \textbf{A}nalysis of \textbf{T}ransits \textbf{O}riginating from \textbf{P}roximate \textbf{S}tars), a new Bayesian tool formulated to validate and vet \tess\ planet candidates.\footnote{Available at \href{https://github.com/stevengiacalone/triceratops}{https://github.com/stevengiacalone/triceratops}.} The procedure calculates the probabilities of a wide range of transit-producing scenarios using the primary transit of the planet candidate, preexisting knowledge of its host and nearby stars, and the current understanding of planet occurrence and stellar multiplicity. 

Our tool is designed to provide fast\footnote{Typical run time of about 5 minutes on a standard 2-core laptop for a single target.} and accurate calculations that can be used to not only validate transiting planet candidates, as validation tools have been used to do in the past, but also to serve as a metric for ranking targets of follow-up programs. Because a majority of \tess\ targets will be bright enough to be followed up with ground-based telescopes, there will inevitably be more planet candidate hosts to observe from the ground than time and resources allow for. We therefore encourage the use of our tool to identify targets that would benefit most from additional vetting.

The layout of this paper is as follows. In Section \ref{sec:2} we present our vetting and validation procedure, including how we determine the possible scenarios for a given target star and calculate the probability of each. In Section \ref{sec:3} we present detailed statistical validation results for a confirmed planet and for a known false positive. In Section \ref{sec:4} we present the results of our calculations for a sample of 68 TOIs that are known planets or false positives, conduct a performance assessment, and define the criteria a TOI must meet in order to be validated. In Section \ref{sec:5} we report observations that identify several TOIs as false positives originating from nearby stars and compare these observations with \texttt{TRICERATOPS} predictions. In Section \ref{sec:6} we apply our tool to 384 unclassified TOIs and statistically validate 12. In Section \ref{sec:7} we provide a discussion of our results, provide suggestions for how our tool can best be utilized, and present features that we plan on implementing in the future. Lastly, we provide concluding remarks in Section \ref{sec:8}. 

\section{Procedure}\label{sec:2}

Our validation procedure is initiated when the user inputs the ID a target star listed in the \tess\ Input Catalog (TIC) that has a transiting planet candidate. Using the MAST module of \texttt{astroquery} \citep{ginsburg2019astroquery}, the tool queries the TIC for all stars within a circle of radius 10 pixels from the target. The positions, \tess\ magnitudes, and available stellar properties of each star are recorded for later use. Next, the user is required to specify the aperture used to extract the \tess\ light curve for each sector in which the target was observed. The remaining steps of the procedure are summarized as follows:
\begin{enumerate}
    \item \texttt{TRICERATOPS} calculates the proportion of flux contributed to the aperture by each star near the target. Using the user-entered transit depth, the algorithm identifies the stars bright enough to produce the observed transit-like signal.
    \item Using the user-entered primary transit of the planet candidate and light curve models of transiting planets and eclipsing binaries, \texttt{TRICERATOPS} calculates the marginal likelihood of each transit-producing scenario.
    \item Given the marginal likelihood and prior probability of each scenario, the algorithm calculates the probability of each scenario.
    \item The algorithm uses these probabilities to determine if the planet candidate can be classified as a validated planet, a likely planet, or a likely nearby false positive.
\end{enumerate}

\subsection{Flux Ratio Calculation}

Initially, each star within 10 pixels of the target is considered a potential origin of the transit-like event. Because each star is contributing a different amount of flux to the aperture, the size that the transiting object must be to produce the observed transit depth is different for each star. Because the transiting object size is important for determining the probability of each scenario, the relative flux contributed by each star in the aperture is essential information.

We calculate the flux ratio contributed by each star using a method similar to that used in \cite{stassun2018tess} to determine the contamination ratios reported for candidate target stars in the TIC. Specifically, we assume the point spread function (PSF) of each star takes the form of a circular 2D Gaussian where the area under each Gaussian (i.e., the total flux) is determined using the \tess\ magnitudes reported in the TIC. We estimate the standard deviation of the Gaussian using the \tess\ pixel response function (PRF) models on MAST.\footnote{\url{https://archive.stsci.edu/missions-and-data/transiting-exoplanet-survey-satellite-tess}} Due to effects relating to the design of the \tess\ optics, the exact PRF for a star is dependent on the location on the CCD on which it is observed. These models allow one to estimate the PRF for a given star by providing the size and shape of the \tess\ PRF at 25 locations on each CCD. We fit each PRF model to a circular 2D Gaussian and record the best-fit standard deviation, finding that it typically ranges between 0.6 and 0.9 pixels. For simplicity, we adopt a standard deviation of 0.75 pixels for all stars, regardless of CCD location. For each star, we integrate the flux in the aperture and divide by the total flux contributed to the aperture by all stars to determine its flux ratio, $X_s$. For targets that are observed in multiple sectors, we assume the flux ratio for a given star is the average of its flux ratios across each sector.

\begin{figure}[t!]
        \begin{center}
            \includegraphics[width=0.48\textwidth]{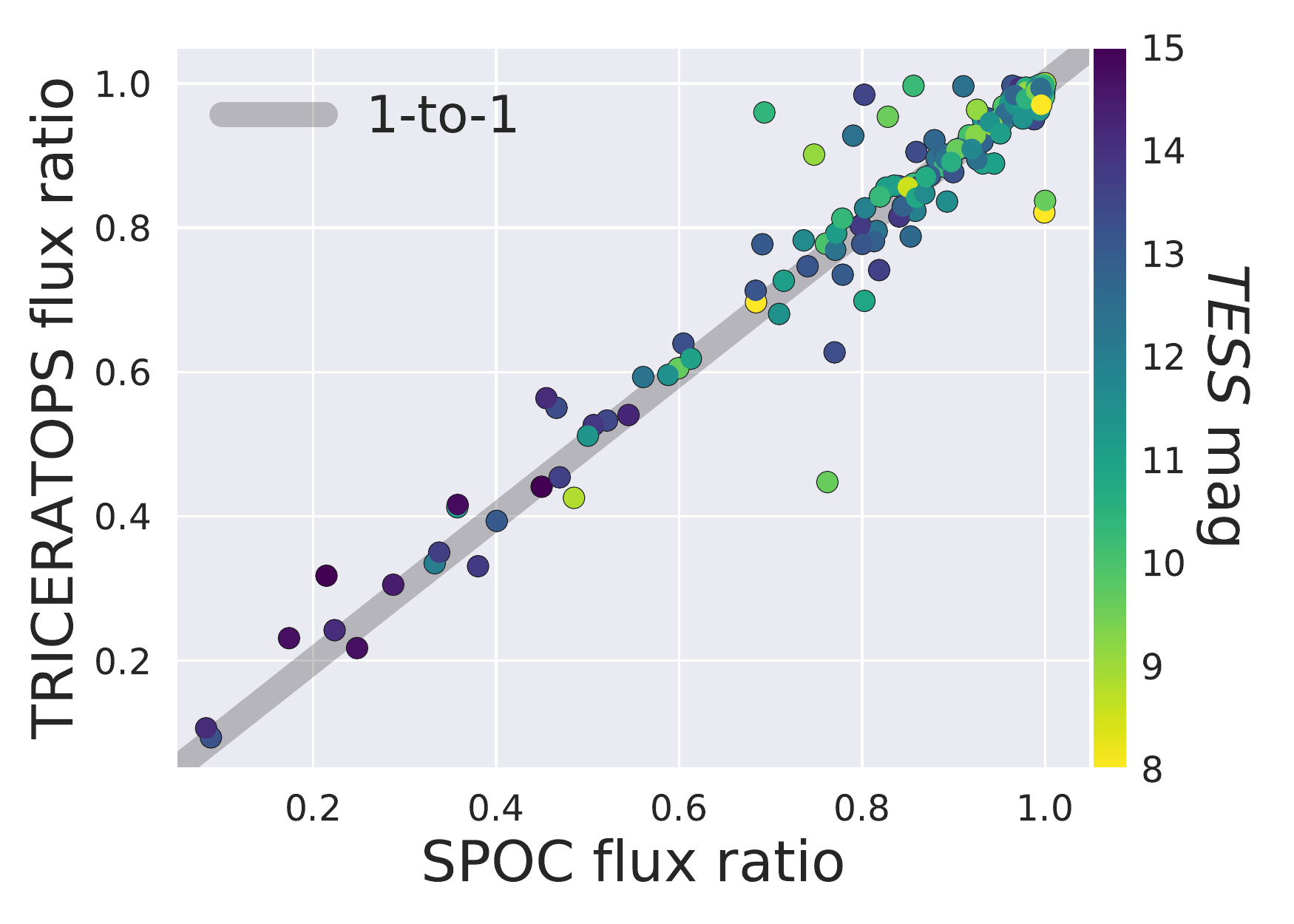}
        \end{center}
    \vspace{-15pt}
    \caption{Comparison of target star flux ratios (i.e., the fraction of the flux in the aperture due to the target star) reported by \texttt{TRICERATOPS} and the \tess\ SPOC pipeline for a 228 TOIs. A 1-to-1 line is also shown for illustrative purposes. The two methods yield consistent results, with slightly larger discrepancies for brighter stars.}
    \label{fig:2_1}
\end{figure}

To ensure that our method provides reliable flux ratios, we compare in Figure \ref{fig:2_1} the target star flux ratios for 228 TOIs obtained using our method with those reported by the \tess\ Science Processing Operations Center (SPOC) pipeline \citep{jenkins2016tess}, which calculates flux ratios using the actual PRF models discussed above.\footnote{Note that the decision to use Gaussian models rather than the actual \tess\ PRFs for our calculation was made in the interest of computational expediency.} Both of these calculations are carried out with the aperture used by the \tess\ SPOC pipeline to extract the light curve of the target star. The figure shows good agreement between the two calculations, with a slightly better agreement for fainter stars.

After flux ratios are determined, we eliminate stars that are too faint to be the source of the observed dimming event. If the observed transit depth is $\delta_\mathrm{obs}$, the relative transit depth for each star is simply $\delta_s = \delta_\mathrm{obs} / X_s$. For stars that contribute relatively little flux to the aperture, it is possible for $\delta_s$ to exceed unity. We exclude these stars from further analysis.

\subsection{Transit Scenario Identification}

After calculating the flux ratio for each star in the aperture, we determine the scenarios that can produce the observed transit-like event. Our procedure considers a total of fifteen scenarios for the target star and an additional three scenarios for each nearby star with $\delta_s < 1$. These scenarios are summarized in Table \ref{tab:Scenarios}.

The fifteen target star scenarios can be classified into three configurations. The first is the case where the target star has no unresolved stellar companion (where we define ``companion'' to encompass both bound and foreground/background stars). In this case, we consider the scenarios of a transiting planet with the reported orbital period around the target star (TP), an EB with the reported orbital period around the target star (EB), and an EB with twice the reported orbital period around the target star (EBx2P). The last of these scenarios is meant to capture the possibility that the observed transit is caused by eclipsing binary stars of roughly equal size, such that the primary and secondary eclipses are mistaken for the primary transit of a smaller object with half the orbital period. The second configuration is that in which the target star has an unresolved bound stellar companion. In this case, we consider the scenarios of a transiting around the target star with the reported orbital period (Primary TP, or PTP), an eclipsing binary with the reported orbital period around the target star (Primary EB, or PEB), an eclipsing binary with twice the reported orbital period around the target star (Primary EBx2P, or PEBx2P), a transiting planet with the reported orbital period around the companion (Secondary TP, or STP), an eclipsing binary around the companion (Secondary EB, or SEB), and an eclipsing binary with twice the reported orbital period around the companion (Secondary EBx2P, or SEBx2P). The third configuration is that in which there there is an unresolved foreground or background star along the line of sight to the target star. In this case, we again consider the scenarios of a transiting planet with the reported orbital period around the target star (Diluted TP, or DTP), an eclipsing binary with the reported orbital period around the target star (Diluted EB, or DEB), an eclipsing binary with twice the reported orbital period around the target star (Diluted EBx2P, or DEBx2P), a transiting planet with the reported orbital period around the companion (Background TP, or BTP), an eclipsing binary with the reported orbital period around the companion (Background EB, or BEB), and an eclipsing binary with twice the reported orbital period around the companion (Background EBx2P, or BEBx2P).\footnote{The BTP and BEB scenarios also include unresolved foreground stars, but the case where a background star is blended with the target star is typically the relevant one.}

For nearby stars with $\delta_s < 1$, we also consider the scenarios of a transiting planet with the reported orbital period around that star (Nearby TP, or NTP), an eclipsing binary with the reported orbital period around that star (Nearby EB, or NEB), and an eclipsing binary with twice the reported orbital period around that star (Nearby EBx2P, or NEBx2P). Each of these scenarios operates under the assumption that the nearby star has no unresolved stellar companion. These scenarios can also be omitted by the calculation if false positives originating from the respective nearby stars have been ruled out through supplementary follow-up.

\begin{figure*}[ht]
    \centering
    \includegraphics[width=1.0\textwidth]{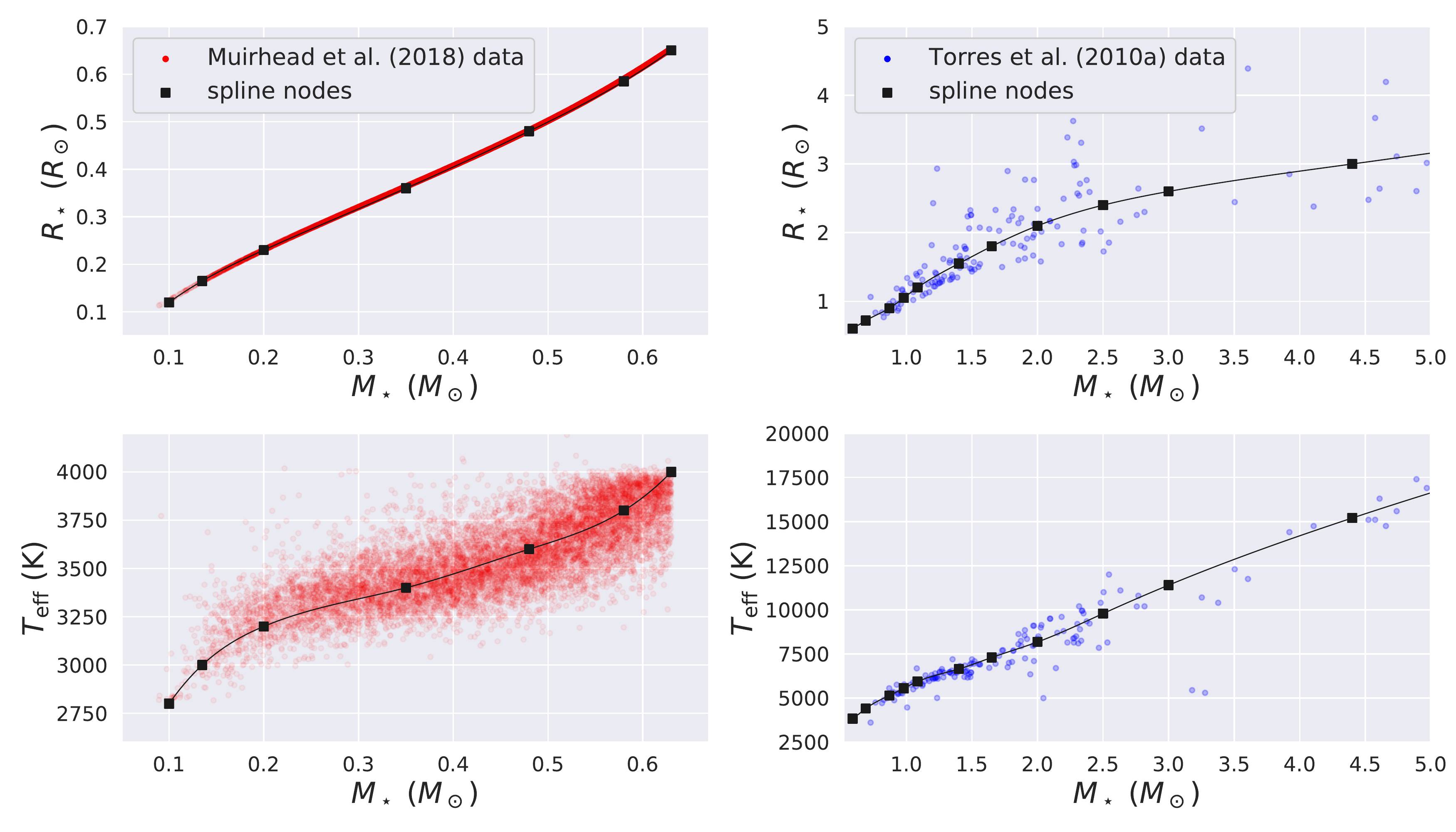}
    \vspace{-15pt}
    \caption{\emph{Left:} $R_\star$ and $T_{\rm eff}$ vs $M_\star$ for stars in the \tess\ Cool dwarf Catalog. Red points are stars from the catalog, and black squares are nodes used to draw spline relations through these points. \emph{Right:} $R_\star$ and $T_{\rm eff}$ vs $M\star$ for stars in \cite{torres2010accurate}. Blue points are stars from \cite{torres2010accurate}, and black squares are nodes used to draw spline relations through these points.}
    \label{fig:2_2}
\end{figure*}

\subsection{Stellar Property Estimation}\label{sec:splines}

Whenever possible, we use the stellar properties listed in the TIC in our calculations. However, for reasons that will be discussed, there are times in our procedure where we must estimate the properties (i.e., mass $M_\star$, radius $R_\star$, and effective temperature $T_{\rm eff}$) of a star in order to determine the probability of the corresponding scenario. We do so using the empirical and semi-empirical relations between stellar properties used to populate these fields in the TIC.

For stars with $M_\star > 0.63 M_\odot$ (corresponding roughly to $T_{\rm eff} > 4000$ K), we determine stellar properties using the results from \cite{torres2010accurate}. Using the same method discussed in Section 3 of \cite{stassun2018tess}, we draw spline curves through the distribution of points in $M_\star - T_{\rm eff}$ and $M_\star - R_\star$ space. For stars with $M_\star \leq 0.63 M_\odot$, we repeat this process using a sample of stars from the specially curated \tess\ Cool dwarf Catalog \citep{muirhead2018catalog}. We select nodal points using the sample such that they are continuous with the curves obtained for hotter stars.

The spline curves and the samples on which they are based are shown in Figure \ref{fig:2_2}. The result of this process is a set of relations that allows us to estimate the $R_\star$ and $T_{\rm eff}$ of a star given $M_\star$.

\subsection{Probability Calculation}

We employ a Bayesian framework in our procedure, and thus make use of Bayes' theorem:
\begin{equation}
    p(S_j | D) \propto p(S_j) p(D | S_j)
\end{equation}
where $p(S_j | D)$ is the posterior probability of the $j$th scenario $S_j$ given the data $D$, $p(S_j)$ is the prior probability of scenario $S_j$, and $p(D | S_j)$ is the marginal likelihood of the data $D$ given the scenario $S_j$ (sometimes also referred to as the global likelihood, or the Bayesian evidence). Because we work with a transit model characterized by the parameter vector $\theta_j$, we express the marginal likelihood as the marginalization of the likelihood $p(D | \theta_j, S_j)$ over $\theta_j$:
\begin{equation}\label{eq:marg}
    p(D | S_j) = \int p(\theta_j | S_j) p(D | \theta_j, S_j) d\theta
\end{equation}
where $p(\theta_j | S_j)$ is the prior distribution of the model parameters. We discuss how these quantities are calculated throughout the remainder of this section.

After calculating $p(S_j | D)$ for each scenario, we determine the relative probability of each scenario using the equation
\begin{equation}\label{prob}
    \mathcal{P}_j = \frac{p(S_j | D)}{\sum\limits_j p(S_j | D)}.
\end{equation}
From here, we define two quantities that are useful for vetting and validation purposes. First, the ``False Positive Probability'' (FPP) is given by
\begin{equation}
    {\rm FPP} = 1 - (\mathcal{P}_{\rm TP} + \mathcal{P}_{\rm PTP} + \mathcal{P}_{\rm DTP}).
\end{equation}
This quantity represents the probability that the observed transit is due to something other than a transiting planet around the target star. Second, the ``Nearby False Positive Probability'' (NFPP) is given by
\begin{equation}
    {\rm NFPP} = \sum (\mathcal{P}_{\rm NTP} + \mathcal{P}_{\rm NEB} + \mathcal{P}_{\rm NEBx2P})
\end{equation}
(i.e., the sum of all scenarios involving nearby stars). This quantity represents the probability that the observed transit originates from a resolved nearby star rather than the target star.

\subsubsection{Scenario Priors}

The scenario prior represents the prior probability of a given scenario before the data is considered. The only scenario prior we employ in our calculation is the probability of a transiting planet or eclipsing binary having the $P_{\rm orb}$ applied to the model. For both transiting planets and eclipsing binaries, we assume the probability distribution of $P_{\rm orb}$ takes the form of a broken power law in the range $0.1-50$ days. Using these probability distributions, we calculate the prior probability of a orbital period $P_{\rm orb}'$ by integrating the probability distribution between $P_{\rm orb}' - 0.1$ and $P_{\rm orb}' + 0.1$:
\begin{equation}\label{Porb_jnt}
    p(P_{\rm orb}') = \int^{P_{\rm orb}' + 0.1}_{P_{\rm orb}' - 0.1} p(P_{\rm orb}) d P_\mathrm{orb} .
\end{equation}

For transiting planets we base the behavior of this distribution on studies of planet occurrence rates as a function of orbital period \citep[e.g.,][]{howard2012planet, dong2013fast, petigura2013prevalence, dressing2015occurrence, mulders2015stellar, mulders2018exoplanet}. We express $p(P_{\rm orb})$ as a broken power law with a break at $P_{\rm orb} = 10$ days and the form
\begin{equation}
    p(P_{\rm orb}) \sim \left\{
        \begin{array}{ll}
            P_{\rm orb}^{1.5} & \quad 0.1 \, {\rm days} \leq P_{\rm orb} \leq 10 \, {\rm days} \\
            P_{\rm orb}^{0.0} & \quad 10 \, {\rm days} < P_{\rm orb} \leq 50 \, {\rm days} \\
        \end{array}
    \right. .
\end{equation}
Note that while planet occurrence is typically expressed as a non-separable function of both planet radius and $P_{\rm orb}$, we treat the two variables as independent in our calculation procedure.

For eclipsing binaries we base the behavior of this distribution on the results of the \kepler\ Eclipsing Binary Catalog \citep{kirk2016kepler}, which contains the properties of thousands of objects that were classified as EBs based on their light curve morphologies. After correcting the catalog for eclipsing binaries that were not detected due to orbital misalignment, we find that $p(P_{\rm orb})$ is best expressed as a broken power law with a break at $P_{\rm orb} = 0.3$ days and the form
\begin{equation}
    p(P_{\rm orb}) \sim \left\{
        \begin{array}{ll}
            P_{\rm orb}^{5.0} & \quad 0.1 \, {\rm days} \leq P_{\rm orb} \leq 0.3 \, {\rm days} \\
            P_{\rm orb}^{0.5} & \quad 0.3 \, {\rm days} < P_{\rm orb} \leq 50 \, {\rm days} \\
        \end{array}
    \right. .
\end{equation}

It is common for validation procedures to also include priors that capture the overall planet occurrence and stellar multiplicity rate. Planet occurrence rate studies have revealed that the probability of a FGKM dwarf hosting a planet with $P_{\rm orb} < 50$ days ranges from $10-100\%$, decreasing as a function of increasing host star mass \citep[e.g.,][]{fressin2013false, petigura2013prevalence, dressing2015occurrence}. Stellar multiplicity rate studies have determined that the probability of a FGKM dwarf hosting a stellar companion with $P_{\rm orb} < 50$ days ranges from $1-10\%$, increasing as a function of increasing host mass \citep{moe2017mind}. This implies that all scenarios involving transiting planets should have a prior probability $10-100 \times$ higher than those involving eclipsing binaries. At first, we included this prior in the algorithm. However, after testing the performance of our tool on known transiting planets and astrophysical false positives (see Section \ref{sec:4}), we concluded that the prior gave transiting planet scenarios too much of an advantage. This advantage often caused an underestimation of ${\rm FPP}$, which led the algorithm to classify astrophysical false positives as transiting planets. To avoid this apparent bias, we omit these priors from our calculation procedure.

\begin{figure*}[ht]
    \centering
    \includegraphics[width=1.0\textwidth]{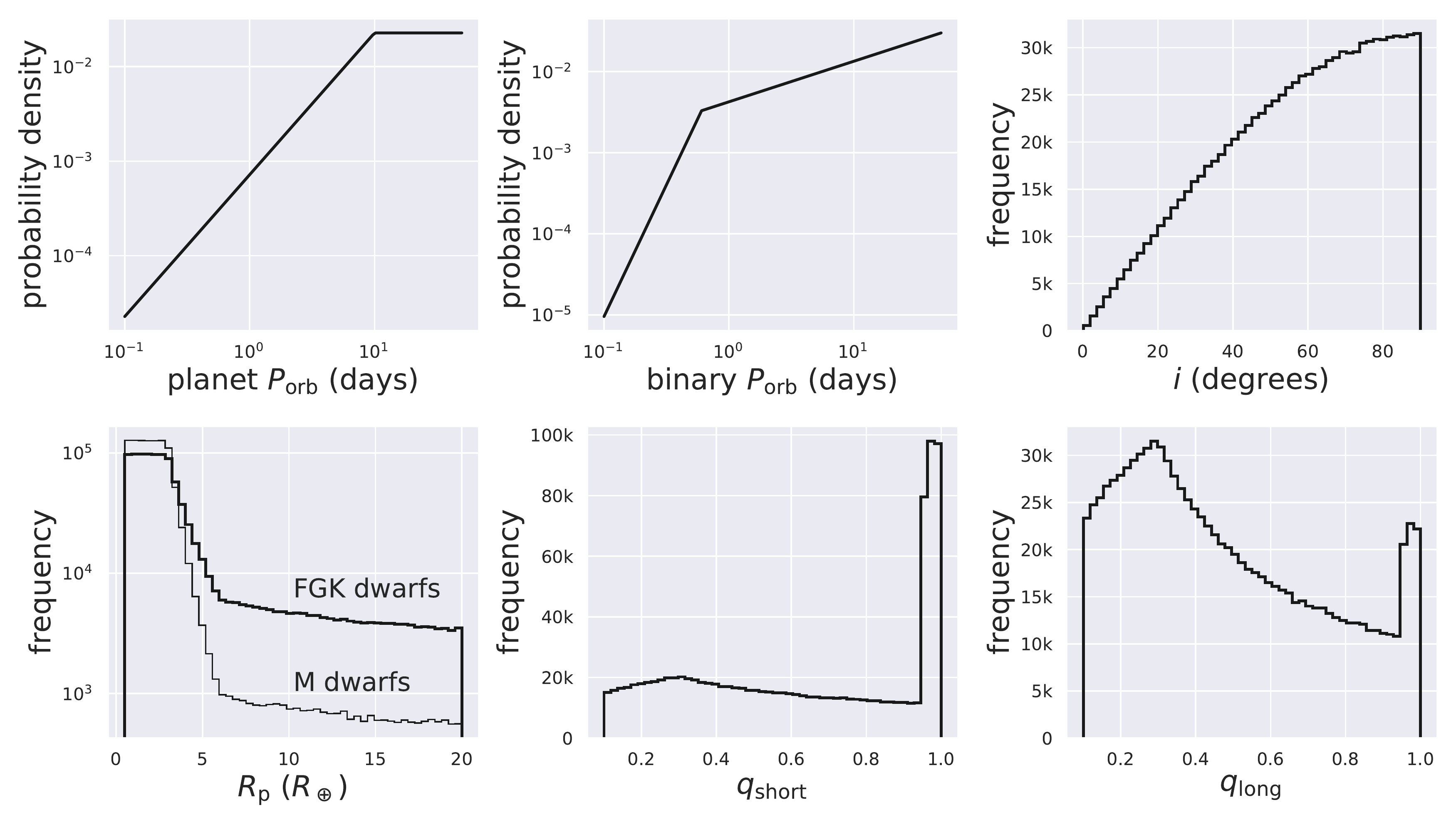}
    \vspace{-15pt}
    \caption{Visualizations of the distributions used to determine model priors and sample parameters in our calculations. \emph{Top left:} The probability density function for the orbital periods of transiting planets. \emph{Top center:} The probability density function for the orbital periods of eclipsing binaries. \emph{Top right:} The parameter prior distribution for inclination. \emph{Bottom left:} The parameter prior distribution for planet radius. \emph{Bottom center:} The parameter prior distribution for short-period stellar companion mass ratio. \emph{Bottom right:} The parameter prior distribution for long-period stellar companion mass ratio. }
    \label{fig:2_3}
\end{figure*}

\subsubsection{Parameter Prior Distributions}\label{sec:dist}

Every scenario we test is associated with a vector $\theta_j$ of parameters that are needed for modeling the light curves of each scenario. The parameters that compose these vectors for each scenario are shown in Table \ref{tab:Scenarios}. To reflect the fact that certain values of these parameters are more common than others, each is associated with a probability distribution. In this section, we define each of these parameters and their respective probability distributions. Examples of these distributions are shown in Figure \ref{fig:2_3} for a sample size of $10^6$.

The parameter $i$ represents the inclination of the orbit of a transiting planet or eclipsing binary. Assuming an isotropic distribution of orbits, the distribution of inclinations takes the form
\begin{equation}
    p(i) \sim \sin{i} .
\end{equation}

The parameter $R_{\rm p}$ represents the radius of a transiting planet. Because this distribution is known to be dependent on host star mass, we use different distributions for M dwarfs and FGK dwarfs. The two distributions differ in the prevalence of giant planets ($R_{\rm p} > 6  R_\oplus$), which are known to be less common around M dwarfs than they are around their more massive counterparts by a factor of $\sim 10$ \citep[e.g.,][]{dressing2013occurrence, fressin2013false, petigura2013prevalence, mulders2015increase}. We express these distributions as broken power laws with breaks at $R_{\rm p} = 3 R_\oplus$ and $R_{\rm p} = 6 R_\oplus$ and a range of $R_{\rm p} = 0.5-20 R_\oplus$ \citep[e.g.,][]{mulders2015increase, mulders2018exoplanet}.\footnote{Note that we do not model the gap in the radius distribution between $1.5-2.0 R_\oplus$ \citep{fulton2017california}.} For M dwarfs the distribution takes the form
\begin{equation}
    p(R_{\rm p}) \sim \left\{
        \begin{array}{ll}
            R_{\rm p}^{0.0} & \quad 0.5 R_\oplus \leq R_{\rm p} \leq 3 R_\oplus \\
            R_{\rm p}^{-7.0} & \quad 3 R_\oplus < R_{\rm p} \leq 6 R_\oplus \\
            R_{\rm p}^{-0.5} & \quad 6 R_\oplus < R_{\rm p} \leq 20 R_\oplus  \\
        \end{array}
    \right.
\end{equation}
and for FGK dwarfs the distribution takes the form
\begin{equation}
    p(R_{\rm p}) \sim \left\{
        \begin{array}{ll}
            R_{\rm p}^{0.0} & \quad 0.5 R_\oplus \leq R_{\rm p} \leq 3 R_\oplus \\
            R_{\rm p}^{-4.0} & \quad 3 R_\oplus < R_{\rm p} \leq 6 R_\oplus \\
            R_{\rm p}^{-0.5} & \quad 6 R_\oplus < R_{\rm p} \leq 20 R_\oplus  \\
        \end{array}
    \right. .
\end{equation}

The parameter $q_{\rm short}$ represents the mass ratio between the host star and a short-period stellar companion (i.e., an eclipsing binary). To calculate this distribution, we extrapolate from the results of \cite{moe2017mind} for Sun-like stars. In the study, $q$ is parameterized as a broken power law with a break at $q=0.3$ and a range of $q = 0.1-1.0$. In addition, the parameterization takes into account the excess of stellar ``twins'' (stellar companions with $q > 0.95$) with a term $\mathcal{F}_{\rm twin}$ (defined as the fraction of stars with $q > 0.3$ that have $q > 0.95$) that boosts the prevalence of these stars in the probability distribution. For short-period stellar companions, the distribution takes the form
\begin{equation}
    p(q_{\rm short}) \sim \left\{
        \begin{array}{ll}
            q_{\rm short}^{0.3} & \quad 0.1 \leq q \leq 0.3 \\
            q_{\rm short}^{-5.0} & \quad 0.3 < q \leq 1.0 \\
        \end{array}
    \right.
\end{equation}
with $\mathcal{F}_{\rm twin} = 0.3$.

The parameter $q_{\rm long}$ represents the mass ratio between the target star and a long-period stellar companion (i.e., an unresolved bound companion). Again, we utilize the parameterization and extrapolate results of \cite{moe2017mind} for Sun-like stars. For long-period stellar companions, the distribution takes the form
\begin{equation}
    p(q_{\rm long}) \sim \left\{
        \begin{array}{ll}
            q_{\rm long}^{0.3} & \quad 0.1 \leq q \leq 0.3 \\
            q_{\rm long}^{-0.95} & \quad 0.3 < q \leq 1.0 \\
        \end{array}
    \right.
\end{equation}
with $\mathcal{F}_{\rm twin} = 0.05$.

The parameter ``simulated star'' represents the properties of a star drawn from a population of stars simulated with TRILEGAL. To determine the properties of blended stars used in DTP, DEB, DEBx2P, BTP, BEB, and BEBx2P scenarios, we simulate a population of stars in a 0.1 deg$^2$ region of the sky centered at the target star. We then produce a distribution of possible foreground/background stars by removing all stars with \tess\ magnitudes brighter than the target and fainter than 21, which typically yields between $300-1000$ stars. When simulating an instance of these scenarios, we draw a star directly from this distribution.

\subsubsection{Marginal Likelihoods}

Because the integral in Equation \ref{eq:marg} is typically impossible to solve analytically, it is common to approximate the integral by sampling $p(\theta_j | S_j)$. This is, in fact, what is done when calculating odds ratios between competing scenarios in the \texttt{PASTIS} and \texttt{VESPA} validation procedures. In this work, we calculate the marginal likelihood using Arithmetic Mean Estimation \citep{kass1995bayes}. This method allows us to calculate the marginal likelihood using Monte Carlo sampling by approximating Equation \ref{eq:marg} as
\begin{equation}
    p(D | S_j) \sim \frac{1}{N} \sum_{n=1}^{N} p(D | \theta_j^{(n)}, S_j)
\end{equation}
where $\theta_j^{(n)}$ is the $n$th sample from the parameter prior distribution and $N$ is the total number of samples. This is typically regarded as the simplest estimator of the marginal likelihood, but it is often avoided because it can produce a large variance in $p(D | S_j)$ if $N$ is not sufficiently high and is relatively inefficient when integrating over a large number of parameters. We take two approaches to combat these drawbacks: (1) we chose a $N$ high enough to produce results that are consistent between consecutive calculations (which we determine to be $N=10^6$), and (2) we make simplifying assumptions in our transiting planet and eclipsing binary models that minimize the number of parameters we must marginalize over.

The first simplifying assumption we make is to assume that the $M_\star$, $R_\star$ and $T_{\rm eff}$ of each resolved star is known precisely. Unless the user provides these parameters, they are assumed to be equal to those listed in the TIC. In addition, any other stars added to our transit model that do not have estimates for these quantities (e.g., eclipsing binaries or unresolved companions) are assumed to be precisely characterized based on their $M_\star$ (see Section \ref{sec:splines}). Because the transit models are sensitive to these parameters, this assumption saves us from having to marginalize over a distribution of target star properties.

The second simplifying assumption we make is to assume a fixed orbital period and zero eccentricity ($e$) in all scenarios considered, which significantly simplifies the orbital solution of the system. There is strong evidence that short-period planets are biased towards lower $e$ \citep[e.g.,][]{kane2012exoplanet,kipping2013parametrizing,shabram2016eccentricity}. According to the NASA Exoplanet Archive,\footnote{\url{https://exoplanetarchive.ipac.caltech.edu/}} 84$\%$ of confirmed planets with $P_{\rm orb} < 30$ days and reported eccentricities have $e < 0.2$.  The same justification can be applied to short-period eclipsing binaries. \cite{moe2017mind}, showed that the $e$ distribution of binary stars with $P_{\rm orb} < 10$ days goes like $e^{-0.8}$. This implies that 72$\%$ of short-period eclipsing binaries have $e < 0.2$. Because a majority of TOIs will have $P_{\rm orb} < 30$ days (due to the $\sim 27$ day intervals in which sectors are observed and the general requirement for at least 2 transits be observed for a system to become a planet candidate), we believe the assumption of circular orbits is justified in most cases. However, users of \texttt{TRICERATOPS} should be aware that this assumption becomes less valid as longer orbital periods are considered.

We calculate $p(D | \theta_j^{(n)}, S_j)$ as the product of two terms:
\begin{equation}
    p(D | \theta_j^{(n)}, S_j) = p(D_{\rm tra} | \theta_j^{(n)}, S_j) \times w^{(n)}
\end{equation}
where the first term is the likelihood of the transit data and $w^{(n)}$ is a weight that encapsulates our ability to rule out unresolved companions near the target star using high-resolution imaging follow-up. This weight is intended to decrease the likelihood of scenarios involving unresolved companions when stronger constraints on the existence such companions are applied.  

The likelihood of the transit data is calculated using the equation
\begin{equation}\label{eq:likelihood}
    p(D_{\rm tra} | \theta_j^{(n)}, S_j) \propto \prod exp \left[ -\frac{1}{2} \left( \frac{y_l - f(t_l|\theta_j^{(n)})}{\sigma} \right)^2 \right]
\end{equation}
where $y_l$ is the flux of the $l$th data point, $f(t_l|\theta_j^{(n)})$ is the flux given by the model for the parameter vector $\theta_j^{(n)}$ at the time of the $l$th data point, and $\sigma$ is the characteristic uncertainty of the flux.

For PTP, PEB, PEBx2P, STP, SEB, and SEBx2P scenarios we calculate $w^{(n)}$ using Equation 23 of \cite{moe2017mind}. Equation 23 of \cite{moe2017mind} provides the frequency of bound stellar companions as a function of primary mass and orbital period. We calculate this quantity for the $n$th sample of the parameter prior distribution using the following steps: (1) determine magnitude difference between the primary and secondary star using the mass of the target and the $n$th draw of $q_{\rm long}$, (2) use the contrast curve obtained from high-resolution imaging to determine the angular separation beyond which the simulated secondary would have been detected, (3) convert this angular separation to an orbital period using the parallax of the target and the masses of the target and simulated secondary, and (4) use this orbital period and Equation 23 of \cite{moe2017mind} to calculate the corresponding frequency of bound stellar companions. If no high-resolution imaging data is available to fold in, the angular separation used in step (2) is assumed to be $2\farcs2$ \citep{brown2018gaia}.

For DTP, DEB, DEBx2P, BTP, BEB, and BEBx2P scenarios we calculate $w^{(n)}$ using the results of the TRILEGAL simulation discussed in Section \ref{sec:dist}. Specifically, we calculate this likelihood as the frequency of unresolved foreground and background stars aligned with the target star in the sky. This calculation is performed with the following steps: (1) determine the magnitude difference between the target star and the $n$th drawn foreground/background star, (2) use the contrast curve obtained from high-resolution imaging to determine the angular separation beyond which the simulated foreground/background star would have been detected, (3) use this separation and the total number of simulated stars to estimate the frequency of unresolved foreground/background stars near the target. As for the previous scenarios, if no high-resolution imaging data is available to fold in, the angular separation used in step (2) is assumed to be $2\farcs2$ \citep{brown2018gaia}.

We set the maximum value of $w^{(n)}$ for each scenario to 1. We also set $w^{(n)}=1$ for TP, EB, EBx2P, NTP, NEB, and NEBx2P scenarios, which do not involve unresolved companions. 

\subsubsection{Light Curve Modeling}

We calculate Equation \ref{eq:likelihood} by modeling light curves using a modified version of \texttt{batman} \citep{kreidberg2015batman}. Here, we describe the steps that go into simulating the transits of each scenario.

The simplest scenario to model is the TP scenario, in which we assume that all of the flux originates from the host star. For this scenario, we use \texttt{batman} in its default form. For this scenario, as well as all other scenarios, we use quadratic limb darkening coefficients chosen based on the $T_{\rm eff}$ and $\log g$ of the host star \citep{claret2018new}.

\begin{figure}[t!]
        \begin{center}
            \includegraphics[width=0.48\textwidth]{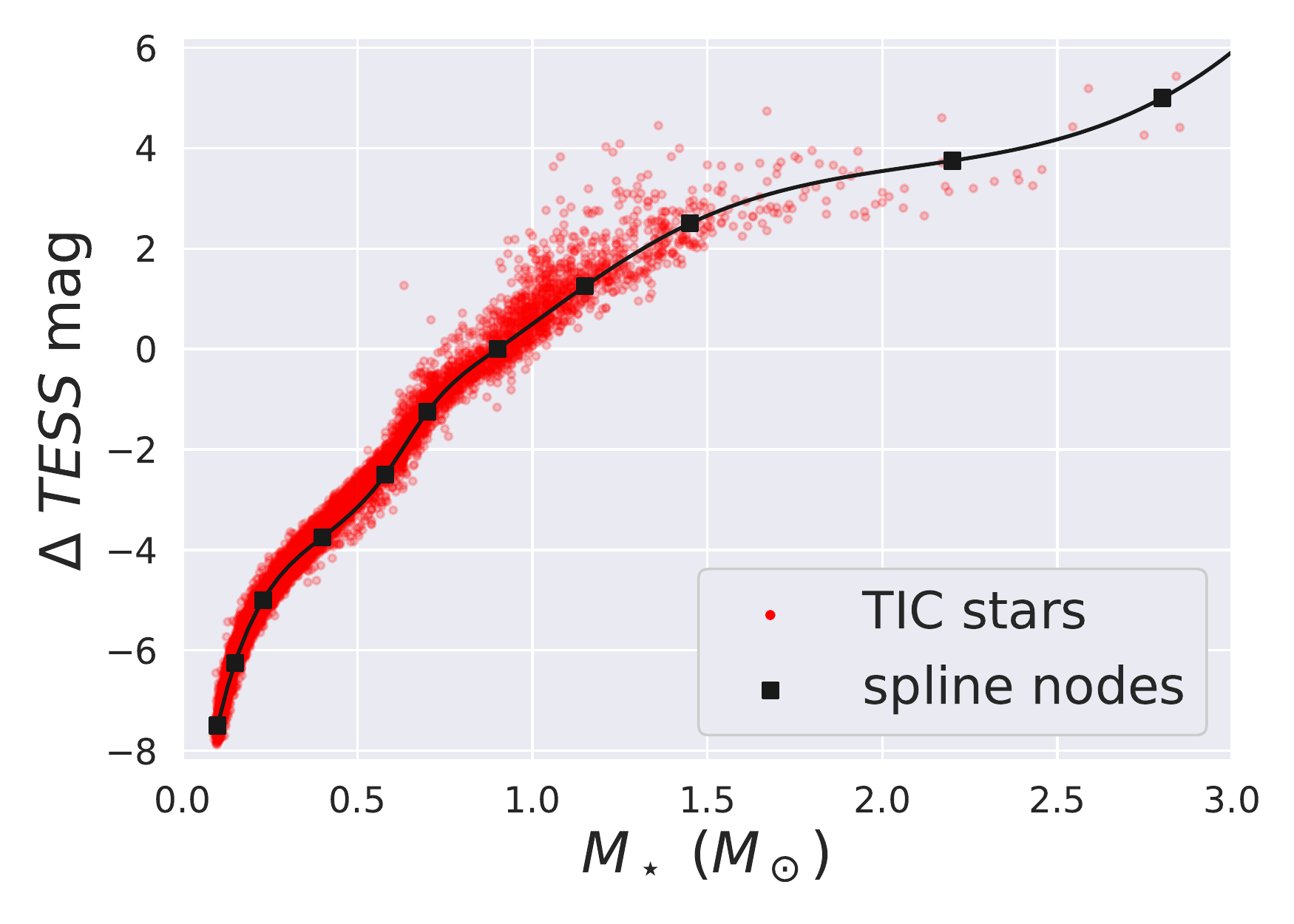}
        \end{center}
    \vspace{-15pt}
    \caption{$\Delta$ \tess\ magnitude between a star of mass $M_\star$ and a 10th magnitude, $1 M_\odot$ star. Red points are stars queried from the TIC located between $99-101$ pc away. Black squares are the nodes of the spline relation used to calculate the \tess\ mag of unresolved stars modeled in our calculations.}
    \label{fig:2_4}
\end{figure}

For all scenarios involving eclipsing binaries, we must account for the fact that the flux is split between the host star and the short-period companion. Doing so requires an estimate for the flux contributed by the eclipsing binary, which we find by determining a relation between $M_\star$ and \tess\ magnitude. We begin by querying the TIC for all stars located a distance between $99-101$ pc away. We then draw a spline curve through the distribution of points in the \tess\ magnitude -- $M_\star$ plane, which is shown in Figure \ref{fig:2_4}. This relation allows us to calculate the \tess\ band flux ratio between two stars given their masses and adjust the in-transit flux of the light curve accordingly.

For scenarios involving unresolved companions, we again must account for the flux dilution from the additional star. For scenarios involving a unresolved bound companion (whose mass is determined by $q_{\rm long}$), we use the spline relation shown in Figure \ref{fig:2_4} to determine the flux contribution of the star. For scenarios involving an unresolved foreground/background star, we use the \tess\ magnitude provided by TRILEGAL to determine the flux contribution of the star.

Lastly, we apply constraints to our transit models for all ``EB'' and ``EBx2P'' scenarios. For the former, we require $q_{\rm short} < 0.95$ and for the expected secondary eclipse depth to be shallower than $1.5 \times$ the scatter of the \tess\ light curve flux (else the secondary eclipse would have been detected and identified as such). For the latter, we require $q_{\rm short} > 0.95$. If the $n$th model light curve does not satisfy these conditions, we set the likelihood of the transit to zero. 

\section{Examples}\label{sec:3}

For illustrative purposes we display here each step of our calculation for two TOIs, one of which has been confirmed as a transiting planet and one of which has been ruled out as a nearby eclipsing binary.

\subsection{TIC 270380593 (TOI 465.01)}\label{subsec:example}

We apply our algorithm on the previously-confirmed TOI 465.01 (WASP-156b, \citealt{demangeon2018discovery}), a $\sim 6 R_\oplus$ planet orbiting a K dwarf with a 3.84 day orbital period. The host star, which has a \tess\ magnitude of 10.73 and is located 122 pc away, was observed with a 2-minute cadence in sector 4.

\begin{figure*}[!ht]
    \centering
    \includegraphics[width=\textwidth]{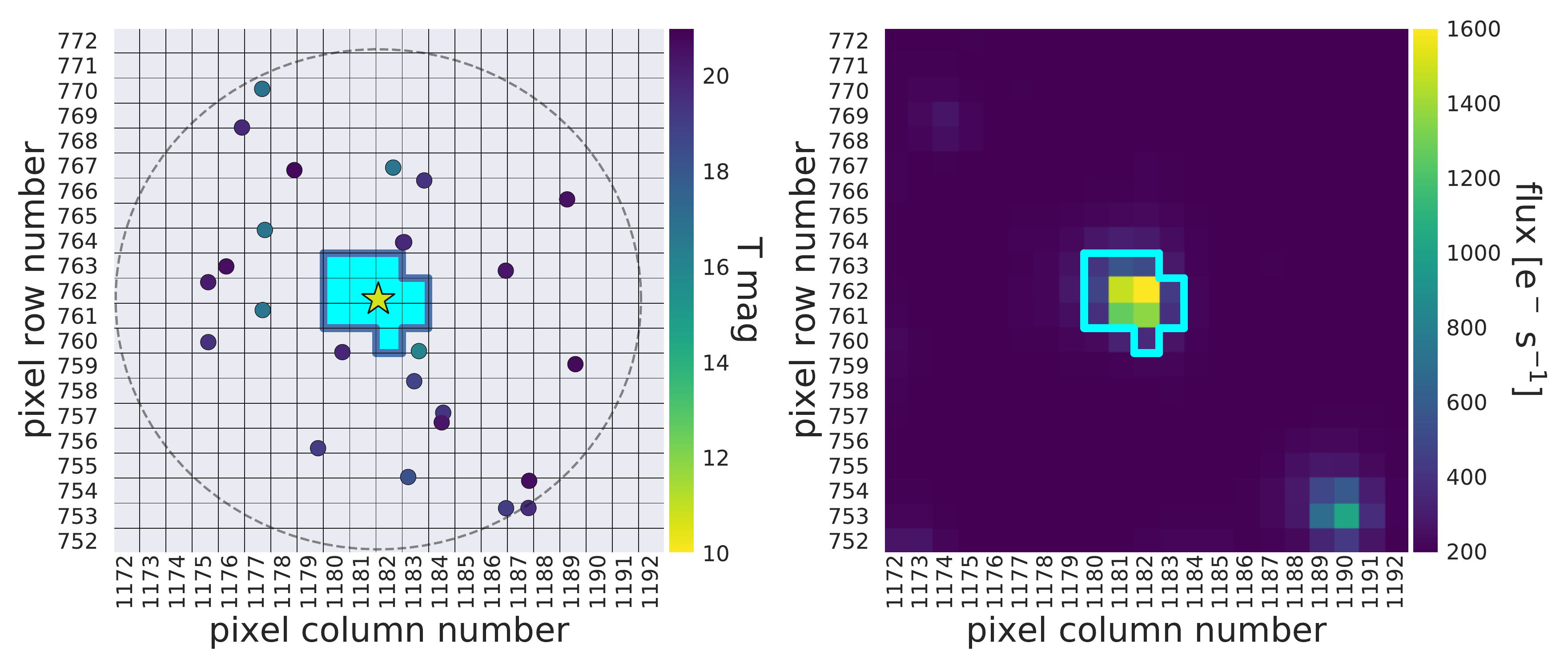}
    \vspace{-15pt}
    \caption{Visualization of TIC querying for TOI 465.01 (TIC 270380593). Left: All stars within 10 pixels of the target star (the limits of which are approximated by the black dashed line). The target star is located in the center pixel and is indicated by a star symbol. The aperture used to extract the light curve is highlighted in blue. Right: Time-averaged \tess\ image of the same pixels, with the same aperture overlaid.}
    \label{fig:3_1}
\end{figure*}

We begin by searching for all other stars within 10 pixels of the target star. This is shown in Figure \ref{fig:3_1}, where the location of each nearby star relative to the local \tess\ pixels is shown on the left and the corresponding \tess\ image is shown on the right. Next, we calculate the flux contribution of each star and determine which contribute enough flux to the aperture to produce a transit with the reported depth. In this case, the target star is the only star bright enough to host the signal. We therefore ignore NTP, NEB, and NEBx2P scenarios for the remainder of this analysis, which leaves 15 scenarios to be considered.

% \newpage

\begin{deluxetable*}{cccccccccc}[t!]
\tabletypesize{\footnotesize}
\tablewidth{\textwidth}
 \tablecaption{ Scenario Probabilities for TOI 465.01  \label{tab:3_1}}
 \tablehead{
 \colhead{Scenario} & \colhead{TIC ID} & \colhead{$M_\star \, (M_\odot)$} & \colhead{$R_\star \, (R_\odot)$} & \colhead{$P_{\rm orb}$ (days)} & \colhead{$i$ (deg)} & \colhead{$R_{\rm p}$ ($R_\oplus$)} & \colhead{$R_{\rm EB}$ ($R_\odot$)} & \colhead{$\mathcal{P}_j$} & \colhead{$\mathcal{P}_j$ with AO}
 }
 \startdata 
 TP     & 270380593 & 0.81 & 0.85 & 3.84 & 87.3 & 6.27  &      & 0.39    & 0.61     \\
 EB     & 270380593 & 0.81 & 0.85 & 3.84 & 85.3 &       & 0.10 & $<0.01$ & $<0.01$  \\
 EBx2P  & 270380593 & 0.81 & 0.85 & 7.67 & 85.3 &       & 0.84 & $<0.01$ & $<0.01$  \\
 PTP    & 270380593 & 0.81 & 0.85 & 3.84 & 87.4 & 6.35  &      & 0.22    & 0.14     \\
 PEB    & 270380593 & 0.81 & 0.85 & 3.84 & 86.4 &       & 0.10 & $<0.01$ & $<0.01$  \\
 PEBx2P & 270380593 & 0.81 & 0.85 & 7.67 & 85.4 &       & 0.83 & $<0.01$ & $<0.01$  \\
 STP    & 270380593 & 0.79 & 0.82 & 3.84 & 87.8 & 8.71  &      & 0.31    & 0.19     \\
 SEB    & 270380593 & 0.63 & 0.65 & 3.84 & 89.8 &       & 0.10 & 0.01    & $<0.01$  \\
 SEBx2P & 270380593 & 0.48 & 0.49 & 7.67 & 87.3 &       & 0.49 & $<0.01$ & $<0.01$  \\
 DTP    & 270380593 & 0.81 & 0.85 & 3.84 & 87.5 & 6.26  &      & 0.06    & 0.06     \\
 DEB    & 270380593 & 0.81 & 0.85 & 3.84 & 85.7 &       & 0.10 & $<0.01$ & $<0.01$  \\
 DEBx2P & 270380593 & 0.81 & 0.85 & 7.67 & 85.3 &       & 0.83 & $<0.01$ & $<0.01$  \\
 BTP    & 270380593 & 0.55 & 0.48 & 3.84 & 89.3 & 19.36 &      & $<0.01$ & $<0.01$  \\
 BEB    & 270380593 & 0.81 & 0.75 & 3.84 & 89.7 &       & 0.19 & $<0.01$ & $<0.01$  \\
 BEBx2P & 270380593 & 0.83 & 1.01 & 7.67 & 85.4 &       & 0.85 & $<0.01$ & $<0.01$  \\ \hline
 TIC\tablenotemark{a} & 270380593 & $0.81^{+0.10}_{-0.10}$  & $0.85^{+0.06}_{-0.06}$ &  &  &  &  &  \\
 WASP-156b\tablenotemark{b} & 270380593 & $0.84^{+0.05}_{-0.05}$  & $0.76^{+0.03}_{-0.03}$ &  &  $89.1^{+0.6}_{-0.9}$ & $5.72^{+0.22}_{-0.22}$ &  & &
 \enddata
%  \vspace{-0.5cm}
 \tablenotetext{a}{Host star properties from version 8 of the TIC \citep{stassun2018tess}.}
 \tablenotetext{b}{Best-fit host star and planet properties from \cite{demangeon2018discovery}.}
 \vspace{-25pt}
\end{deluxetable*}

\begin{figure*}[!ht]
        \begin{center}
            \includegraphics[width=1.0\textwidth]{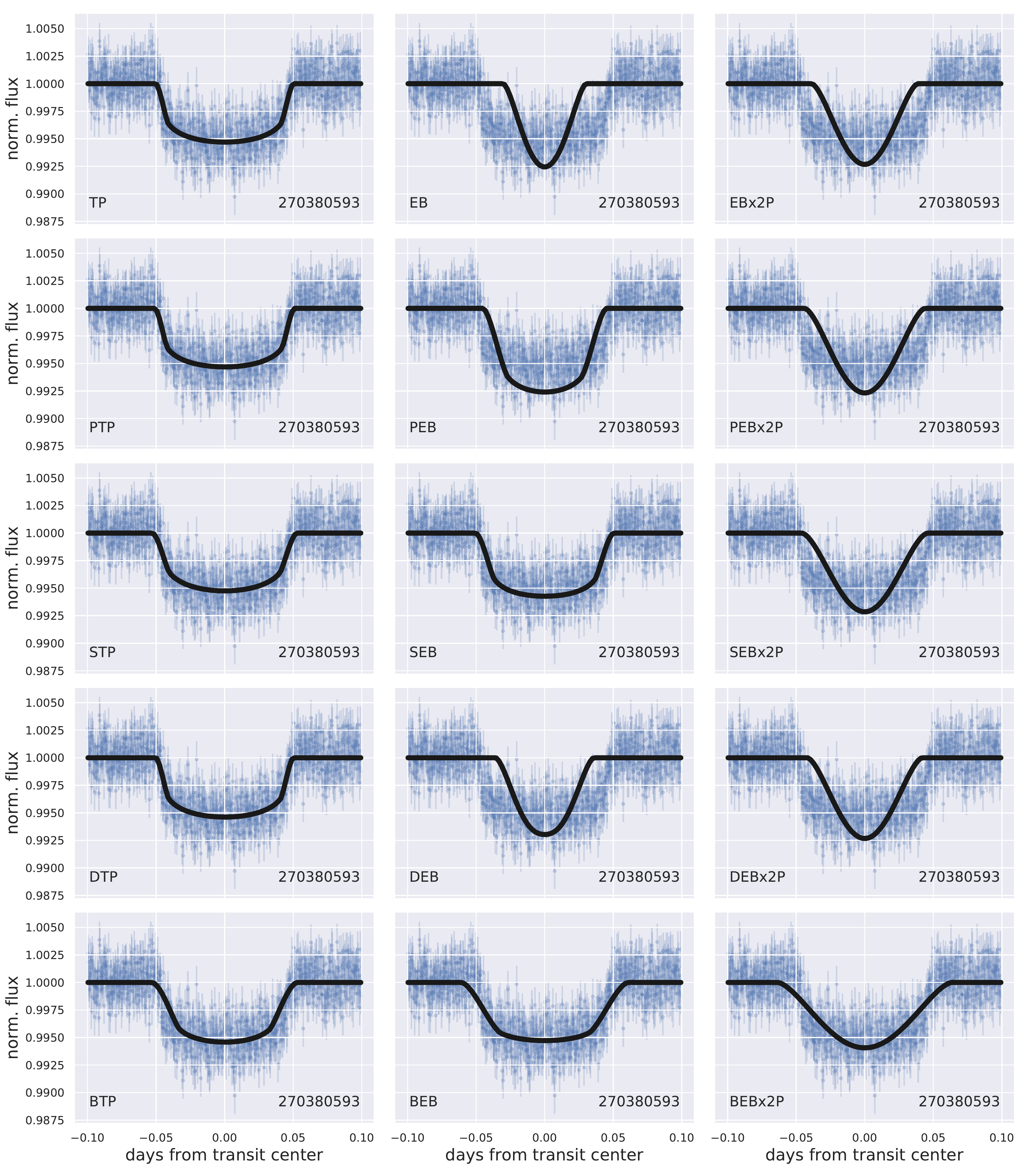}
        \end{center}
    \vspace{-15pt}
    \caption{Fit of each transit scenario for TOI 465.01. The purple points are 2-minute cadence \tess\ data, while the black curves are the best-fit light curves. The scenario being fit for is in the bottom left of each panel, and the TIC ID of the star being fit for is in the bottom right of each panel.}
    \label{fig:3_3}
\end{figure*}

\clearpage

\begin{figure}[t]
    \centering
    \includegraphics[width=0.48\textwidth]{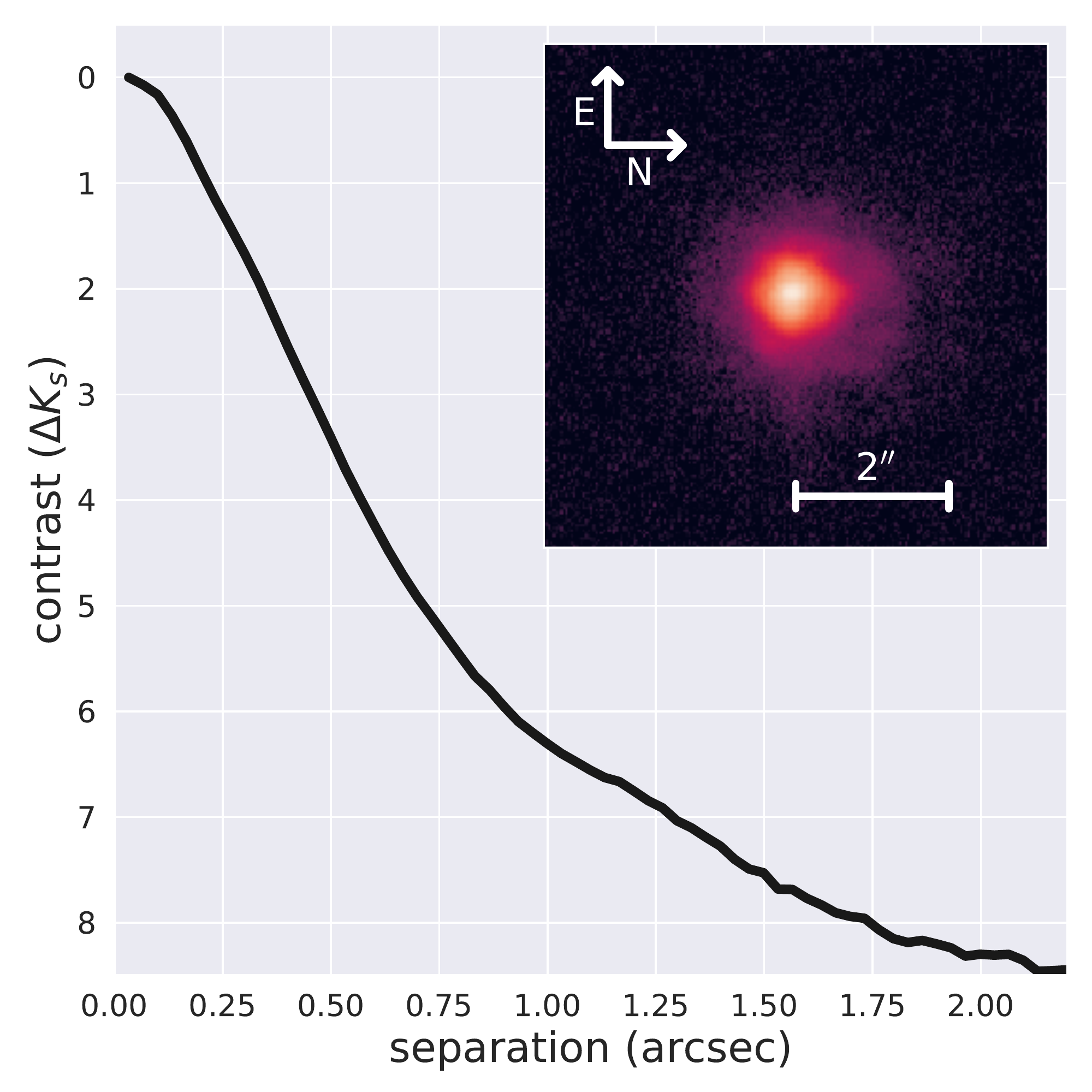}
    \vspace{-15pt}
    \caption{High-resolution image of TOI 465 obtained with ShARCS/ShaneAO in $K_s$ band and corresponding contrast curve.}
    \label{fig:3_2}
\end{figure}

Next, we determine the best-fit model parameters for each of the 15 scenarios considered. The results of this step are displayed in Figure \ref{fig:3_3} and Table \ref{tab:3_1}. Figure \ref{fig:3_3} shows the best-fit transit models for each scenario compared to the extracted \tess\ light curve. Table \ref{tab:3_1} shows the best-fit values for several transit model parameters. We see in both of these that the best-fitting scenario is the TP scenario.

The final step in the procedure is to calculate the relative probability of each scenario using Equation \ref{prob}. These probabilities are shown in the right-most columns of Table \ref{tab:3_1}. For this TOI, we find that ${\rm FPP} = 0.33$ and ${\rm NFPP} = 0.0$.

The above calculation was done assuming unresolved companions near the target star can be ruled out beyond $2\farcs2$. However, if one is able to further constrain the separation beyond which an unresolved star could exist, this number can be decreased to that new separation. On 2019 July 10, we obtained adaptive optics (AO) assisted high-resolution images of this TOI with ShARCS/ShaneAO on the Shane 3-meter telescope at Lick Observatory, shown in Figure \ref{fig:3_2}. These images were reduced using the steps outlined in \citet{hirsch2019discovery} and Savel et al. (in prep), which we refer the reader to for more information. With these observations, we produce a contrast curve (also shown in Figure \ref{fig:3_2}) that can be folded in to the \texttt{TRICERATOPS} analysis in order to further constrain the probabilities of scenarios involving unresolved companions. 

To show how this changes the results of our tool, we repeat the calculation with this constraint applied. The impact that this AO follow-up has on the probability of each scenario is shown in the right-most column of Table \ref{tab:3_1}, which now yields ${\rm FPP} = 0.19$.

\subsection{TIC 438490744 (TOI 529.01)}

We also apply our algorithm on TOI 529.01, a candidate with a 1.67 day orbital period that has been ruled out as a NEB around the nearby star TIC 438490748 (see Section \ref{sec:5} for more details). The originally proposed host star is an M dwarf with a \tess\ magnitude of 14.14 and a distance of 63 pc away. This TOI was observed with a 2-minute cadence in sector 6.

\begin{figure*}[!ht]
    \centering
    \includegraphics[width=\textwidth]{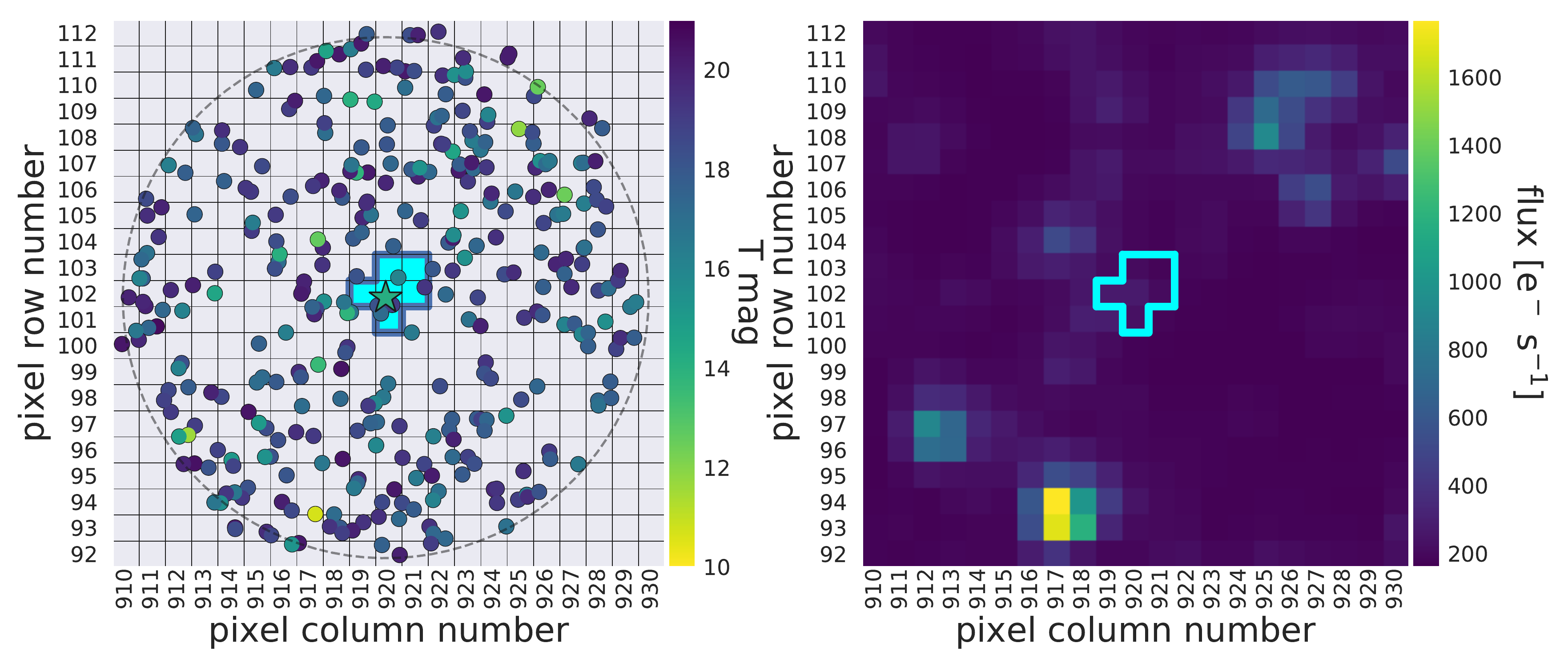}
    \vspace{-15pt}
    \caption{Visualization of TIC querying for TOI 529.01 (TIC 438490744). Left: All stars within 10 pixels of the target star (the limits of which are approximated by the black dashed line). The target star is located in the center pixel and is indicated by a star symbol. The aperture used to extract the light curve is highlighted in blue. Right: Time-averaged \tess\ image of the same pixels, with the same aperture overlaid.}
    % \vspace{40pt}
    \label{fig:3_4}
\end{figure*}

We again begin by searching for all other stars within 10 pixels of the target star, as shown in Figure \ref{fig:3_4}. After calculating the flux contribution due to each star, it is determined that two nearby stars, TIC 438490736 and TIC 438490748, contribute enough light to the aperture for them to host the observed transit. As a result, there are 21 scenarios to be considered for this TOI.

\begin{figure*}[b]
        \begin{center}
            \includegraphics[width=1.0\textwidth]{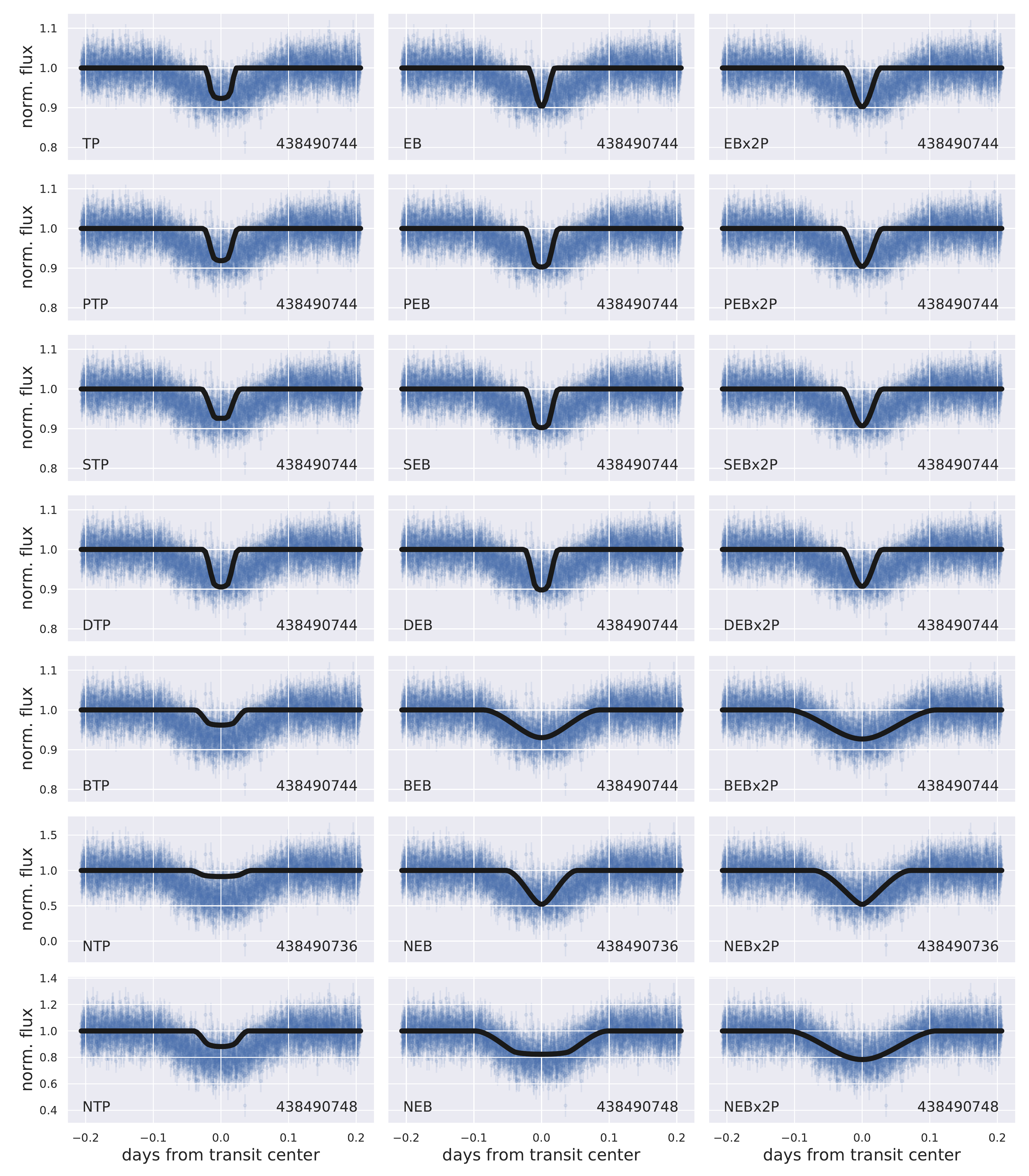}
        \end{center}
    \vspace{-15pt}
    \caption{Fit of each transit scenario for TOI 529.01. The purple points are 2-minute cadence \tess\ data, while the black curves are the best-fit light curves. The scenario being fit for is in the bottom left of each panel, and the TIC ID of the star being fit for is in the bottom right of each panel.}
    \label{fig:3_5}
\end{figure*}

Figure \ref{fig:3_5} and Table \ref{tab:3_2} show the best-fit transits and transit model parameters for these scenarios, respectively. According to these results, the most probable scenario is the NEBx2P scenario around the nearby star TIC 438490748. In fact, the preference for this scenario is so strong that this TOI has ${\rm FPP} > 0.99$ and ${\rm NFPP} > 0.99$.

\section{Planet Vetting and Validation}\label{sec:4}

In this section, we analyze the performance of \texttt{TRICERATOPS} by running it on several classified TOIs observed with both 2-minute cadence and 30-minute cadence observations. Using these results, we define the conditions a TOI must meet to be vetted and validated.

\subsection{2-minute Cadence Data}

We begin by running our code on TOIs identified in 2-minute cadence data collected by \tess. In the first two years of the \tess\ mission, these observations were collected for $\sim $200,000 nearby dwarfs stars across nearly the entire sky. These observations are processed by the \tess\ Science Processing Operations Center (SPOC) pipeline \citep{jenkins2016tess}, which identifies TCEs and generates data validation reports that contain information useful for further vetting. These stars are then subjected to manual vetting by the \tess\ Science Office to compile a set of TOIs that consist of the TCEs with the best chances of being actual planets.

% \newpage

\begin{deluxetable*}{ccccccccc}[!ht]
\tabletypesize{\footnotesize}
\tablewidth{\textwidth}
 \tablecaption{ Scenario Probabilities for TOI 529.01  \label{tab:3_2}}
 \tablehead{
 \colhead{Scenario} & \colhead{TIC ID} & \colhead{$M_\star \, (M_\odot)$} & \colhead{$R_\star \, (R_\odot)$} & \colhead{$P_{\rm orb}$ (days)} & \colhead{$i$ (deg)} & \colhead{$R_{\rm p}$ ($R_\oplus$)} & \colhead{$R_{\rm EB}$ ($R_\odot$)} & \colhead{$\mathcal{P}_j$}
 }
 \startdata 
 TP     & 438490744 & 0.21 & 0.24 & 1.67 & 89.9 & 6.89  &      & $<0.01$  \\
 EB     & 438490744 & 0.21 & 0.24 & 1.67 & 86.6 &       & 0.10 & $<0.01$  \\
 EBx2P  & 438490744 & 0.21 & 0.24 & 3.33 & 87.3 &       & 0.24 & $<0.01$  \\
 PTP    & 438490744 & 0.21 & 0.24 & 1.67 & 90.0 & 8.61  &      & $<0.01$  \\
 PEB    & 438490744 & 0.21 & 0.24 & 1.67 & 89.5 &       & 0.10 & $<0.01$  \\
 PEBx2P & 438490744 & 0.21 & 0.24 & 3.33 & 87.7 &       & 0.24 & $<0.01$  \\
 STP    & 438490744 & 0.09 & 0.10 & 1.67 & 89.2 & 19.70 &      & $<0.01$  \\
 SEB    & 438490744 & 0.18 & 0.22 & 1.67 & 89.7 &       & 0.10 & $<0.01$  \\
 SEBx2P & 438490744 & 0.48 & 0.24 & 3.33 & 87.7 &       & 0.24 & $<0.01$  \\
 DTP    & 438490744 & 0.21 & 0.24 & 1.67 & 89.2 & 9.82  &      & $<0.01$  \\
 DEB    & 438490744 & 0.21 & 0.24 & 1.67 & 89.5 &       & 0.10 & $<0.01$  \\
 DEBx2P & 438490744 & 0.21 & 0.24 & 3.33 & 87.7 &       & 0.24 & $<0.01$  \\
 BTP    & 438490744 & 0.51 & 0.45 & 1.67 & 89.8 & 19.92 &      & $<0.01$  \\
 BEB    & 438490744 & 1.05 & 1.42 & 1.67 & 89.6 &       & 1.05 & $<0.01$  \\
 BEBx2P & 438490744 & 0.93 & 1.67 & 3.33 & 84.4 &       & 0.97 & $<0.01$  \\ 
 NTP    & 438490736 & 0.67 & 0.69 & 1.67 & 89.5 & 19.94 &      & $<0.01$  \\
 NEB    & 438490736 & 0.67 & 0.69 & 1.67 & 88.1 &       & 0.56 & $<0.01$  \\
 NEBx2P & 438490736 & 0.67 & 0.69 & 3.33 & 89.5 &       & 0.69 & $<0.01$  \\
 NTP    & 438490748 & 0.51 & 0.45 & 1.67 & 89.7 & 19.98 &      & $<0.01$  \\
 NEB    & 438490748 & 1.12 & 1.75 & 1.67 & 89.8 &       & 0.76 & 0.06     \\
 NEBx2P & 438490748 & 1.08 & 1.54 & 3.33 & 85.2 &       & 1.16 & 0.94     \\ \hline
 TIC\tablenotemark{a} & 438490744 & $0.21^{+0.02}_{-0.02}$  & $0.24^{+0.01}_{-0.01}$ &  &  &  &  &  \\
 \enddata
%  \vspace{-0.5cm}
 \tablenotetext{a}{Host star properties from version 8 of the TIC \citep{stassun2018tess}.}
%  \vspace{10pt}
\end{deluxetable*}

\clearpage

\begin{figure*}[!ht]
    \includegraphics[width=0.5\textwidth]{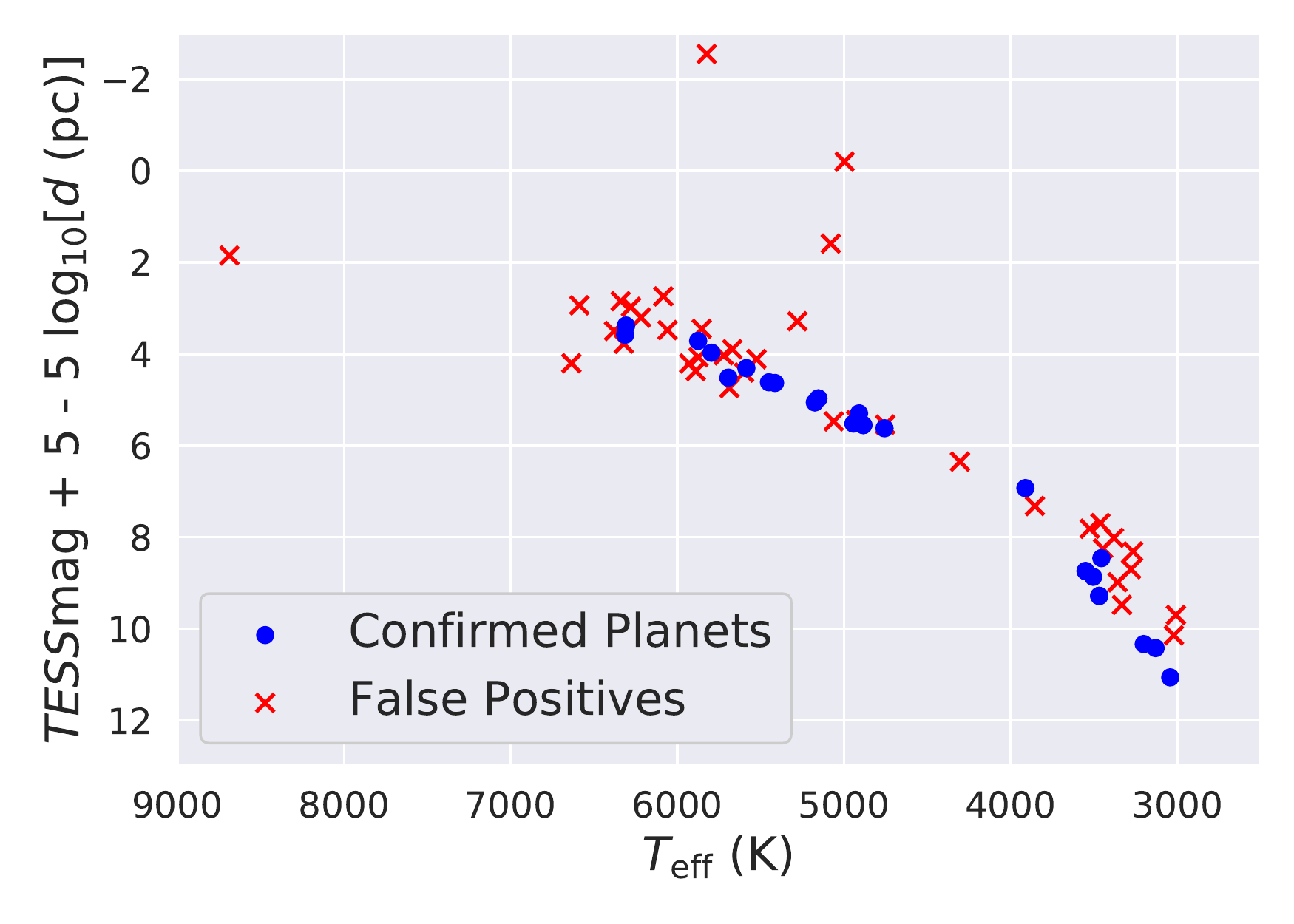}
    \includegraphics[width=0.5\textwidth]{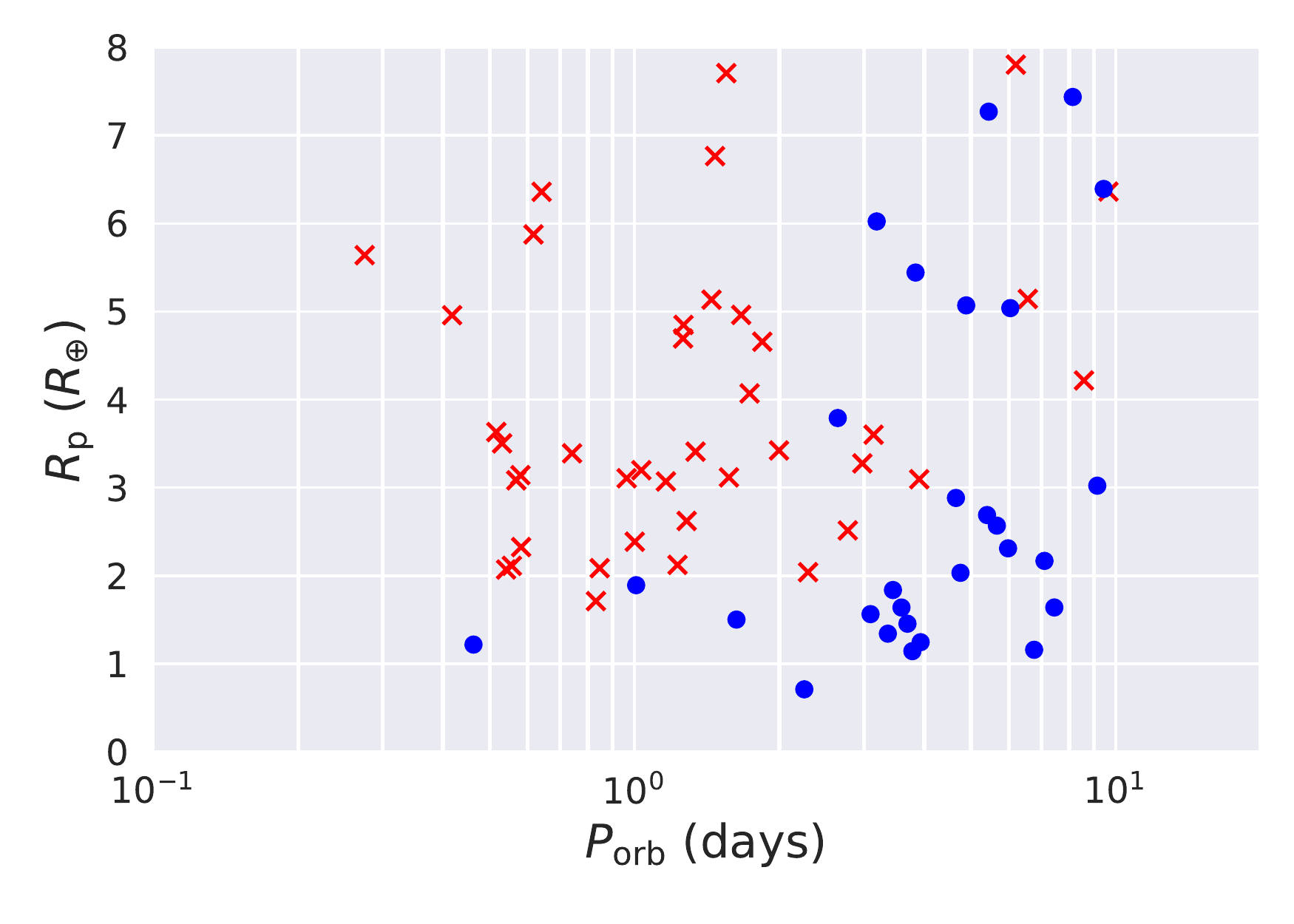}
    \vspace{-15pt}
    \caption{Host star (\emph{left}) and planet (\emph{right}) properties of confirmed planets and false positives used in our performance analysis. The sample includes systems with a diversity of host spectral types, planet orbital periods, and predicted planet radii (i.e., the best-fit radii from the TP scenario).}
    \label{fig:4_1_1}
\end{figure*}

\begin{figure*}[!ht]
    \includegraphics[width=1.02\textwidth]{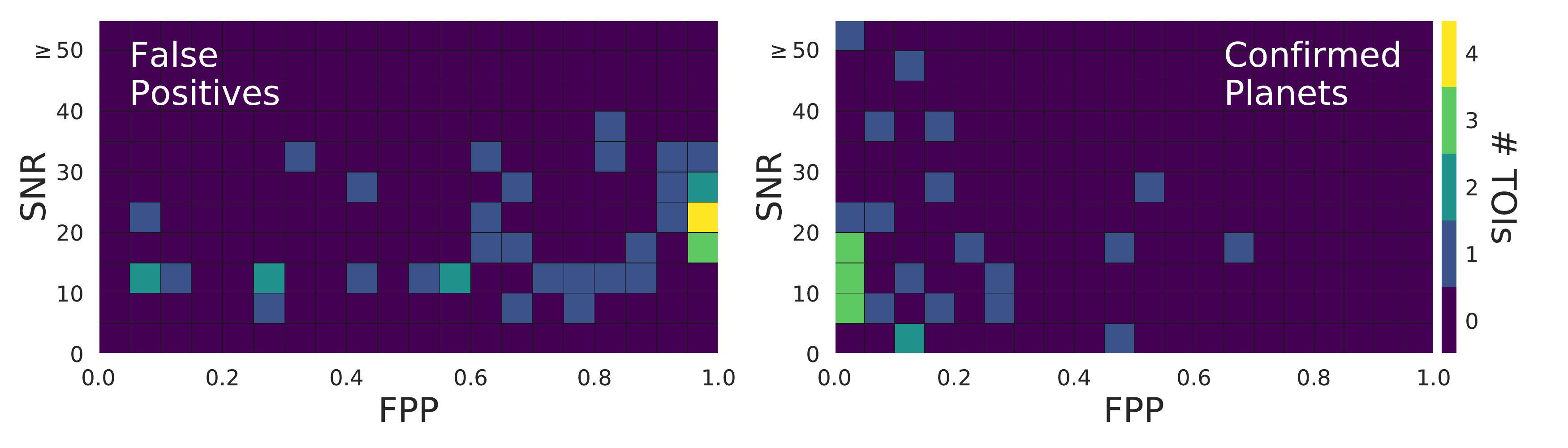}
    \vspace{-15pt}
    \caption{SNR vs FPP for all false positives (\emph{left}) and confirmed planets (\emph{right}) used in our performance analysis. Our tool performs better for TOIs with higher SNRs. \texttt{TRICERATOPS} performs best when ${\rm SNR} > 15$.}
    \label{fig:4_1_2}
\end{figure*}

We use publicly available information from the \tess\ Follow-up Observation Program (TFOP) website\footnote{https://exofop.ipac.caltech.edu/tess/index.php} and 2-minute cadence \tess\ light curves from MAST to obtain the phase-folded light curves and apertures that we input into \texttt{TRICERATOPS} for each TOI. Because a key function of our algorithm is the identification of TOIs that are false positives around nearby stars, we use light curves extracted using simple aperture photometry instead of those processed with pre-data-conditioning step of the SPOC pipeline, which removes contamination and variability originating from nearby stars. In order to recreate the conditions under which one would use our tool on new TOIs, we only use data from the first sector in which each TOI is observed and restrict the analysis to TOIs with at least 3 transits.

In order to have a ground truth with which to compare the results of our algorithm, we restrict our sample of TOIs to those that have been designated as confirmed planets (CPs) and those that have been designated as false positives (FPs) by the TFOP. We also discard TOIs that have been designated FPs due to instrumental false alarms (which our tool does not test for), TOIs without estimates for $M_\star$, $R_\star$, and $T_{\rm eff}$ in the TIC, and TOIs for which we are unable to feasibly recover a transit with the purported orbital parameters. Lastly, we only include planets with best-fit planet radii $R_{\rm p} < 8 R_\oplus$ under the TP scenario. This radius corresponds roughly to the minimum radius of a brown dwarf \citep[e.g.,][]{sorahana2013radii} and has been used as an upper limit in the size of objects that can be validated in past validation studies \citep[e.g.,][]{mayo2018275}, due to the fact that giant planets, brown dwarfs, and low-mass stars are typically indistinguishable based on radius alone. This leaves 68 TOIs in total, 28 of which are confirmed planets and 40 of which are false positives. The system properties of these TOIs are displayed in Figure \ref{fig:4_1_1}.

After generating light curves for these TOIs, we calculate the FPP and NFPP for each to determine the limits within which \texttt{TRICERATOPS} can be used reliably. First, we explore how our predictions depend on the signal-to-noise ratio (SNR) of the data. We define the SNR as
\begin{equation}\label{eq:SNR}
    {\rm SNR} = \frac{\delta_{\rm obs}}{\sigma_{\rm CDPP}} \sqrt{n_{\rm tra}}
\end{equation}
where $\delta_{\rm obs}$ is the observed transit depth (i.e., not corrected for dilution from nearby stars), $\sigma_{\rm CDPP}$ is the combined differential photometric precision (CDPP, \citealt{christiansen2012derivation}) of the 2-minute cadence data, and $n_{\rm tra}$ is the number of observed transits. We calculate $\sigma_{\rm CDPP}$ by applying the \texttt{estimate$\_$cdpp} method of \texttt{lightkurve} \citep{2018ascl.soft12013L} over the duration of the transit. Because this quantity incorporates our confidence in the size of a transiting object and the overall density of data points in-transit, is should correlate with the ability of \texttt{TRICERATOPS} to characterize the shape of a given transit.

The results of this analysis are shown in Figure \ref{fig:4_1_2}. For both CPs and FPs, \texttt{TRICERATOPS} generally has more accurate predictions when SNR is higher. Specifically, FPP alone does not appear to be a reliable predictor of TOI disposition when ${\rm SNR} < 15$, where FPs are frequently assigned low values of FPP that would ideally be reserved for CPs.

\begin{figure*}[b]
    \includegraphics[width=0.5\textwidth]{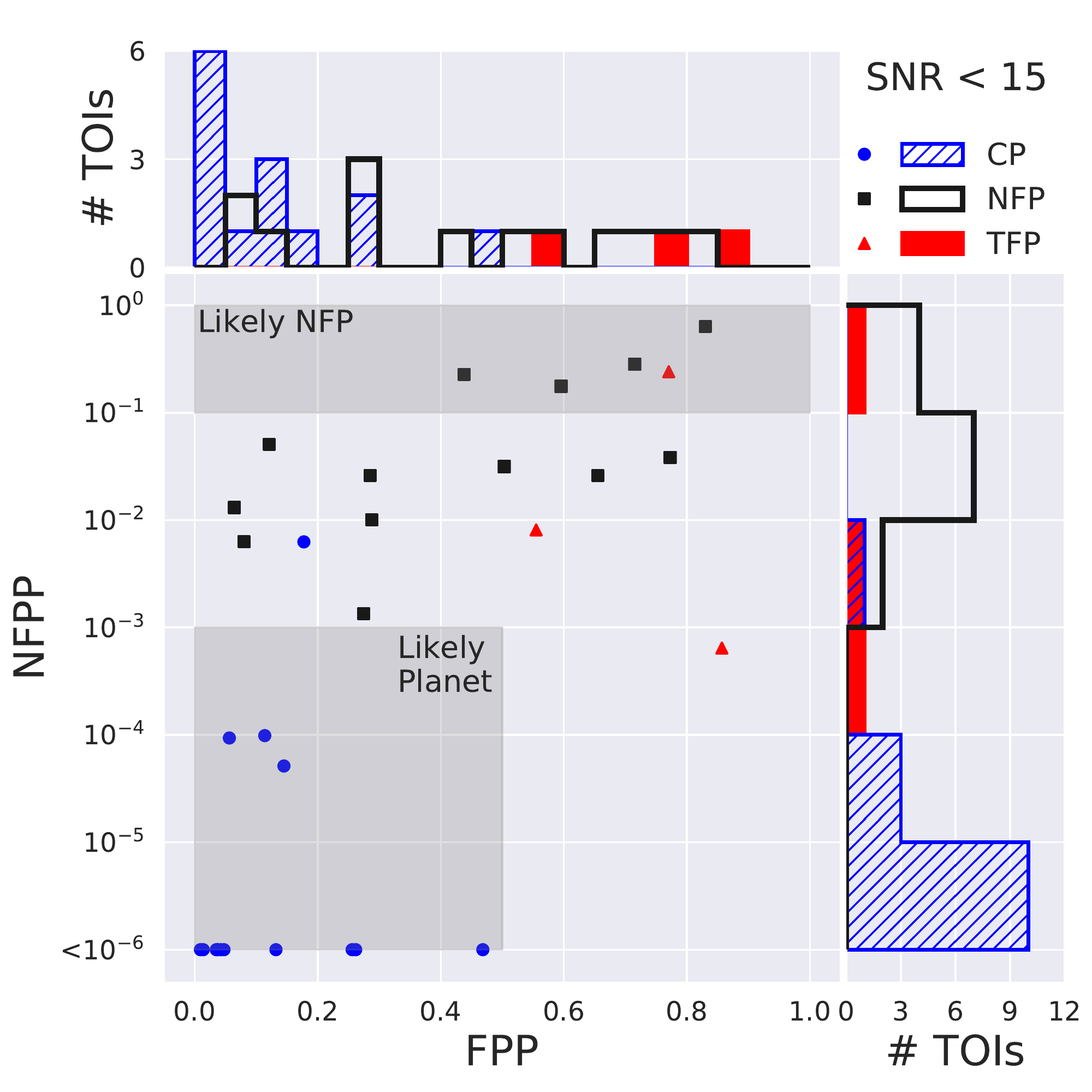}
    \includegraphics[width=0.5\textwidth]{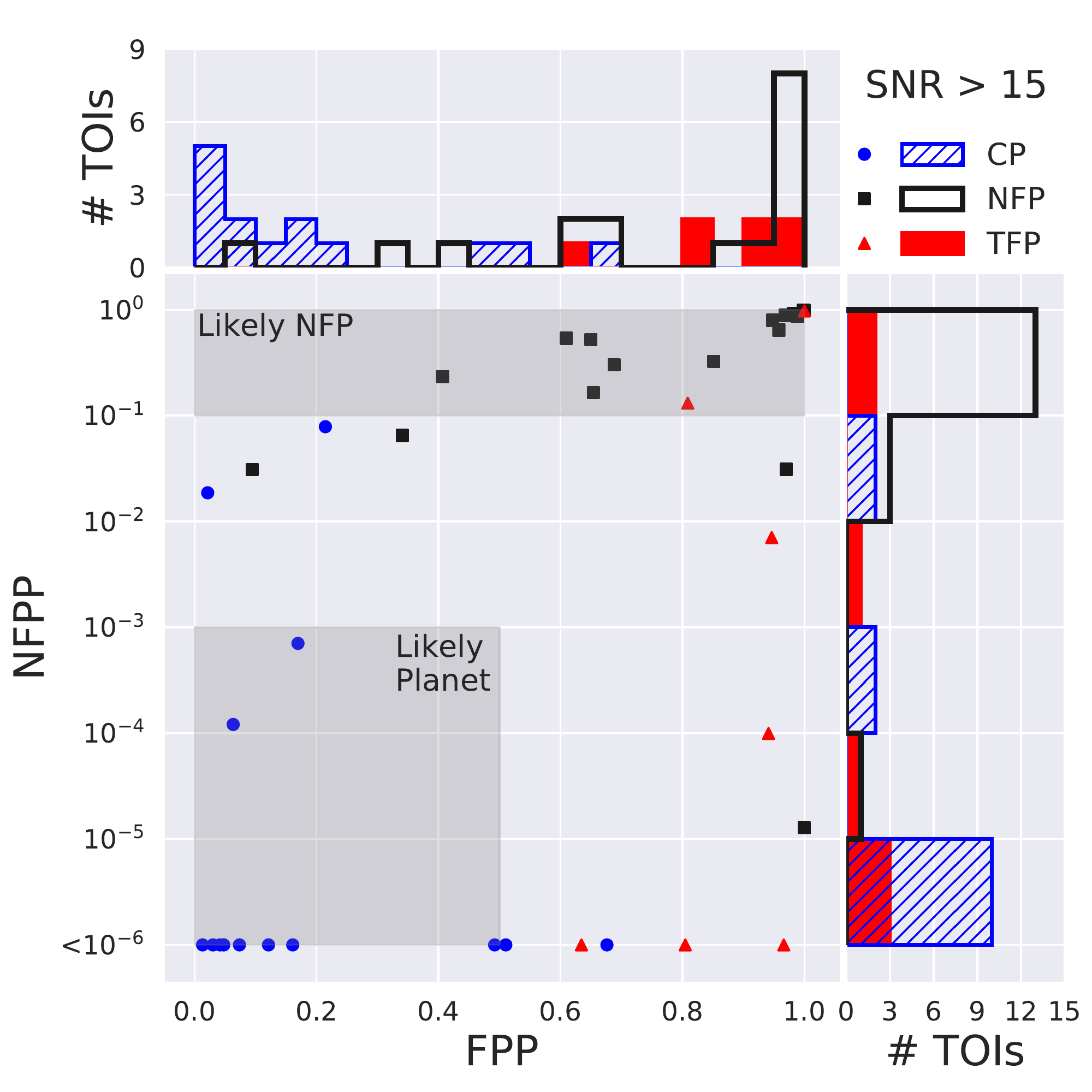}
    \vspace{-15pt}
    \caption{NFPP vs FPP for ${\rm SNR} < 15$ (\emph{left}) and ${\rm SNR} > 15$ (\emph{right}). We designate TOIs with ${\rm NFPP} < 10^{-3}$ and ${\rm FPP} < 0.5$ as likely planets. For TOIs with ${\rm NFPP} < 10^{-3}$, and ${\rm FPP} \leq 0.01$, we are able to rule out FPs with a high enough confidence to consider them validated. Lastly, we are able to identify TOIs that are NFPs with high confidence when ${\rm NFPP} > 10^{-1}$.}
    \label{fig:4_1_3}
\end{figure*}

Second, we explore how our algorithm performs when NFPP is also considered. Figure \ref{fig:4_1_3} shows the distribution of the TOIs in NFPP--FPP space for ${\rm SNR} < 15$ (on the left) and ${\rm SNR} > 15$ (on the right). In the figure, we differentiate TOIs that are CPs, TOIs that have been ruled out as FPs around nearby stars (nearby false positives, or NFPs), and TOIs that have been ruled out as FPs originating from the immediate vicinity of the target star (target false positives, or TFPs). The most salient feature of this figure is the region defined by ${\rm NFPP < 10^{-3}}$ and ${\rm FPP < 0.5}$ that contains nearly all of the CPs, none of the NFPs or TFPs, and is independent of SNR. We designate TOIs that exist within this region as likely planets. 

Another visible feature of Figure \ref{fig:4_1_3} is the pile-up of CPs in the region defined by ${\rm NFPP < 10^{-3}}$ and ${\rm FPP < 0.05}$. Because this region is representative of TOIs with the best chances of being bona fide planets, we use it as a guide in defining our criteria for validating planets. Typically, the standard for validating planets (e.g., with \texttt{VESPA}) is to achieve a FPP below $1 \%$. We therefore define validated planets as TOIs with ${\rm NFPP < 10^{-3}}$ and ${\rm FPP < 0.015}$ (or ${\rm FPP \leq 0.01}$, when rounding to the nearest percent).

\begin{figure*}[b]
    \includegraphics[width=0.5\textwidth]{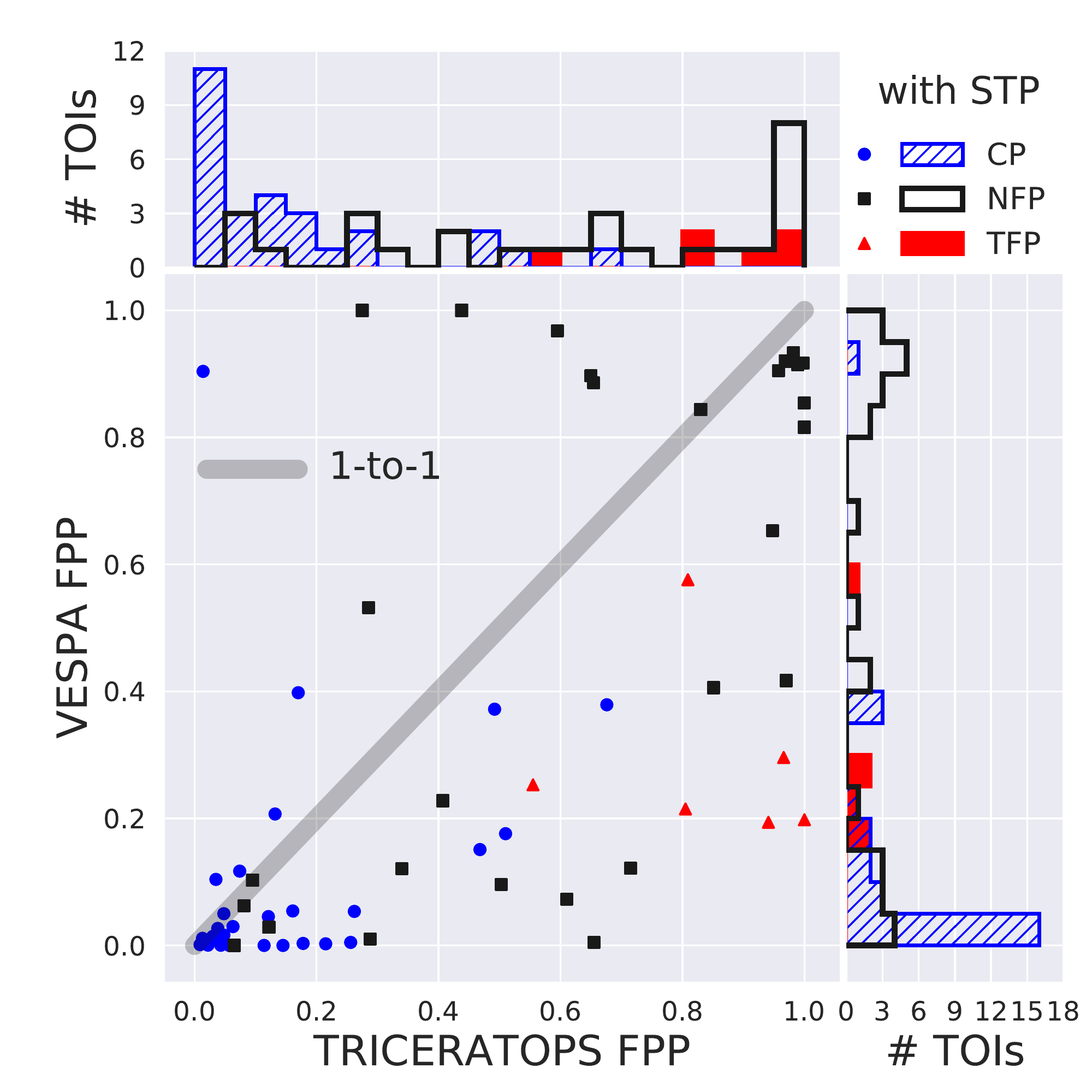}
    \includegraphics[width=0.5\textwidth]{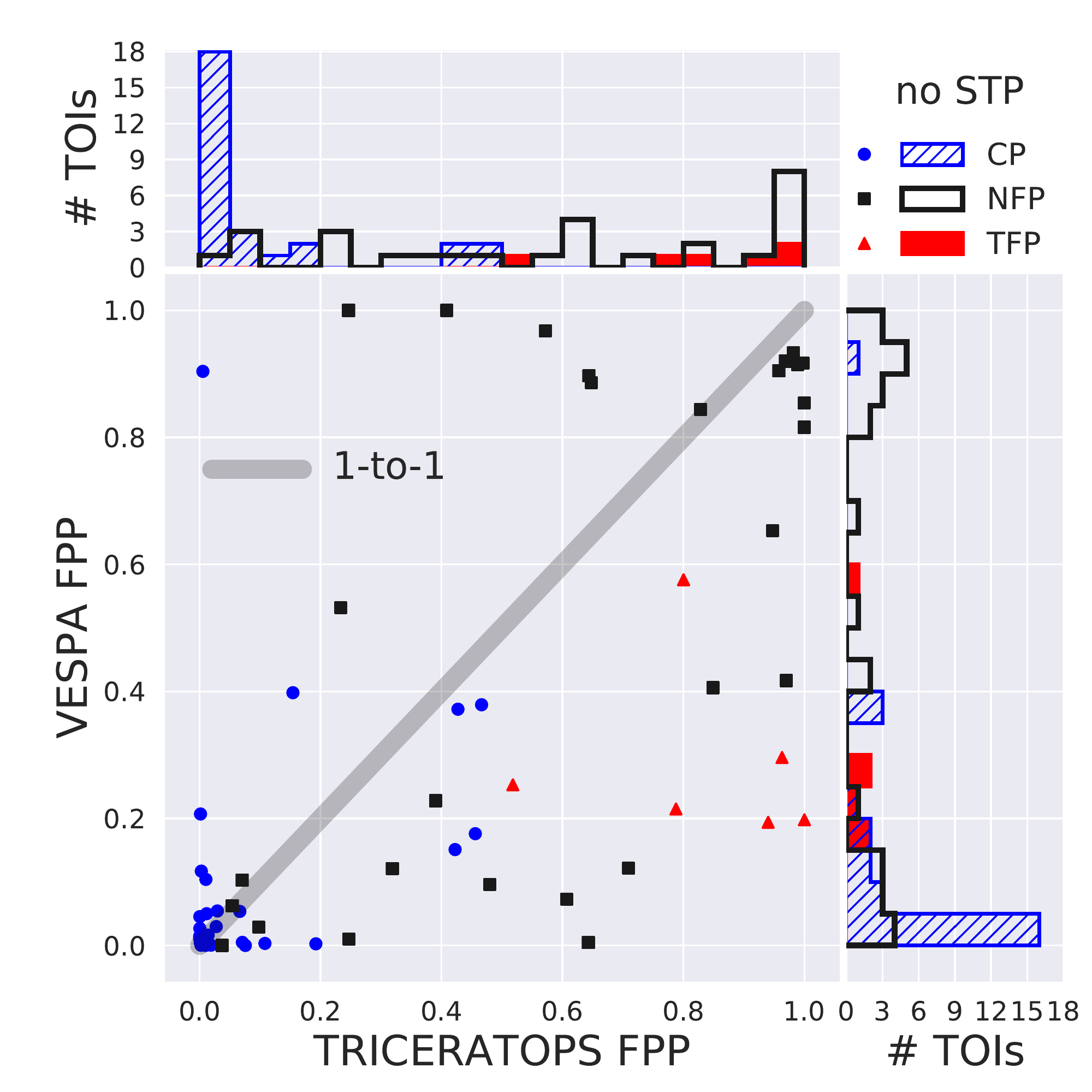}
    \vspace{-15pt}
    \caption{\texttt{VESPA} FPP vs \texttt{TRICERATOPS} FPP for the TOIs in Figure \ref{fig:4_1_3}. \emph{Left:} Comparison with the STP scenario included in the \texttt{TRICERATOPS} calculation. \emph{Right:} Comparison without the STP scenario included in the \texttt{TRICERATOPS} calculation. CPs that score a low FPP with \texttt{TRICERATOPS} tend to also score a low FPP with \texttt{VESPA}. This agreement is stronger when the STP scenario (which is not considered in \texttt{VESPA}) is omitted in \texttt{TRICERATOPS}. Conversely, FPs (and in particular, NFPs) generally score higher FPPs with \texttt{TRICERATOPS} than with \texttt{VESPA} due to the ability of the former to consider nearby stars as potential sources.}
    \label{fig:4_1_4}
\end{figure*}

As a cross-check of our definition of a validated planet, we calculate the FPP of the TOIs in Figure \ref{fig:4_1_3} using \texttt{VESPA}. We run \texttt{VESPA} using the coordinates, stellar photometry (\tess mag, $B$mag, $V$mag, $J$mag, $H$mag, and $K$mag), $T_{\rm eff}$, $\log g$, and parallax listed for each TOI in the TIC. We use the same transit data used in our \texttt{TRICERATOPS} runs and assume a maximum unresolved star separation of $2\farcs2$. The FPPs obtained with \texttt{VESPA} are compared to the FPPs obtained with \texttt{TRICERATOPS} in Figure \ref{fig:4_1_4}. According to the figure, TOIs that score a a low FPP with \texttt{TRICERATOPS} generally score a similar FPP with \texttt{VESPA}. When it comes to FPs, and NFP in particular, \texttt{TRICERATOPS} typically assigns higher FPPs than \texttt{VESPA} does. This is a reflection of our calculation procedure, which considers each star that contributes flux to the target aperture as a potential source of the observed transit. One might also note that there are a few NFPs that are scored low FPPs with both tools. However, because of our condition that a TOI have ${\rm NFPP} < 10^{-3}$ to be classified as a validated planet or a likely planet, \texttt{TRICERATOPS} would not identify these candidates as planets. Conversely, because \texttt{VESPA} explicitly requires the assumption that no contaminating stars exist within a specified radius of the target star, it could classify these candidates as planets if all nearby stars are not ruled out as transit sources prior to the analysis. To avoid outcomes like this, \texttt{VESPA} requires a separate calculation of the probability that the transit originates from the target star prior to its FPP calculation \citep[e.g.,][]{morton2016false}.

It is also worth noting that the calculation procedures between the tools are not identical. An important difference is that \texttt{TRICERATOPS} takes into account the STP scenario, which involves a planet transiting an unresolved bound companion, whereas \texttt{VESPA} does not. This false positive scenario typically has a non-negligible probability of being the ground truth and therefore inflates the FPP obtained with \texttt{TRICERATOPS} relative to that of \texttt{VESPA}. To test how this impacts the FPP comparison, we calculate the \texttt{TRICERATOPS} FPP for each TOI both using (left-hand panel of Figure \ref{fig:4_1_4}) and omitting (right-hand panel of Figure \ref{fig:4_1_4}) the STP scenario. We see that when this scenario is included, there are several TOIs that score a validation-worthy FPP with \texttt{VESPA} that do not with \texttt{TRICERATOPS}. However, when this scenario is omitted from the calculation, the two tools return more consistent results. This suggests that \texttt{TRICERATOPS} is more conservative when validating TOIs and will oftentimes rely on supplementary follow-up observations to achieve ${\rm FPP} \le 0.01$.

\begin{figure*}[!ht]
    \includegraphics[width=1.02\textwidth]{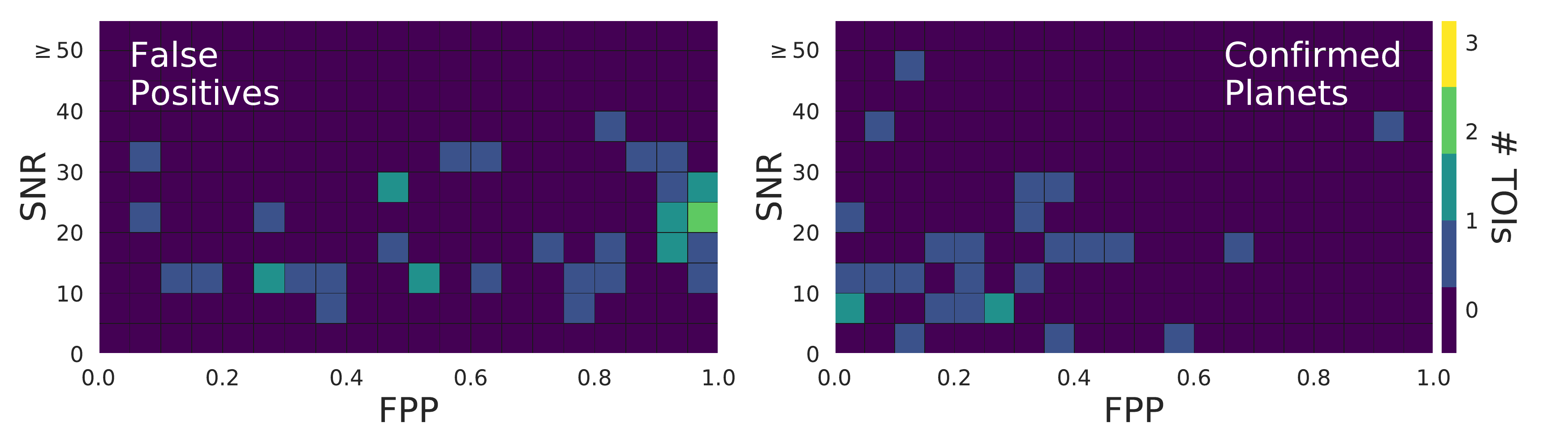}
    \vspace{-15pt}
    \caption{SNR vs FPP for the same false positives (\emph{left}) and confirmed planets (\emph{right}) shown in Figure \ref{fig:4_1_2}, but calculated using light curves extracted from 30-minute cadence \tess\ data. While there still appears to be a correlation between SNR and performance, it is less clear here than in Figure \ref{fig:4_1_2}.}
    \label{fig:4_2_1}
\end{figure*}

\begin{figure*}[!ht]
    \includegraphics[width=0.5\textwidth]{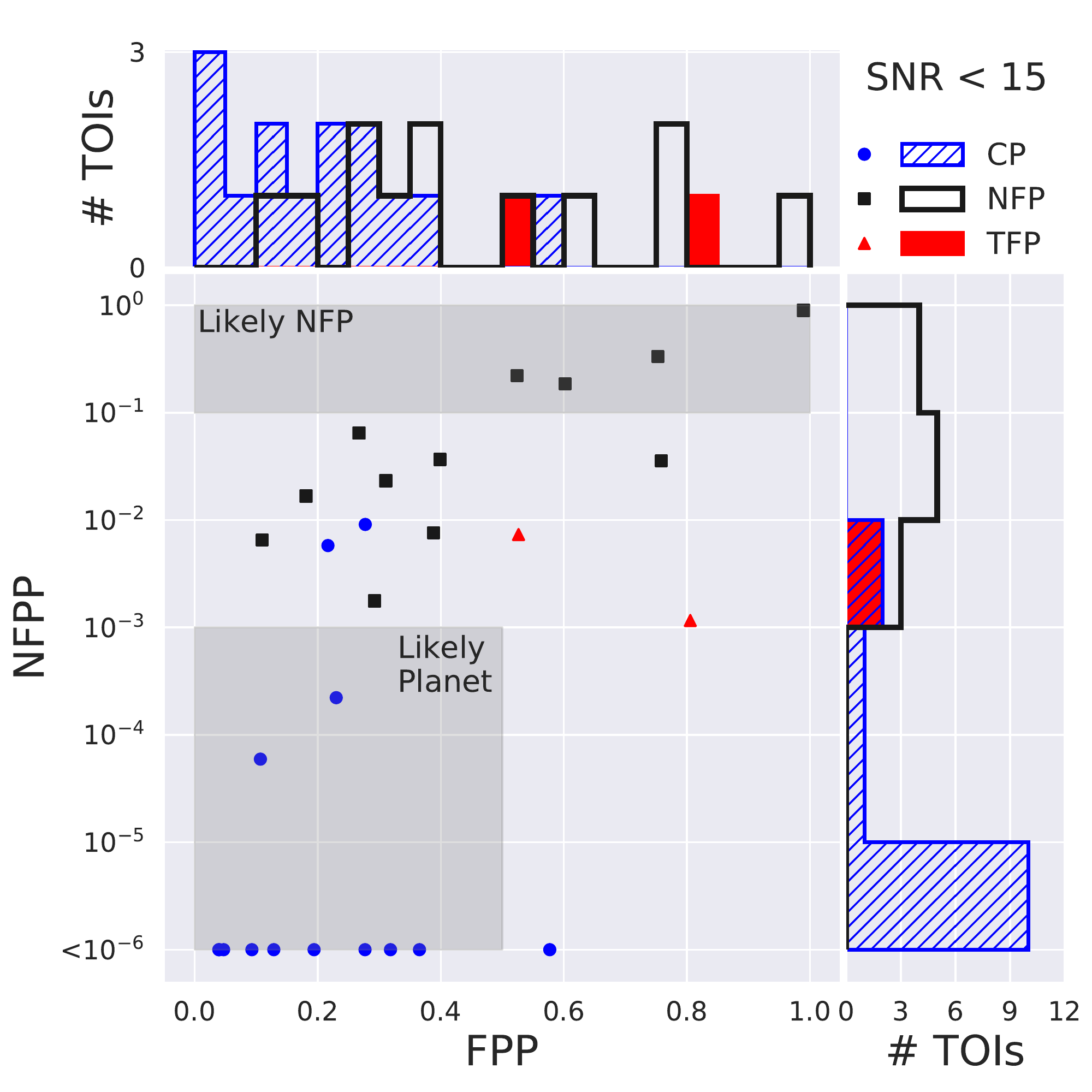}
    \includegraphics[width=0.5\textwidth]{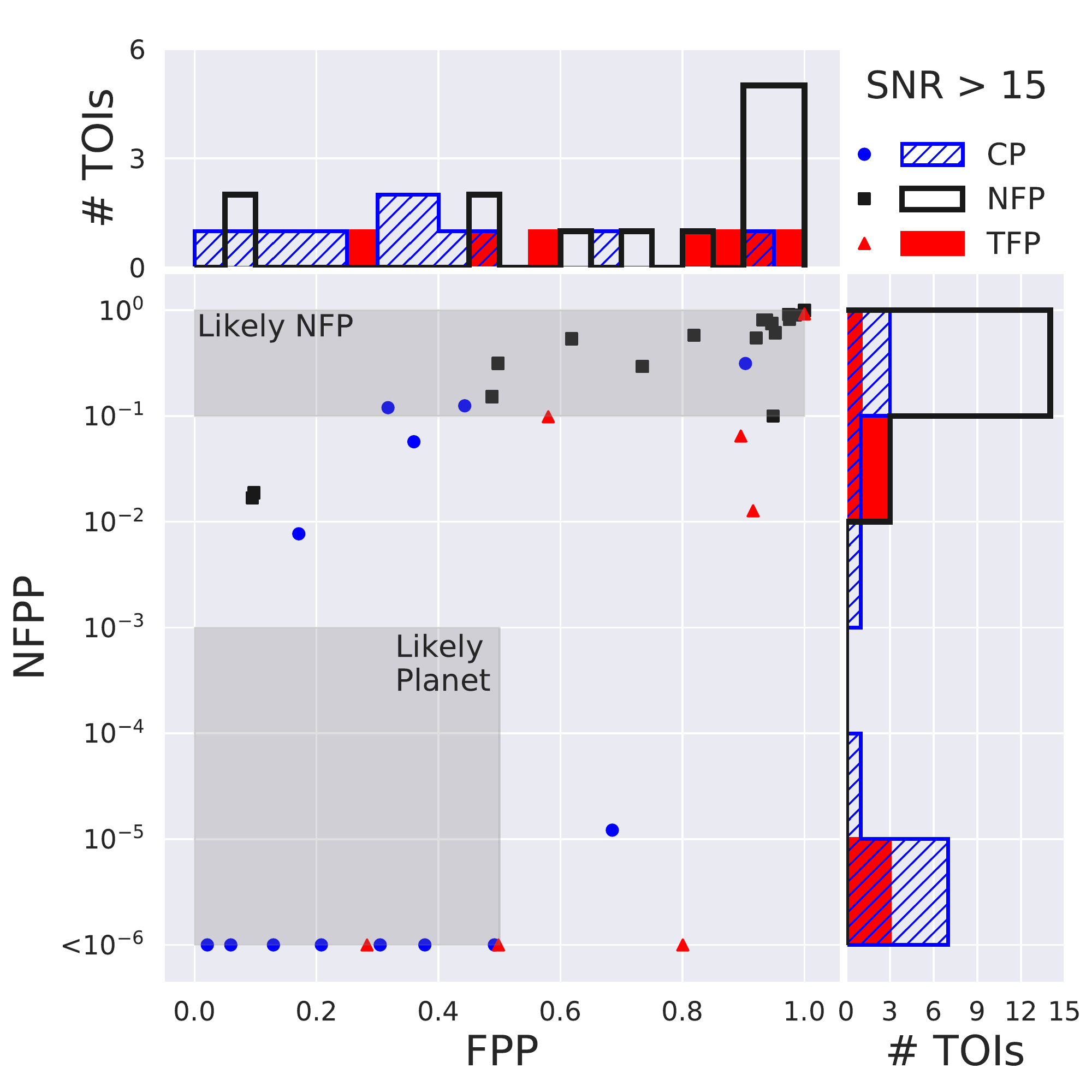}
    \vspace{-15pt}
    \caption{NFPP vs FPP for ${\rm SNR} < 15$ (\emph{left}) and ${\rm SNR} > 15$ (\emph{right}), but calculated using 30-minute cadence \tess\ data. While we are unable to identify a region in which we can validate TOIs, we can still designate TOIs with ${\rm NFPP} < 10^{-3}$ and ${\rm FPP} < 0.5$ as likely planets. Additionally, we are still able to identify TOIs that are NFPs with high confidence when ${\rm NFPP} > 10^{-1}$.}
    \label{fig:4_2_2}
\end{figure*}

% \vspace{-40pt}

\subsection{30-minute Cadence Data}

One might expect our code to have a more difficult time distinguishing CPs from FPs when using data with a longer cadence, as they would yield fewer points with which to characterize the shape of the transit. To test this, we also run our code on 30-minute cadence light curves of the same TOIs. We use \texttt{eleanor} \citep{feinstein2019eleanor} to extract these light curves from \tess\ Full Frame Images (FFIs) within the same sectors and apertures used to obtain the 2-minute cadence light curves.\footnote{More precisely, we run our code on 67 of the 68 TOIs analyzed in the previous section. We were unable to recover the FFI data for TOI 1796.01 (the TOI with the highest SNR in Figure \ref{fig:4_1_2}) due to a bug in \texttt{eleanor}, which returns a error claiming that the TOI has not yet been observed upon searching for its data.}

In Figure \ref{fig:4_2_1}, we show how SNR affects the new FPP calculations. As in the previous section, \texttt{TRICERATOPS} is able to correctly identify CPs and FPs more frequently when SNR is high, but the correlation is weaker overall. Specifically, the FPPs of CPs are less concentrated near zero here than those calculated with the 2-minute data.

In Figure \ref{fig:4_2_2}, we reproduce the NFPP vs FPP analysis from the previous section using the 30-minute cadence data. We again see that most CPs are contained within a region defined by ${\rm NFPP < 10^{-3}}$ and ${\rm FPP < 0.5}$, with very few FPs also falling within this region. Specifically, the region contains 18 CPs and only 2 FPs. In addition, almost no CPs have a ${\rm FPP} > 0.7$ (with the exception of one, which is mistaken for a nearby false positive), which implies that a high FPP is still indicative of actual FPs. We thus again designate TOIs with ${\rm NFPP < 10^{-3}}$ and ${\rm FPP < 0.5}$ as likely planets. However, unlike the results obtained with the 2-minute cadence data, there does not appear to be a region of parameter space in which planets can be confidently validated. Nonetheless, \texttt{TRICERATOPS} results involving long cadence \tess\ data are useful for vetting TOIs and prioritizing them for follow-up observations to further investigate the nature of the signal.

\section{Nearby False Positive Identification}\label{sec:5}

In addition to its ability to identify likely planets and validate TOIs, \texttt{TRICERATOPS} is proficient at identifying NFPs. In Figures \ref{fig:4_1_3} and \ref{fig:4_2_2}, TOIs with a ${\rm NFPP} > 10^{-1}$ are NFPs 85$\%$ and 82$\%$ of the time, respectively. Additionally, the region defined by ${\rm NFPP} > 10^{-1}$ contains over half of the NFPs in our sample for calculations conducted using both 2-minute and 30-minute data. These results suggest that \texttt{TRICERATOPS} can be used to predict which TOIs are NFPs and to determine which nearby stars have the highest probability of hosting the observed transit. We therefore classify TOIs in this region of parameter space as likely NFPs.

As an additional step to assess the ability of our tool to identify NFPs, we compile a set of observations collected by members of \tess\ Follow-up Observing Program (TFOP)\footnote{https://tess.mit.edu/followup} Sub Group 1 (SG1) that rule out 30 TOIs as NFPs. The follow-up observations were scheduled using the {\tt TESS Transit Finder}, which is a customized version of the {\tt Tapir} software package \citep{Jensen:2013}. Below, we outline these observations and compare the empirical results with the NFPPs predicted by \texttt{TRICERATOPS} using 2-minute cadence data. A summary of these targets is given in Table \ref{table:tfop} and details about the facilities used are given in Table \ref{table:observatories}.

Another method of discerning NFPs is by searching for centroid offsets in the \tess\ pixels encompassing a TOI. Often times, the true source of a NFP can be identified using the magnitude and direction of these offsets. In addition to the observations collected by TFOP SG1, we compare our \texttt{TRICERATOPS} predictions with the difference image centroiding analyses for these TOIs in their SPOC data validation reports \citep{twicken2018kepler}.

With these comparisons, we display that \texttt{TRICERATOPS} often yields similar results to both follow-up observations and predictions made using centroid offsets. For several of these TOIs (17/30), our tool assigns a NFPP high enough to classify them as likely NFPs. For those that do not meet this criterion, FPP and NFPP are high enough to rule out the possibility of the TOI being a planet. Lastly, in cases where there are several NFP candidates (of which there are 28), \texttt{TRICERATOPS} is frequently (10/28) able to predict which nearby star is the true host of the transit signal.

\begin{deluxetable*}{rrcrccccc}
\tablecaption{TFOP SG1 false-positive identification compared to \texttt{TRICERATOPS} predictions\label{table:tfop}}
\tablehead{\colhead{Target} & \colhead{TOI} & \colhead{TFOP SG1} & \colhead{True Host} & \colhead{FPP} & \colhead{Total} & \colhead{True Host} & \colhead{$\#$ NFP} & \colhead{True Host} \\[-8pt]
 \colhead{TIC ID} & \colhead{Number} & \colhead{Disposition} & \colhead{TIC ID} & & \colhead{NFPP} & \colhead{NFPP} & \colhead{Candidates} & \colhead{Rank}}
\startdata
260043723 & 217.01 & NEB &  260043722 & 0.0806 & 0.0063 & 0.0059 & 2 & 1 \\
279740441 & 273.01 & NEB &  279740439 & 0.6095 & 0.5377 & 0.2041 & 2 & 2 \\
250386181 & 390.01 & NEB &  250386182 & 0.9703 & 0.0311 & 0.0311 & 2 & 1 \\
219388773 & 399.01 & NEB & 219388775  & 0.2882 & 0.0101 & 0.0101 & 1 & 1 \\
176778112 & 408.01 & NEB & 176778114  & 0.3405 & 0.0650 & 0.0438 & 2 & 1 \\
20178111 & 467.01 & NEB & 20178112    & 0.4065 & 0.2332 & 0.1638 & 3 & 1 \\
427352241 & 485.01 & NEB & 427352247  & 0.6498 & 0.5219 & 0.0693 & 3 & 2 \\
108645766 & 497.01 & NEB & 108645800  & 0.8299 & 0.6361 & - & 4 & - \\
274138511 & 506.01 & NEB & 760244235  & 0.1215 & 0.0507 & 0.0030 & 10 & 6 \\
431999925 & 513.01 & NEB & 431999916  & 0.9819 & 0.9230 & 0.1482 & 8 & 2 \\
438490744 & 529.01 & NEB & 438490748  & 1.0000 & 0.9938 & 0.9938 & 2 & 1 \\
302895996 & 531.01 & NEB & 302895984  & 0.9477 & 0.7971 & 0.0509 & 5 & 3 \\
53593457 & 543.01 & NEB & 53593470    & 0.9580 & 0.6436 & - & 5 & - \\
59003115 & 556.01 & NEB & 59003118    & 0.2854 & 0.0258 & 0.0050 & 2 & 2 \\
1133072 & 566.01 & NEB & 830310300    & 0.9687 & 0.8854 & 0.0124 & 9 & 6 \\
146463781 & 636.01 & NEB & 146463868  & 0.9887 & 0.8640 & - & 3 & - \\
432008938 & 643.01 & NEB & 432008934  & 0.9996 & 0.00001 & - & 2 & - \\
54085154 & 662.01 & NEB & 54085149    & 0.2747 & 0.0013 & 0.0008 & 2 & 1 \\
147660201 & 670.01 & NPC & 147660207  & 0.6543 & 0.1652 & 0.0868 & 9 & 1 \\
391821647 & 708.01 & NEB & $\sim35\arcsec$ W & 0.5955 & 0.1760 & - & 141 & - \\
373424049 & 742.01 & NEB & 373424060  & 0.4377 & 0.2268 & 0.0006 & 31 & 23 \\
271596418 & 868.01 & NEB & 271596416  & 0.6551 & 0.0259 & 0.0078 & 7 & 1 \\
364107753 & 909.01 & NEB & 1310226289 & 0.0645 & 0.0131 & 0.0068 & 4 & 1 \\
253990973 & 1061.01 & NEB & 253985122 & 0.5030 & 0.0315 & 0.0037 & 9 & 4 \\
308034948 & 1206.01 & NEB & unknown   & 0.7727 & 0.0383 & - & 108 & - \\
274762761 & 1256.01 & NEB & 274762865 & 0.9981 & 0.9869 & - & 6 & - \\
267561446 & 1284.01 & NEB & 267561450 & 0.7151 & 0.2818 & 0.0235 & 13 & 4 \\
274662200 & 1285.01 & NEB & 274662220 & 0.6880 & 0.3031 & 0.0501 & 21 & 2 \\
408203470 & 1289.01 & NEB & 408203452 & 0.8512 & 0.3258 & 0.1435 & 10 & 1 \\
233681149 & 1340.01 & NEB & 233681148 & 0.0947 & 0.0309 & 0.0309 & 1 & 1 \\
\enddata
\tablecomments{``Total NFPP'' is the total NFPP for the TOI. ``True Host NFPP'' is the NFPP for only the true host of the signal. ``$\#$ NFP Candidates'' is the number of nearby sources bright enough to host the signal. ``True Host Rank'' is the rank of the true host NFPP, compared to the NFPPs of all other NFP candidates (where a rank of 1 corresponds to the highest NFPP).}
\end{deluxetable*}

\begin{deluxetable*}{llDDc}
\tablecaption{Facilities used for TFOP SG1 followup\label{table:observatories}}
\tablehead{\colhead{Observatory} & \colhead{Location} & \twocolhead{Aperture} & \twocolhead{Pixel scale} & \colhead{FOV}\\[-2mm]
 & & \twocolhead{(m)} & \twocolhead{(arcsec)} & \colhead{(arcmin)}
}
\decimals
\startdata
Univ.\ Louisville Moore Obs.\ / CDK20N & Louisville, KY, US & 0.51 & 0.54 & $37 \times 37$ \\
Univ.\ Louisville Manner Telescope & Mt. Lemmon, AZ, US & 0.61 & 0.39 & $26 \times 26$ \\
Mt. Kent Observatory / CDK700 & Toowoomba, Australia & 0.7 & 0.4 & $27 \times 27$ \\
Hazelwood Observatory & Churchill, Victoria, Australia & 0.318 & 0.55 & $20 \times 13.5$ \\
LCOGT 0.4m &  (various) & 0.4 & 0.57 & $29.2 \times 19.5$ \\
LCOGT 1.0m & (various) & 1.0 & 0.39 & $26.5 \times 26.5$ \\
Fred L. Whipple Obs. / MEarth-North & Amado, AZ, USA & 0.4 & 0.76 & $26 \times 26$ \\
Tel.\ Carlos S\'{a}nchez / MuSCAT2  & Teide Obs., Tenerife, Spain & 1.52 & 0.44 & $7.4 \times 7.4 $ \\
El Sauce Observatory  & Coquimbo Province, Chile & 0.36 & 1.47 & $18.8 \times 12.5$ \\
Perth Exoplanet Survey Telescope (PEST) & Perth, Australia & 0.3 & 1.2 & $31 \times 21$ \\
HATNet & (various) & 0.11 & 14 & $492 \times 492$  \\
HAT-South & (various) & 0.18 & 3.7 & $492 \times 492$ \\
TRAPPIST-South & La Silla, Chile & 0.6 & 0.6 & $ 22 \times 22$ \\
Steward Observatory Phillips Telescope &  Mt. Lemmon, AZ, US & 0.6 & 0.38  & $ 26 \times 26$ \\
\enddata
\end{deluxetable*}

\subsection{TIC 260043723 (TOI 217.01)}

TFOP SG1 confirms the true host of the signal is TIC 260043722. Previous HAT South data suggested that this TOI is a NEB, which was confirmed by PEST Observatory $R_C$-band observations with a depth of 200 ppt. This star was also correctly identified as the host of the signal by the SPOC centroid offset analysis. \texttt{TRICERATOPS} identifies 2 nearby sources other than the target star bright enough to host the signal, one of which is TIC 260043722. The total NFPP calculated by \texttt{TRICERATOPS} is 0.0063. TIC 260043722 has a NFPP of 0.0059, making it the most probable NFP host. This NFPP is too low to classify the TOI as a likely NFP and too high to classify the TOI as a likely planet. In addition, the calculated FPP of 0.0806 is too high to classify the TOI as a validated planet.

\subsection{TIC 279740441 (TOI 273.01)}

TFOP SG1 confirms the true host of the signal is TIC 279740439. The signal was a nearby planet candidate (signal not on the original TOI, but still possibly planetary) based on observations from the TRAPPIST telescope that show a depth of 40 ppt in a custom $I+z$-band filter.  Later observations with LCOGT \citep{Brown:2013} showed a $V$-band depth of 30 ppt on the nearby candidate; the wavelength-dependent eclipse depth indicates that it is an eclipsing binary. This star was also correctly identified as the host of the signal by the SPOC centroid offset analysis. \texttt{TRICERATOPS} identifies 2 nearby sources other than the target star bright enough to host the signal, one of which is TIC 279740439. The total NFPP calculated by \texttt{TRICERATOPS} is 0.5377. TIC 279740439 has a NFPP of 0.2041, making it the 2nd most probable NFP host. This NFPP is high enough to classify the TOI as a likely NFP. In addition, the calculated FPP of 0.6095 is too high to classify the TOI as a likely planet or validated planet.

\subsection{TIC 250386181 (TOI 390.01)}

TFOP SG1 confirms the true host of the signal is TIC 250386182. The TOI is a NEB, based on LCOGT observations in the PanSTARRS $zs$ filter showing a depth of roughly 350 ppt. This star was also correctly identified as the host of the signal by the SPOC centroid offset analysis. \texttt{TRICERATOPS} identifies 2 nearby sources other than the target star bright enough to host the signal, one of which is TIC 250386182. The total NFPP calculated by \texttt{TRICERATOPS} is 0.0311. TIC 250386182 has a NFPP of 0.0311, making it the most probable NFP host. This NFPP is too low to classify the TOI as a likely NFP. However, the calculated FPP of 0.9703 is too high to classify the TOI as a likely planet or validated planet.

\subsection{TIC 219388773 (TOI 399.01)}

TFOP SG1 confirms the true host of the signal is TIC 219388775. The TOI is a NEB with depth of 130 ppt, based on LCOGT $zs$ observations. This star was also correctly identified as the host of the signal by the SPOC centroid offset analysis. \texttt{TRICERATOPS} identifies 1 nearby source other than the target star bright enough to host the signal, which is TIC 219388775. The total NFPP calculated by \texttt{TRICERATOPS} is 0.0101. TIC 219388775 has a NFPP of 0.0101, making it the most probable NFP host. This NFPP is too low to classify the TOI as a likely NFP and too high to classify the TOI as a likely planet. In addition, the calculated FPP of 0.2882 is too high to classify the TOI as a validated planet.

\subsection{TIC 176778112 (TOI 408.01)}

TFOP SG1 confirms the true host of the signal is TIC 176778114. The TOI is a NEB with primary and secondary eclipse depths of $\sim 430$ ppt and $\sim 300$ ppt in LCOGT $r'$ observations. This star was also correctly identified as the host of the signal by the SPOC centroid offset analysis. \texttt{TRICERATOPS} identifies 2 nearby sources other than the target star bright enough to host the signal, one of which is TIC 176778114. The total NFPP calculated by \texttt{TRICERATOPS} is 0.0650. TIC 176778114 has a NFPP of 0.0438, making it the most probable NFP host. This NFPP is too low to classify the TOI as a likely NFP and too high to classify the TOI as a likely planet. In addition, the calculated FPP of 0.3405 is too high to classify the TOI as a validated planet.

\subsection{TIC 20178111 (TOI 467.01)}

TFOP SG1 confirms the true host of the signal is TIC 20178112. The TOI is a NEB, based on PEST Observatory $R_C$ observations that show a  $\sim 55$ ppt eclipse on TIC 20178112, which Gaia shows as two stars with magnitudes $G=14.2$ and $G=15.9$. This star was also correctly identified as the host of the signal by the SPOC centroid offset analysis. \texttt{TRICERATOPS} identifies 3 nearby sources other than the target star bright enough to host the signal, one of which is TIC 20178112. The total NFPP calculated by \texttt{TRICERATOPS} is 0.2332. TIC 20178112 has a NFPP of 0.1638, making it the most probable NFP host. This NFPP is high enough to classify the TOI as a likely NFP. In addition, the calculated FPP of 0.4065 is too high to classify the TOI as a validated planet.

\subsection{TIC 427352241 (TOI 485.01)}

TFOP SG1 confirms the true host of the signal is TIC 427352247. The TOI is a NEB, based on LCOGT $r'$ observations that show a 200 ppt, V-shaped eclipse. This star was also correctly identified as the host of the signal by the SPOC centroid offset analysis. \texttt{TRICERATOPS} identifies 3 nearby sources other than the target star bright enough to host the signal, one of which is TIC 427352247. The total NFPP calculated by \texttt{TRICERATOPS} is 0.5219. TIC 427352247 has a NFPP of 0.0693, making it the 2nd most-probably NFP host. This NFPP is high enough to classify the TOI as a likely NFP. In addition, the calculated FPP of 0.6498 is too high to classify the TOI as a likely planet or validated planet.

\subsection{TIC 108645766 (TOI 497.01)}

TFOP SG1 confirms the true host of the signal is TIC 108645800. The TOI is a NEB, based on LCOGT $r'$ observations with a depth of at least 100 ppt, and confirmed by archival HAT South data. This star was also correctly identified as the host of the signal by the SPOC centroid offset analysis. \texttt{TRICERATOPS} identifies 4 nearby sources other than the target star bright enough to host the signal, one of which is TIC 108645800. The total NFPP calculated by \texttt{TRICERATOPS} is 0.6361, but the NFPP around TIC 108645800 was not calculated due to unknown stellar parameters. This NFPP is high enough to classify the TOI as a likely NFP. In addition, the calculated FPP of 0.8299 is too high to classify the TOI as a likely planet or validated planet.

\subsection{TIC 274138511 (TOI 506.01)}

TFOP SG1 confirms the true host of the signal is TIC 760244235. The TOI is a NEB with depth of at least 200 ppt, based on LCOGT $r'$ observations. This star was also correctly identified as the host of the signal by the SPOC centroid offset analysis. \texttt{TRICERATOPS} identifies 10 nearby sources other than the target star bright enough to host the signal, one of which is TIC 760244235. The total NFPP calculated by \texttt{TRICERATOPS} is 0.0507. TIC 760244235 has a NFPP of 0.0030, making it the 6th most probable NFP host. This NFPP is too low to classify the TOI as a likely NFP and too high to classify the TOI as a likely planet. In addition, the calculated FPP of 0.1215 is too high to classify the TOI as a validated planet.

\subsection{TIC 431999925 (TOI 513.01)}

TFOP SG1 confirms the true host of the signal is TIC 431999916. The TOI is a NEB with depth of at least 90 ppt, based on LCOGT $i'$ observations. This star was also correctly identified as the host of the signal by the SPOC centroid offset analysis. \texttt{TRICERATOPS} identifies 8 nearby sources other than the target star bright enough to host the signal, one of which is TIC 431999916. The total NFPP calculated by \texttt{TRICERATOPS} is 0.9230. TIC 431999916 has a NFPP of 0.1482, making it the 2nd most probable NFP host. This NFPP is high enough to classify the TOI as a likely NFP. In addition, the calculated FPP of 0.9819 is too high to classify the TOI as a likely planet or validated planet.

\subsection{TIC 438490744 (TOI 529.01)}

TFOP SG1 confirms the true host of the signal is TIC 438490748. The TOI is a NEB with depth of $\sim 80$ ppt, based on K2 and HAT-South data.  TIC 438490748 (the source of the signal) is a pair of stars in Gaia, so the true depth may be deeper. This star was also correctly identified as the host of the signal by the SPOC centroid offset analysis. \texttt{TRICERATOPS} identifies 2 nearby sources other than the target star bright enough to host the signal, one of which is TIC 438490748. The total NFPP calculated by \texttt{TRICERATOPS} is 0.9938. TIC 438490748 has a NFPP of 0.9938, making it the most probable NFP host. This NFPP is high enough to classify the TOI as a likely NFP. In addition, the calculated FPP of 1.0 is too high to classify the TOI as a likely planet or validated planet.

\subsection{TIC 302895996 (TOI 531.01)}

TFOP SG1 confirms the true host of the signal is TIC 302895984. The TOI is a NEB with a depth of 200 ppt in the $I$ band from LCOGT observations. This star was also correctly identified as the host of the signal by the SPOC centroid offset analysis. \texttt{TRICERATOPS} identifies 5 nearby sources other than the target star bright enough to host the signal, one of which is TIC 302895984. The total NFPP calculated by \texttt{TRICERATOPS} is 0.7971. TIC 302895984 has a NFPP of 0.0509, making it the 3rd most probable NFP host. This NFPP is high enough to classify the TOI as a likely NFP. In addition, the calculated FPP of 0.9477 is too high to classify the TOI as a likely planet or validated planet.

\subsection{TIC 53593457 (TOI 543.01)}

TFOP SG1 confirms the true host of the signal is TIC 53593470. The TOI is a NEB with a depth of $\sim 250$ ppt in both $g'$ and $i'$ in LCOGT observations. This star was also correctly identified as the host of the signal by the SPOC centroid offset analysis. \texttt{TRICERATOPS} identifies 5 nearby sources other than the target star bright enough to host the signal, one of which is TIC 53593470. The total NFPP calculated by \texttt{TRICERATOPS} is 0.6436, but the NFPP around TIC 53593470 was not calculated due to unknown stellar parameters. This NFPP is high enough to classify the TOI as a likely NFP. In addition, the calculated FPP of 0.9580 is too high to classify the TOI as a likely planet or validated planet.

\subsection{TIC 59003115 (TOI 556.01)}

TFOP SG1 confirms the true host of the signal is TIC 59003118. This is K2-78b (EPIC 210400751) \citep{Crossfield2016}, which was later shown to be an NEB \citep{Cabrera2017}. This star was also correctly identified as the host of the signal by the SPOC centroid offset analysis. \texttt{TRICERATOPS} identifies 2 nearby sources other than the target star bright enough to host the signal, one of which is TIC 59003118. The total NFPP calculated by \texttt{TRICERATOPS} is 0.0258. TIC 59003118 has a NFPP of 0.0050, making it the 2nd most probable NFP host. This NFPP is too low to classify the TOI as a likely NFP and too high to classify the TOI as a likely planet. In addition, the calculated FPP of 0.2854 is too high to classify the TOI as a validated planet.

\subsection{TIC 1133072 (TOI 566.01)}

TFOP SG1 confirms the true host of the signal is TIC 830310300. The TOI is a NEB, based on observations from LCOGT and  Mt.\ Kent Observatory in $i'$, and El Sauce Observatory in $R_C$.  The depth is at least 500 ppt in $i'$. In this case, the SPOC centroid offset analysis failed to identify the presence of a background source at the $3\sigma$ level of significance. \texttt{TRICERATOPS} identifies 9 nearby sources other than the target star bright enough to host the signal, one of which is TIC 830310300. The total NFPP calculated by \texttt{TRICERATOPS} is 0.8854. TIC 830310300 has a NFPP of 0.0124, making it the 6th most probable NFP host. This NFPP is high enough to classify the TOI as a likely NFP. In addition, the calculated FPP of 0.9687 is too high to classify the TOI as a likely planet or validated planet.

\subsection{TIC 146463781 (TOI 636.01)}

TFOP SG1 confirms the true host of the signal is TIC 146463868. The TOI is a NEB, based on LCOGT $I_C$-band observations with a depth of 300 ppt. This star was also correctly identified as the host of the signal by the SPOC centroid offset analysis. \texttt{TRICERATOPS} identifies 3 nearby sources other than the target star bright enough to host the signal, one of which is TIC 146463868. The total NFPP calculated by \texttt{TRICERATOPS} is 0.8640, but the NFPP around TIC 146463868 was not calculated due to unknown stellar parameters. This NFPP is high enough to classify the TOI as a likely NFP. In addition, the calculated FPP of 0.9887 is too high to classify the TOI as a likely planet or validated planet.

\subsection{TIC 432008938 (TOI 643.01)}

TFOP SG1 confirms the true host of the signal is TIC 432008934. The TOI is a NEB, based on the centroid offset from the SPOC S01-S09 vetting report. \texttt{TRICERATOPS} identifies 2 nearby sources other than the target star bright enough to host the signal, but neither is TIC 432008934. The total NFPP calculated by \texttt{TRICERATOPS} is $1e-5$. This NFPP is too low to classify the TOI as a likely NFP. However, the calculated FPP of 0.9996 is too high to classify the TOI as a likely planet or validated planet.

\subsection{TIC 54085154 (TOI 662.01)}

TFOP SG1 confirms the true host of the signal is TIC 54085149. The TOI is a NEB, based on LCOGT $i'$ observations that show a depth of 400 ppt at two different epochs. In this case, the SPOC centroid offset analysis found a significant offset, but the offset did not point directly to the true host. \texttt{TRICERATOPS} identifies 2 nearby sources other than the target star bright enough to host the signal, one of which is TIC 54085149. The total NFPP calculated by \texttt{TRICERATOPS} is 0.0013. TIC 54085149 has a NFPP of 0.0008, making it the most probable NFP host. This NFPP is too low to classify the TOI as a likely NFP and too high to classify the TOI as a likely planet. In addition, the calculated FPP of 0.2747 is too high to classify the TOI as a validated planet.

\subsection{TIC 147660201 (TOI 670.01)}

TFOP SG1 confirms the true host of the signal is TIC 147660207. This candidate was retired from SG1 as nearby planet candidate.  Observations show the true source of the signal to be a $\sim 4$ ppt event in the nearby star TIC 147660207, which is still an active planet candidate as of this writing.  The event was seen in $R_C$ from El Sauce Observatory, and in $i'$ from Mt.\ Kent and Hazelwood Observatories. This star was also correctly identified as the host of the signal by the SPOC centroid offset analysis. \texttt{TRICERATOPS} identifies 9 nearby sources other than the target star bright enough to host the signal, one of which is TIC 147660207. The total NFPP calculated by \texttt{TRICERATOPS} is 0.1652. TIC 147660207 has a NFPP of 0.0868, making it the most probable NFP host. This NFPP is high enough to classify the TOI as a likely NFP. In addition, the calculated FPP of 0.6543 is too high to classify the TOI as a likely planet or validated planet.

\subsection{TIC 391821647 (TOI 708.01)}

TFOP SG1 confirms the TOI is a NFP. The TOI is a NEB, based on large scatter in the image centroid from sector to sector in a very crowded field, and a  possible secondary eclipse.  From the SPOC S01--S09 report, this is a clear NEB $\sim35\arcsec$  west. Although the exact source of the NEB is not clear from the SPOC centroid offset analysis, it is likely too faint, and thus the event is too deep to be planetary. \texttt{TRICERATOPS} identifies 141 nearby sources other than the target star bright enough to host the signal. The total NFPP calculated by \texttt{TRICERATOPS} is 0.1760. This NFPP is high enough to classify the TOI as a likely NFP. In addition, the calculated FPP of 0.5955 is too high to classify the TOI as a likely planet or validated planet.

\subsection{TIC 373424049 (TOI 742.01)}

TFOP SG1 confirms the true host of the signal is TIC 373424060. The TOI is a NEB, based on LCOGT observations that show a depth of $\sim 200$ ppt in the $zs$ filter. This star was also correctly identified as the host of the signal by the SPOC centroid offset analysis. \texttt{TRICERATOPS} identifies 31 nearby sources other than the target star bright enough to host the signal, one of which is TIC 373424060. The total NFPP calculated by \texttt{TRICERATOPS} is 0.2268. TIC 373424060 has a NFPP of 0.0006, making it the 23rd most probable NFP host. This NFPP is high enough to classify the TOI as a likely NFP. In addition, the calculated FPP of 0.4377 is too high to classify the TOI as a validated planet.

\subsection{TIC 271596418 (TOI 868.01)}

TFOP SG1 confirms the true host of the signal is TIC 271596416. The TOI is a NEB, based on LCOGT observations that show a depth of 70--100 ppt in $zs$ and $\sim 30$ ppt in $i'$. This star was also correctly identified as the host of the signal by the SPOC centroid offset analysis. \texttt{TRICERATOPS} identifies 7 nearby sources other than the target star bright enough to host the signal, one of which is TIC 271596416. The total NFPP calculated by \texttt{TRICERATOPS} is 0.0259. TIC 271596416 has a NFPP of 0.0078, making it the most probable NFP host. This NFPP is too low to classify the TOI as a likely NFP and too high to classify the TOI as a likely planet. In addition, the calculated FPP of 0.6551 is too high to classify the TOI as a validated planet.

\subsection{TIC 364107753 (TOI 909.01)}

TFOP SG1 confirms the true host of the signal is TIC 1310226289. The TOI is a NEB, based on LCOGT observations that show a depth of at least 75 ppt in $zs$. This star was also correctly identified as the host of the signal by the SPOC centroid offset analysis. \texttt{TRICERATOPS} identifies 4 nearby sources other than the target star bright enough to host the signal, one of which is TIC 1310226289. The total NFPP calculated by \texttt{TRICERATOPS} is 0.0131. TIC 1310226289 has a NFPP of 0.0068, making it the most probable NFP host. This NFPP is too low to classify the TOI as a likely NFP and too high to classify the TOI as a likely planet. In addition, the calculated FPP of 0.0645 is too high to classify the TOI as a validated planet.

\subsection{TIC 253990973 (TOI 1061.01)}

TFOP SG1 confirms the true host of the signal is TIC 253985122. The TOI is a NEB, based on PEST Observatory $R_C$ band observations with a depth of $\sim 600$ ppt. In this case, the SPOC centroid offset analysis failed to identify the presence of a background source at the $3\sigma$ level of significance. \texttt{TRICERATOPS} identifies 9 nearby sources other than the target star bright enough to host the signal, one of which is TIC 253985122. The total NFPP calculated by \texttt{TRICERATOPS} is 0.0315. TIC 253985122 has a NFPP of 0.0037, making it the 4th most probable NFP host. This NFPP is too low to classify the TOI as a likely NFP and too high to classify the TOI as a likely planet. In addition, the calculated FPP of 0.5030 is too high to classify the TOI as a likely planet or validated planet.

\subsection{TIC 308034948 (TOI 1206.01)}

TFOP SG1 confirms the TOI is a NEB.  Stellar parameters from Gaia and TIC indicate $R_* > 40\ R_\sun$, but the orbital period of $< 1$ day would place the companion's orbit inside the star if it were on target. The SPOC centroid offset suggest that the signal originates from a star to the south. \texttt{TRICERATOPS} identifies 108 nearby sources other than the target star bright enough to host the signal. The total NFPP calculated by \texttt{TRICERATOPS} is 0.0383. This NFPP is too low to classify the TOI as a likely NFP and too high to classify the TOI as a likely planet. In addition, the calculated FPP of 0.7727 is too high to classify the TOI as a likely planet or validated planet.

\subsection{TIC 274762761 (TOI 1256.01)}

TFOP SG1 confirms the true host of the signal is TIC 274762865. The TOI is a NEB, based on archival MEarth-North \citep{nutzman2008design, irwin2015mearth} observations that show no event on target, and eclipses at the \tess\ ephemeris in a neighboring star. SPOC difference image analysis correctly identified this star as the true host. \texttt{TRICERATOPS} identifies 6 nearby sources other than the target star bright enough to host the signal, one of which is TIC 274762865. The total NFPP calculated by \texttt{TRICERATOPS} is 0.9869, but the NFPP around TIC 274762865 was not calculated due to unknown stellar parameters. This NFPP is high enough to classify the TOI as a likely NFP. In addition, the calculated FPP of 0.9981 is too high to classify the TOI as a likely planet or validated planet.

\subsection{TIC 267561446 (TOI 1284.01)}

TFOP SG1 confirms the true host of the signal is TIC 267561450. The TOI is a NEB, based on observations by the University of Louisville Manner Telescope and MuSCAT2 at Teide Observatory in $g'$, $r'$, $i'$, and $z'$ that show a $\sim 200$ ppt eclipse. This star was also correctly identified as the host of the signal by the SPOC centroid offset analysis. \texttt{TRICERATOPS} identifies 13 nearby sources other than the target star bright enough to host the signal, one of which is TIC 267561450. The total NFPP calculated by \texttt{TRICERATOPS} is 0.2818. TIC 267561450 has a NFPP of 0.0235, making it the 4th most probable NFP host. This NFPP is high enough to classify the TOI as a likely NFP. In addition, the calculated FPP of 0.7151 is too high to classify the TOI as a likely planet or validated planet.

\subsection{TIC 274662200 (TOI 1285.01)}

TFOP SG1 confirms the true host of the signal is TIC 274662220. The TOI is a NEB, based on observations at the University of Louisville Manner Telescope that show a depth of 150 ppt in $r'$. This star was also correctly identified as the host of the signal by the SPOC centroid offset analysis. \texttt{TRICERATOPS} identifies 21 nearby sources other than the target star bright enough to host the signal, one of which is TIC 274662220. The total NFPP calculated by \texttt{TRICERATOPS} is 0.3031. TIC 274662220 has a NFPP of 0.0501, making it the 2nd most probable NFP host. This NFPP is high enough to classify the TOI as a likely NFP. In addition, the calculated FPP of 0.6880 is too high to classify the TOI as a likely planet or validated planet.

\subsection{TIC 408203470 (TOI 1289.01)}

TFOP SG1 confirms the true host of the signal is TIC 408203452. The TOI is a NEB, based on observations in a long-pass GG495 filter at the Steward Observatory Phillips 0.6m Telescope on Mount Lemmon that show a 35 ppt eclipse.  Observations at the University of Louisville Moore Observatory show a depth of 60 ppt in $r'$. This star was also correctly identified as the host of the signal by the SPOC centroid offset analysis. \texttt{TRICERATOPS} identifies 10 nearby sources other than the target star bright enough to host the signal, one of which is TIC 408203452. The total NFPP calculated by \texttt{TRICERATOPS} is 0.3258. TIC 408203452 has a NFPP of 0.1435, making it the most probable NFP host. This NFPP is high enough to classify the TOI as a likely NFP. In addition, the calculated FPP of 0.8512 is too high to classify the TOI as a likely planet or validated planet.

\subsection{TIC 233681149 (TOI 1340.01)}

TFOP SG1 confirms the true host of the signal is TIC 233681148. The TOI is a NEB, based on SPOC S14--S16 reports that show a centroid offset to the closest star SW. Single pixel photometry on the \tess\ FFIs supports this conclusion. \texttt{TRICERATOPS} identifies 1 nearby source other than the target star bright enough to host the signal, which is TIC 233681148. The total NFPP calculated by \texttt{TRICERATOPS} is 0.0309. TIC 233681148 has a NFPP of 0.0309, making it the most probable NFP host. This NFPP is too low to classify the TOI as a likely NFP and too high to classify the TOI as a likely planet. In addition, the calculated FPP of 0.0947 is too high to classify the TOI as a validated planet.

\section{Results}\label{sec:6}

\begin{figure*}[b!]
    \includegraphics[width=1.02\textwidth]{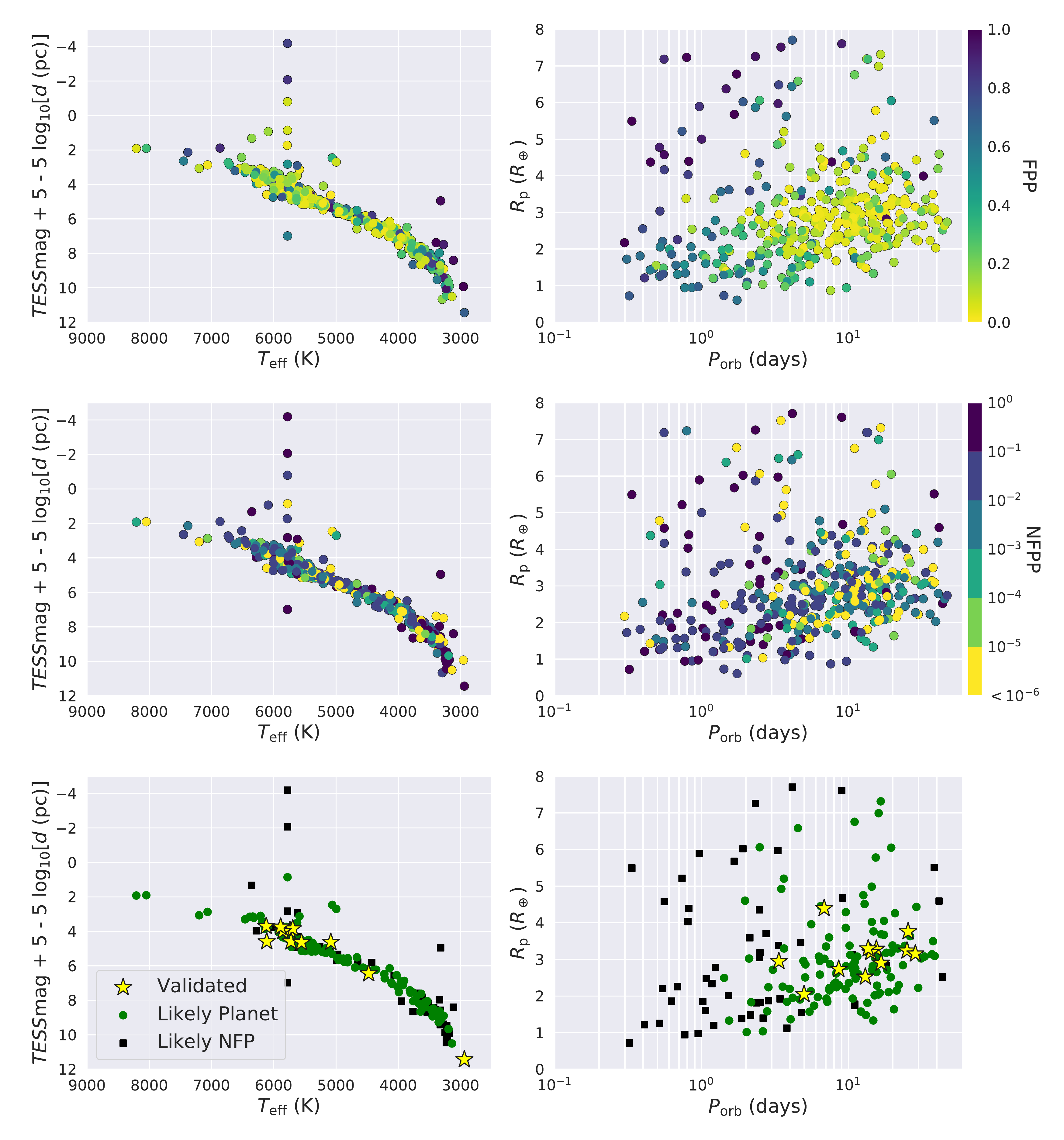}
    \vspace{-15pt}
    \caption{Host star properties (\emph{left}) and planet properties (\emph{right}) of analyzed TOI systems. In the top panels, we color each system according to its FPP. In the center panels, we color each system according to its NFPP. In the bottom panels, we distinguish TOIs that have been classified as validated planets, likely planets, and likely NFPs. In general, planets with smaller radii and longer orbital periods are more likely to be identified as planets. The vertical stack of stars at $T_{\rm eff} \sim 6000$ K are stars with unknown surface temperatures that were assigned a Solar $T_{\rm eff}$ on the TFOP website.}
    \label{fig:6_1}
\end{figure*}

We apply our code to 384 SPOC TOIs that have neither been confirmed as bona fide planets nor rejected as false positives by TFOP. We again restrict our analysis to TOIs with $R_{\rm p} < 8 R_\oplus$, TOIs with host stars that are well characterized in the TIC, and TOIs for which we are able to recover a transit with the purported orbital parameters. However, unlike the sample used in Section \ref{sec:4}, we permit TOIs with orbital periods up to 50 days and extract their light curves using data from all sectors in which they were observed. The results of these calculations are displayed in Figure \ref{fig:6_1} and Table \ref{tab:6_1}.

 In the top panels of Figure \ref{fig:6_1}, we show the host star and planet properties of all TOIs color-coded by FPP. In these panels, we see that TOIs with smaller radii and longer orbital periods tend to have lower FPPs. In the center panels of the figure, we show the same data color-coded by NFPP. In these panels, we again see a propensity for TOIs with smaller radii and longer orbital periods to have lower NFPPs. Nonetheless, there are several TOIs with large radii and short orbital periods that have low NFPP values, which generally represent TOIs without nearby stars bright enough to produce their observed transits. Additionally, we see that neither FPP nor NFPP is closely tied to host spectral type.

In the bottom panels of Figure \ref{fig:6_1}, we present the properties of TOIs that have been classified as validated planets, likely planets, and likely NFPs by our analysis. In total, we statistically validate 12 TOIs, identify 125 TOIs as likely planets, and identify 52 likely NFPs. Our sample of validated TOIs have host stars with a variety of spectral types and planets with radii ranging from 1--5 $R_\oplus$ and orbital periods ranging from 3--30 days. The details for all tested TOIs are given in Table \ref{tab:6_1}. 

The TOI numbers of the planet candidates statistically validated in this analysis are presented in Table \ref{tab:6_2}. Of these, 9 are newly validated and 3 have already been empirically validated via a combination of follow-up observations. The agreement of our statistical validation and the empirical validation of these planet candidates is encouraging for the efficacy of both methods. In addition, we include the FPP calculated by \texttt{TRICERATOPS} in Table \ref{tab:6_2}. Because FPP is expected to have some scatter across runs, we perform the calculation 20 times for each validated TOI and list the mean and standard deviation of the resulting distribution. In doing so, we affirm that our original FPP calculation that validated the planet candidate was not an outlier.

\begin{deluxetable}{ccl}
\tablecaption{Statistically Validated TOIs\label{tab:6_2}}
\tablehead{\colhead{TOI Number} & \colhead{FPP} & \colhead{Original Validation Paper} }
\decimals
\startdata
261.01  & $0.0067 \pm 0.0004$ & this work \\
261.02  & $0.0009 \pm 0.0002$ & this work \\
469.01  & $0.0133 \pm 0.0016$ & this work \\
682.01  & $0.0069 \pm 0.0020$ & this work \\
736.01  & $0.0092 \pm 0.0005$ & \cite{crossfield2019super} \\
836.01  & $0.0141 \pm 0.0019$ & this work \\
1054.01 & $0.0115 \pm 0.0008$ & this work \\
1203.01 & $0.0125 \pm 0.0011$ & this work \\
1230.01 & $0.0132 \pm 0.0005$ & this work \\
1233.01 & $0.0135 \pm 0.0012$ & \cite{daylan2020tess} \\
1339.02 & $0.0127 \pm 0.0011$ & \cite{badenas2020hd} \\
1774.01 & $0.0133 \pm 0.0010$ & this work \\
\enddata
\end{deluxetable}

\section{Discussion}\label{sec:7}

In Figure \ref{fig:6_1} we present the results of \texttt{TRICERATOPS} runs for 384 TOIs, 189 of which are assigned classifications of validated planet, likely planet, or likely NFP. In this figure, a number of patterns emerge that could have broader implications for the population of planets detected by \tess\ and the \tess\ FP rate. As we noted previously, TOIs classified as validated planets or likely planets generally have smaller radii and longer orbital periods. One could interpret this as meaning planets are more common in this region of parameter space. However, we would be remiss if we did not acknowledge that this result is in part due to our choice of $R_{\rm p}$ and $P_{\rm orb}$ priors, which prefer transiting planet scenarios in this region of parameter space. We realize that this effect could be concerning for those who wish to use \texttt{TRICERATOPS} for large-scale statistical studies of planets detected by \tess\, especially in the case where the true underlying prior distributions are unknown, because it could bias their results to agree with previous planet occurrence rate studies. We therefore plan to add alternative prior distributions, such as a uniform prior, that the user can select when they wish their results to be free of such a bias.

To test the extent to which our results are biased by our prior distribution for $R_{\rm p}$, we reran our code on all 384 TOIs with a uniform $R_{\rm p}$ prior. Because our original $R_{\rm p}$ prior penalizes planet candidates with $R_{\rm p} > 5 R_\oplus$, one might expect more of these planet candidates to be classified as validated planets or likely planets when the uniform prior is applied. With the uniform prior, the number of validated planet decreased from 12 to 2 (the number of which with $R_{\rm p} > 5 R_\oplus$ increased from 0 to 1), the number of likely planets decreased from 125 to 93 (the number of which with $R_{\rm p} > 5 R_\oplus$ increased from 8 to 9), and the number of likely NFPs increased from 52 to 93 (the number of which with $R_{\rm p} > 5 R_\oplus$ did not change). These results show that the chance of a planet candidate being classified as a validated planet or a likely NFP is strongly dependent on the choice of $R_{\rm p}$ prior. However, as we do not see a large change in the number of classifications for TOIs with $R_{\rm p} > 5 R_\oplus$, we cannot conclude that our original $R_{\rm p}$ prior significantly biases our results against these TOIs.

Another notable feature of Figure \ref{fig:6_1} is the large number of ultra-short-period planet (i.e., planets with $P_{\rm orb} < 1$ day) TOIs, of which there are 41 with $R_{\rm p} < 8 R_\oplus$. Past studies have found that this type of planet only occurs around $< 1 \%$ of stars \citep{sanchis2014study, adams2016ultra}, but the true rate could be higher if all of these candidates are actual planets. However, this interpretation is dependent on the actual false positive rate of these TOIs. The fact that \texttt{TRICERATOPS} classifies none of these USP candidates as likely planets and many as NFPs suggests that this false positive rate is high. To ensure that this prediction is not an artifact of the aforementioned $P_{\rm orb}$ prior (which is biased towards eclipsing binary scenarios in this region of parameter space), we also repeated our calculations without this prior. Upon removing the prior the number of likely planets increased from 125 to 127, while the number of validated planets and likely NFPs remained the same. The increase can be attributed to three ultra-short-period planet candidates (TOIs 460.01, 561.02, and 864.01) whose classifications were changed from likely NFP to likely planet. This small increase in the number of likely planets suggests that our results are only moderately affected by our $P_{\rm orb}$ prior, and that most ultra-short-period planet candidates are in fact false positives.

\begin{figure}[t!]
        \begin{center}
            \includegraphics[width=0.48\textwidth]{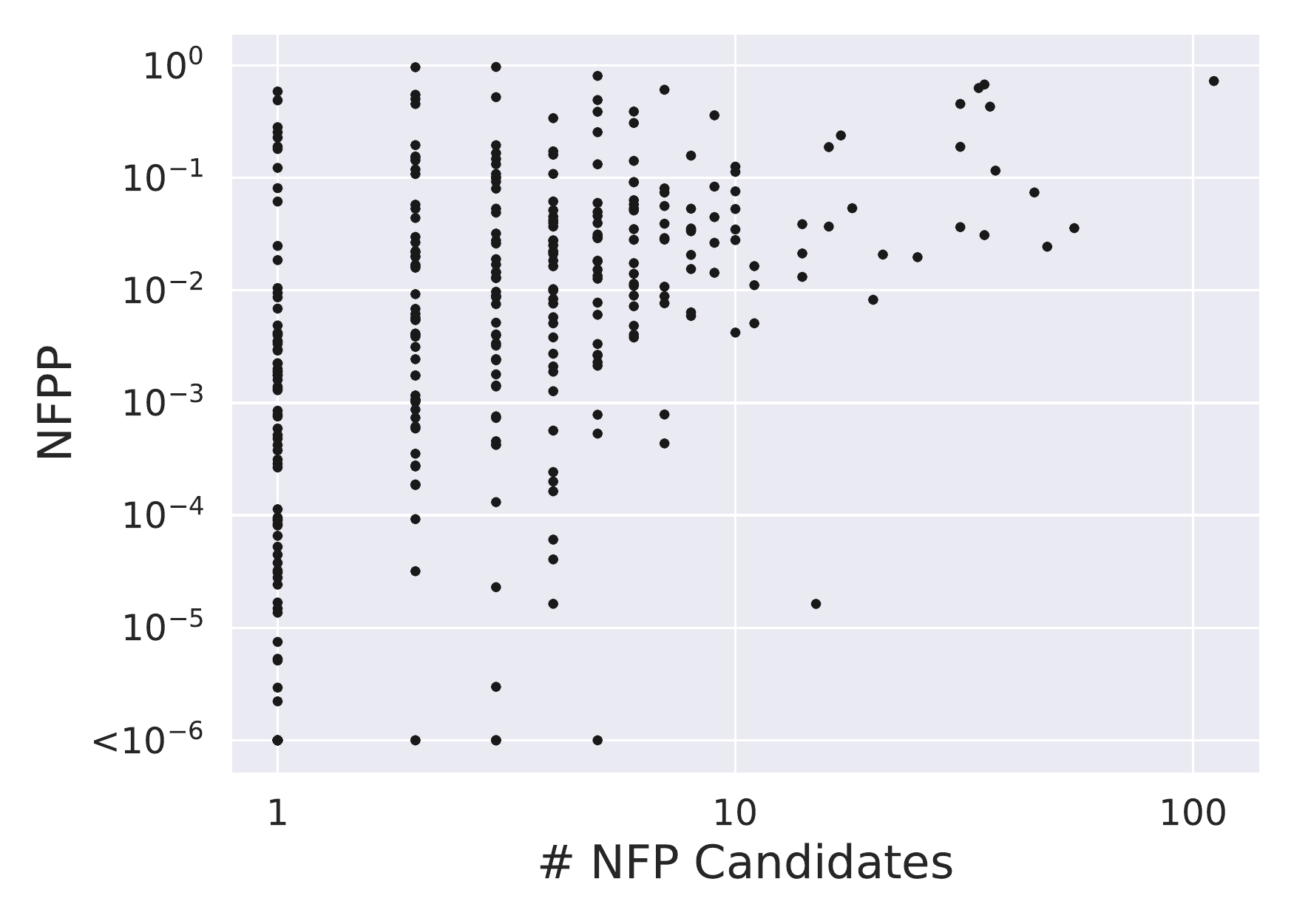}
        \end{center}
    \vspace{-15pt}
    \caption{NFPP versus number of nearby stars bright enough to be a NFP for each of the 384 TOIs tested in Section \ref{sec:6}. TOIs with no potential NFPs are omitted. While the average NFPP increases as the number of possible NFPs increases, \texttt{TRICERATOPS} is able to classify TOIs with any number of nearby host candidates as likely NFPs (${\rm NFPP} > 10^{-1}$).}
    \label{fig:7_1}
\end{figure}

In addition to a statistical validation tool, \texttt{TRICERATOPS} can be used as a vetting tool to prioritize follow-up observations of TOIs. Consider candidates that are classified as a likely planets, but with a FPPs just above the validation threshold. Several TOIs we classify as likely planets match this description, and some (e.g., TOI 1055, Bedell et al. in prep) have been confirmed concurrently with this paper. These TOIs make would ideal targets for high-resolution imaging follow-up, because the resulting data products can be incorporated to achieve a lower FPP and validate the planet candidate. In addition, we displayed in Section \ref{sec:5} that \texttt{TRICERATOPS} is proficient at identifying NFPs, and is often able to predict which nearby star hosts the observed signal. By prioritizing nearby stars with high probabilities of hosting NFPs, observers can increase the rate of true NFP identification. Doing so would allow other members of the \tess\ follow-up community to focus on TOIs that are more likely to be bona fide planets. To display the broad applicability of this prioritization method (i.e., to show that it is not only relevant for TOIs in very crowded fields), we show in Figure \ref{fig:7_1} the NFPP as a function of the number of nearby stars bright enough to be NFPs for the 384 TOIs in our analysis. As one might predict, the expected NFPP increases in more crowded fields. Nonetheless, TOIs with as few as one NFP candidate can be classified as likely NFPs. In other words, \texttt{TRICERATOPS} provides information pertaining to the probability of a given TOI being a NFP beyond what can be gathered from the crowdedness of the surrounding field.

Our tool can also be combined with other validation and vetting tools to provide even more robust validation analyses. As of now, \texttt{TRICERATOPS} is the only validation tool compatible with \tess\ data that models transits from nearby contaminant stars. Seeing as identifying NFPs is one of the strengths of our tool, it can be used as the first step in such an analysis. For example, one could use \texttt{TRICERATOPS} to identify TOIs with sufficiently low NFPPs, and then use tools like \texttt{VESPA} \citep{morton2012efficient, 2015ascl.soft03011M} and \texttt{DAVE} \citep{kostov2019discovery} to further constrain the FPP of the planet candidate around the target star. Additionally, comparing the results of several tools would allow one to build a stronger statistical argument for or against the existence of a planet.

To improve the utility of \texttt{TRICERATOPS}, we plan on adding features that will make the procedure more efficient and robust. First, we will add a feature that searches for in-transit centroid offsets to constrain the probabilities of NFPs. Second, we will improve our priors by expanding to more dimensions that affect planet occurrence rates, such as planet multiplicity. In this vein, it has been shown that planet candidates that are members of systems with multiple planet candidates are almost always bona fide planets \citep[e.g.,][]{lissauer2012almost}. This in and of itself is strong evidence that candidate multi-planet systems with validated planets (including TOIs 736, 836, 1233, and 1339) actually host multiple transiting planets. Third, we will make our tool compatible with additional follow-up constraints, such as time-series photometry that rules out signals around nearby stars and spectroscopic observations that provide limits on eclipsing binary properties, to improve its ability to validate planet candidates. Lastly, we will add additional astrophysical scenarios to our calculation procedure, such as that involving a non-circular orbit and that involving an eclipsing binary where only the secondary eclipse is detected.

\section{Conclusions}\label{sec:8}

We present a new tool, \texttt{TRICERATOPS}, designed for rapid validation of \tess\ Objects of Interest. Using a Bayesian framework, this tool calculates the probabilities of various transit-producing scenarios for a given TOI in order to provide a false positive probability (FPP) and a nearby false positive probability (NFPP). Our tool is also able to fold in information from follow-up observations as additional constraints in these calculations.

We test our tool on 68 TOIs that have been designated as either confirmed planets or astrophysical false positives by members of the \tess\ Observation Follow-up Program (TFOP) based on follow-up observations. We define three classifications based on the results of this analysis. For a TOI to be validated, it must have high cadence observations, ${\rm NFPP} < 10^{-3}$, and ${\rm FPP} < 0.015$. For a TOI to be classified as a likely planet, it must have ${\rm NFPP} < 10^{-3}$ and ${\rm FPP} < 0.5$. Lastly, for a TOI to be classified as a likely nearby false positive (NFP), it must have ${\rm NFPP} > 10^{-1}$. To display the proficiency of our tool in identifying NFPs, we also compare our predictions to TOIs that have been identified as actual NFPs by TFOP.

We apply our tool to 384 TOIs with 2-minute cadence observations that have not yet been classified as confirmed planets or rejected as false positives. We statistically validate 12 TOIs, classify 125 TOIs as likely planets, and classify 52 TOIs as likely NFPs.

In addition to planet validation, we recommend using \texttt{TRICERATOPS} to identify TOIs with high probabilities of being planets or NFPs and prioritizing these candidates as targets for further vetting via follow-up observations. When used in combination with other vetting tools, such as \texttt{VESPA} and \texttt{DAVE}, our tool can also be utilized to perform even more thorough validation analyses of planet candidates. We hope this tool will be a valuable resource in the search for planets with \tess.

\acknowledgements
 We thank the referee Timothy D. Morton (the creator of the \texttt{VESPA} analysis) for his careful analysis and review, in response to which both the \texttt{TRICERATOPS} analysis and this paper have been significantly improved. We thank Stephen T. Bryson, Jack J. Lissauer, Arjun B. Savel, and David W. Latham for helpful conversations and guidance that improved this paper. We thank Jonathan M. Irwin for the collection and contribution of MEarth data used in this paper. We also thank the \tess\ follow-up community for making this work possible through their efforts to vet planet candidates. 
 
 We thank the NASA \tess\ Guest Investigator Program for supporting this work through grant 80NSSC18K1583 (awarded to CDD). SG and CDD also appreciate and acknowledge support from the Hellman Fellows Fund, the Alfred P. Sloan Foundation, the David and Lucile Packard Foundation, and the NASA Exoplanets Research Program (XRP) through grant 80NSSC20K0250. 
 
 This work makes use of observations from the LCOGT network. This material is based upon work supported by the National Science Foundation Graduate Research Fellowship Program under Grant No. DGE-1650115. The research leading to these results has received funding from  the ARC grant for Concerted Research Actions, financed by the Wallonia-Brussels Federation. TRAPPIST is funded by the Belgian Fund for Scientific Research (Fond National de la Recherche Scientifique, FNRS) under the grant FRFC 2.5.594.09.F, with the participation of the Swiss National Science Fundation (SNF). MG and EJ are F.R.S.-FNRS Senior Research Associates. The MEarth Team gratefully acknowledges the David and Lucile Packard Fellowship for Science and Engineering (awarded to D. C.), continued support by the NSF mostly recently under grant AST-1616624, and support by NASA under grant 80NSSC18K0476 (XRP Program). This work is made possible by a grant from the John Templeton Foundation. The opinions expressed in this publication are those of the authors and do not necessarily reflect the views of the John Templeton Foundation. JNW thanks the Heising-Simons Foundation for support. 
 
 Funding for the TESS mission is provided by NASA's Science Mission directorate. This research has made use of the Exoplanet Follow-up Observation Program website, which is operated by the California Institute of Technology, under contract with the National Aeronautics and Space Administration under the Exoplanet Exploration Program. This paper includes data collected by the TESS mission, which are publicly available from the Mikulski Archive for Space Telescopes (MAST). We acknowledge the use of public TESS Alert data from pipelines at the TESS Science Office and TESS Science Processing Operations Center. Resources supporting this work were provided by the NASA High-End Computing (HEC) Program through the NASA Advanced Supercomputing (NAS) Division at Ames Research Center for the production of the SPOC data products.

\facilities{\tess, Shane (ShARCS infrared camera), CDK20N, Univ. Louisville Manner Telescope, CDK700, Hazelwood Observatory, LCOGT, NEarth-North, MuSCAT2, El Sauce Observatory, PEST, HATNet, HAT-South, TRAPPIST-South, Steward Observatory Phillips Telescope}

\software{\texttt{NumPy} \citep{oliphant2006guide}, \texttt{SciPy} \citep{2019arXiv190710121V}, \texttt{pandas} \citep{mckinney2010data} \texttt{matplotlib} \citep{hunter2007matplotlib}, \texttt{astropy} \citep{price2018astropy}, \texttt{beautifulsoup4} \citep{richardson2007beautiful}, \texttt{batman} \citep{kreidberg2015batman}, \texttt{lightkurve} \citep{2018ascl.soft12013L}, \texttt{eleanor} \citep{feinstein2019eleanor}, \texttt{VESPA} \citep{morton2012efficient,2015ascl.soft03011M}}

\clearpage

% \bibliography{paper}
\bibliography{parxiv.bbl}

\startlongtable
\begin{deluxetable*}{ccccccccc}
\tabletypesize{\footnotesize}
\tablewidth{\textwidth}
 \tablecaption{ \texttt{TRICERATOPS} Predictions for Undesignated TOIs  \label{tab:6_1}}
 \tablehead{
 \colhead{TIC ID} & \colhead{TOI} & \colhead{$R_{\rm p}$} & \colhead{$P_{\rm orb}$} & \colhead{SNR} & \colhead{FPP} & \colhead{NFPP} & \colhead{$\#$ NFP} & \colhead{Classification} \\[-8pt] 
 \colhead{} & \colhead{Number} & \colhead{($R_\oplus$)} & \colhead{(days)} &  & \colhead{} & \colhead{} & \colhead{Candidates} & \colhead{}
 }
 \startdata 
  278683844 &     119.01 &          2.13 &         5.54 &   8.3 &   0.04 &  9.25e-05 &                 2 &  Likely Planet \\
  278683844 &     119.02 &          1.93 &        10.69 &   7.0 &   0.06 &  1.88e-04 &                 2 &  Likely Planet \\
  231702397 &     122.01 &          2.51 &         5.08 &   6.6 &   0.06 &  2.79e-05 &                 1 &  Likely Planet \\
   52368076 &     125.03 &          3.38 &        19.98 &   4.0 &   0.04 &  0.00e+00 &                 0 &  Likely Planet \\
  391949880 &     128.01 &          3.06 &         4.94 &   6.9 &   0.10 &  2.76e-02 &                 4 &                \\
  263003176 &     130.01 &          2.32 &        14.34 &   4.3 &   0.04 &  4.02e-03 &                 2 &                \\
   89020549 &     132.01 &          3.02 &         2.11 &  12.4 &   0.03 &  8.12e-05 &                 1 &  Likely Planet \\
  219338557 &     133.01 &          2.37 &         8.20 &  10.5 &   0.05 &  0.00e+00 &                 1 &  Likely Planet \\
  234994474 &     134.01 &          1.49 &         1.40 &  17.3 &   0.11 &  7.62e-03 &                 4 &                \\
   62483237 &     139.01 &          2.93 &        11.06 &  13.0 &   0.03 &  0.00e+00 &                 0 &  Likely Planet \\
  425997655 &     174.03 &          1.57 &        12.16 &   1.6 &   0.05 &  5.16e-04 &                 1 &  Likely Planet \\
  425997655 &     174.04 &          1.12 &         3.98 &   1.8 &   0.30 &  4.19e-03 &                 1 &                \\
  262530407 &     177.01 &          2.24 &         2.85 &  21.9 &   0.08 &  5.91e-04 &                 2 &  Likely Planet \\
  251848941 &     178.01 &          2.87 &         6.56 &  12.1 &   0.04 &  4.11e-03 &                 1 &                \\
  251848941 &     178.02 &          3.14 &        10.35 &  12.1 &   0.07 &  1.76e-03 &                 1 &                \\
  251848941 &     178.03 &          2.43 &         9.96 &   9.9 &   0.07 &  9.49e-03 &                 1 &                \\
  207141131 &     179.01 &          2.98 &         4.14 &  21.1 &   0.02 &  0.00e+00 &                 0 &  Likely Planet \\
   76923707 &     181.01 &          6.58 &         4.53 &  38.7 &   0.24 &  1.63e-04 &                 4 &  Likely Planet \\
  183985250 &     193.01 &          7.23 &         0.79 &  35.6 &   0.99 &  1.75e-03 &                 2 &                \\
   12421862 &     198.01 &          1.64 &        20.43 &  11.6 &   0.13 &  1.48e-05 &                 1 &  Likely Planet \\
  350618622 &     201.02 &          1.74 &         5.85 &   1.9 &   0.45 &  0.00e+00 &                 0 &  Likely Planet \\
  281781375 &     204.01 &          2.52 &        43.83 &   1.1 &   0.27 &  1.66e-01 &                 3 &     Likely NFP \\
  281575427 &     205.01 &          3.03 &         4.25 &   8.8 &   0.09 &  1.42e-03 &                 3 &                \\
   55650590 &     206.01 &          5.22 &         0.74 &   6.2 &   0.73 &  1.60e-01 &                 4 &     Likely NFP \\
  314865962 &     208.01 &          2.97 &        22.45 &   4.0 &   0.02 &  2.45e-03 &                 3 &                \\
   52204645 &     209.01 &          3.02 &         4.38 &   7.1 &   0.07 &  2.73e-03 &                 4 &                \\
  141608198 &     210.01 &          7.61 &         9.01 &   7.2 &   0.92 &  1.32e-01 &                 5 &     Likely NFP \\
  206609630 &     212.01 &          5.49 &         0.34 &  54.2 &   1.00 &  5.02e-01 &                 2 &     Likely NFP \\
  234345288 &     213.01 &          2.84 &        23.52 &  10.9 &   0.20 &  4.24e-04 &                 3 &  Likely Planet \\
  167415965 &     214.02 &          0.94 &         9.70 &   2.1 &   0.34 &  7.41e-02 &                45 &                \\
  231912935 &     215.01 &          3.63 &        26.30 &   7.9 &   0.23 &  3.18e-05 &                 2 &  Likely Planet \\
  150098860 &     220.01 &          3.53 &        10.70 &   9.2 &   0.10 &  2.81e-02 &                 6 &                \\
  316937670 &     221.01 &          1.86 &         0.62 &   8.1 &   0.69 &  1.95e-01 &                 2 &     Likely NFP \\
  326453034 &     223.01 &          4.02 &        14.45 &   3.1 &   0.36 &  0.00e+00 &                 1 &  Likely Planet \\
  160074939 &     230.01 &          7.19 &        13.34 &  12.9 &   0.12 &  1.86e-02 &                 1 &                \\
  415969908 &     233.01 &          2.60 &        11.67 &  11.3 &   0.06 &  0.00e+00 &                 0 &  Likely Planet \\
  305048087 &     237.01 &          1.57 &         5.43 &   3.3 &   0.08 &  0.00e+00 &                 0 &  Likely Planet \\
    9006668 &     238.01 &          1.61 &         1.27 &   7.2 &   0.35 &  5.77e-02 &                 2 &                \\
  101948569 &     240.01 &          3.26 &        19.47 &  11.5 &   0.03 &  9.71e-03 &                 3 &                \\
  118327550 &     244.01 &          3.60 &         7.40 &  11.4 &   0.50 &  0.00e+00 &                 0 &  Likely Planet \\
  201793781 &     248.01 &          2.85 &         5.99 &  11.5 &   0.02 &  2.44e-03 &                 2 &                \\
  179985715 &     249.01 &          2.64 &         6.61 &   9.0 &   0.06 &  1.30e-02 &                 3 &                \\
  224225541 &     251.01 &          2.96 &         4.94 &   8.1 &   0.02 &  6.57e-05 &                 1 &  Likely Planet \\
  237924601 &     252.01 &          5.00 &         1.00 &  11.9 &   0.93 &  8.02e-02 &                 3 &                \\
  322063810 &     253.01 &          1.21 &         3.52 &  11.9 &   0.22 &  2.23e-02 &                 2 &                \\
   37749396 &     260.01 &          1.81 &        13.47 &   3.7 &   0.20 &  0.00e+00 &                 0 &  Likely Planet \\
   63898957 &     261.01 &          2.95 &         3.36 &   5.8 &$<$0.01 &  0.00e+00 &                 0 &      Validated \\
   63898957 &     261.02 &          2.52 &        13.04 &   1.3 &$<$0.01 &  0.00e+00 &                 0 &      Validated \\
   70513361 &     262.01 &          2.75 &        11.15 &  12.9 &   0.06 &  0.00e+00 &                 0 &  Likely Planet \\
  120916706 &     263.01 &          4.58 &         0.56 &  20.7 &   0.97 &  4.88e-01 &                 1 &     Likely NFP \\
  164767175 &     266.01 &          2.72 &        10.77 &   8.9 &   0.03 &  0.00e+00 &                 0 &  Likely Planet \\
  164767175 &     266.02 &          1.97 &         6.19 &   8.3 &   0.05 &  0.00e+00 &                 0 &  Likely Planet \\
  259511357 &     271.01 &          4.35 &         2.48 &  12.1 &   0.73 &  2.54e-01 &                 1 &     Likely NFP \\
  281979481 &     274.01 &          2.21 &         0.54 &  12.1 &   0.67 &  1.41e-01 &                 6 &     Likely NFP \\
  439456714 &     277.01 &          4.23 &         3.99 &  18.6 &   0.14 &  8.09e-02 &                 1 &                \\
  244161191 &     278.01 &          2.17 &         0.30 &  21.2 &   1.00 &  0.00e+00 &                 0 &                \\
  122613513 &     279.01 &          3.08 &        11.49 &  10.9 &   0.05 &  0.00e+00 &                 0 &  Likely Planet \\
   42054565 &     280.01 &          2.87 &        10.18 &  10.0 &   0.06 &  1.39e-03 &                 1 &                \\
   38696105 &     281.01 &          4.09 &         5.57 &  12.7 &   0.19 &  5.49e-03 &                 2 &                \\
   29781292 &     282.02 &          3.03 &        31.32 &   3.6 &   0.23 &  7.36e-04 &                 2 &  Likely Planet \\
  382626661 &     283.01 &          2.08 &        17.62 &   5.8 &   0.05 &  1.76e-03 &                 1 &                \\
  220459976 &     285.01 &          2.65 &        32.33 &   3.0 &   0.06 &  1.87e-03 &                 1 &                \\
  150030205 &     286.01 &          1.51 &         4.51 &   5.9 &   0.22 &  8.38e-03 &                 4 &                \\
  150030205 &     286.02 &          2.03 &        39.36 &   6.2 &   0.06 &  2.98e-03 &                 1 &                \\
  153065527 &     406.01 &          1.48 &        13.17 &  10.2 &   0.02 &  2.66e-04 &                 1 &  Likely Planet \\
  100990000 &     411.01 &          2.19 &         9.57 &   8.3 &   0.05 &  0.00e+00 &                 0 &  Likely Planet \\
  100990000 &     411.02 &          1.36 &         4.04 &   4.6 &   0.23 &  0.00e+00 &                 0 &  Likely Planet \\
   94986319 &     421.01 &          6.99 &        16.07 &  23.2 &   0.10 &  2.77e-04 &                 2 &  Likely Planet \\
   31374837 &     431.02 &          1.55 &         0.49 &  16.1 &   0.14 &  1.60e-03 &                 1 &                \\
   44647437 &     435.01 &          6.48 &         3.35 &  12.0 &   0.79 &  5.91e-04 &                 1 &                \\
  179034327 &     444.01 &          2.95 &        17.96 &   9.0 &   0.01 &  1.06e-03 &                 2 &                \\
  153077621 &     454.01 &          2.83 &        18.08 &  23.8 &   0.97 &  9.61e-01 &                 2 &     Likely NFP \\
   89256802 &     457.01 &          2.34 &         1.18 &  24.7 &   0.72 &  1.80e-01 &                 1 &     Likely NFP \\
   64071894 &     458.01 &          3.05 &        17.53 &   5.1 &   0.03 &  1.34e-03 &                 1 &                \\
    9804616 &     460.01 &          4.78 &         0.52 &  44.5 &   0.92 &  0.00e+00 &                 0 &                \\
    4646810 &     461.01 &          2.71 &        14.49 &   5.3 &   0.10 &  0.00e+00 &                 0 &  Likely Planet \\
  420049884 &     462.01 &          2.51 &         4.11 &   6.2 &   0.14 &  4.51e-02 &                 4 &                \\
  398733009 &     464.01 &          4.40 &         0.82 &  15.4 &   0.99 &  5.85e-01 &                 1 &     Likely NFP \\
   33692729 &     469.01 &          3.29 &        13.63 &  12.7 &   0.01 &  8.32e-12 &                 2 &      Validated \\
   37770169 &     470.01 &          4.50 &        12.19 &  12.5 &   0.11 &  1.89e-03 &                 4 &                \\
  100608026 &     475.01 &          2.62 &         8.26 &  12.5 &   0.03 &  7.83e-04 &                 5 &  Likely Planet \\
  317548889 &     480.01 &          3.02 &         6.87 &  18.7 &   0.04 &  1.36e-05 &                 1 &  Likely Planet \\
  427348923 &     484.01 &          3.45 &         4.73 &  12.0 &   0.83 &  8.05e-01 &                 5 &     Likely NFP \\
  260708537 &     486.01 &          0.60 &         1.74 &   4.6 &   0.62 &  3.50e-02 &                 6 &                \\
   31852980 &     487.01 &          2.44 &        24.33 &   3.5 &   0.14 &  5.73e-03 &                 2 &                \\
  452866790 &     488.01 &          1.20 &         1.20 &   9.3 &   0.24 &  2.18e-02 &                 2 &                \\
   19025965 &     493.01 &          3.95 &         5.95 &  16.3 &   0.10 &  6.15e-02 &                 1 &                \\
   19519368 &     494.01 &          2.46 &         1.70 &   8.5 &   0.28 &  3.35e-02 &                 8 &                \\
  123702439 &     499.01 &          4.12 &         8.52 &  21.0 &   0.13 &  4.19e-02 &                 4 &                \\
  134200185 &     500.01 &          1.32 &         0.55 &  12.6 &   0.74 &  3.47e-02 &                10 &                \\
  453211454 &     509.01 &          3.18 &        18.12 &  13.1 &   0.02 &  0.00e+00 &                 1 &  Likely Planet \\
  238086647 &     510.01 &          3.57 &         1.35 &  1.2 &   0.64 &  3.69e-02 &                 4 &                \\
  119292328 &     512.01 &          1.84 &         7.19 &  10.3 &   0.07 &  4.55e-04 &                 3 &  Likely Planet \\
  264979636 &     518.01 &          4.51 &        17.88 &   2.3 &   0.67 &  9.21e-05 &                 1 &                \\
  148479278 &     520.01 &          2.05 &         0.52 &   6.3 &   0.58 &  2.79e-02 &                10 &                \\
   27649847 &     521.01 &          1.33 &         1.54 &  12.8 &   0.34 &  0.00e+00 &                 0 &  Likely Planet \\
   19451711 &     522.01 &          2.55 &         0.40 &  12.5 &   0.77 &  8.85e-03 &                 7 &                \\
   71512186 &     525.01 &          3.76 &        14.82 &   6.8 &   0.02 &  8.36e-05 &                 1 &  Likely Planet \\
  200593988 &     526.01 &          4.38 &         7.70 &  23.4 &   0.99 &  2.92e-02 &                 5 &                \\
  144700903 &     532.01 &          5.87 &         2.33 &  18.7 &   0.58 &  8.05e-02 &                 7 &                \\
  309791156 &     533.01 &          6.05 &        19.57 &  15.8 &   0.45 &  2.30e-05 &                 3 &  Likely Planet \\
  237751146 &     538.01 &          5.68 &         1.67 &  24.0 &   1.00 &  9.67e-01 &                 3 &     Likely NFP \\
  238004786 &     539.01 &          1.72 &         0.31 &  11.7 &   0.64 &  7.58e-02 &                10 &                \\
   50618703 &     544.01 &          1.94 &         1.55 &  19.4 &   0.11 &  5.77e-03 &                 4 &                \\
  161477033 &     553.02 &          2.51 &        11.93 &   9.7 &   0.04 &  2.01e-03 &                 1 &                \\
  161477033 &     553.03 &          2.79 &        40.90 &   6.3 &   0.04 &  2.25e-03 &                 1 &                \\
  407966340 &     554.01 &          3.35 &         7.05 &   9.1 &   0.14 &  4.80e-04 &                 1 &  Likely Planet \\
   55488511 &     557.01 &          2.56 &         3.35 &   1.4 &   0.20 &  3.47e-03 &                 1 &                \\
  101011575 &     560.01 &          2.95 &         6.40 &  16.0 &   0.02 &  0.00e+00 &                 1 &  Likely Planet \\
  377064495 &     561.01 &          2.86 &        10.78 &  10.0 &   0.02 &  0.00e+00 &                 0 &  Likely Planet \\
  377064495 &     561.02 &          1.43 &         0.45 &   8.7 &   0.58 &  0.00e+00 &                 0 &                \\
  377064495 &     561.03 &          2.08 &        16.37 &   4.1 &   0.05 &  0.00e+00 &                 0 &  Likely Planet \\
   37575651 &     568.01 &          3.86 &         9.60 &   9.7 &   0.07 &  1.37e-17 &                 1 &  Likely Planet \\
  126733133 &     570.01 &          6.38 &         1.47 &  24.2 &   0.94 &  7.87e-04 &                 7 &                \\
  296739893 &     620.01 &          2.86 &         5.10 &  17.7 &   0.07 &  0.00e+00 &                 1 &  Likely Planet \\
  133334108 &     637.01 &          1.87 &         2.85 &  11.6 &   0.26 &  1.13e-01 &                10 &     Likely NFP \\
   22221375 &     652.01 &          2.20 &         3.98 &  12.0 &   0.07 &  7.49e-06 &                 1 &  Likely Planet \\
   35009898 &     654.01 &          2.01 &         1.53 &  23.5 &   0.35 &  1.89e-01 &                 1 &     Likely NFP \\
  124573851 &     669.01 &          3.63 &         3.95 &  13.3 &   0.03 &  3.97e-03 &                 3 &                \\
  151825527 &     672.01 &          5.20 &         3.63 &  26.8 &   0.08 &  5.68e-07 &                 2 &  Likely Planet \\
  158588995 &     674.01 &          4.60 &         1.98 &  43.1 &   0.03 &  0.00e+00 &                 1 &  Likely Planet \\
  294395926 &     678.01 &          4.18 &        11.32 &  18.4 &   0.12 &  5.79e-02 &                 6 &                \\
  429304876 &     682.01 &          4.39 &         6.84 &  17.3 &$<$0.01 &  0.00e+00 &                 0 &      Validated \\
   77156829 &     696.01 &          0.95 &         0.86 &   8.1 &   0.60 &  6.17e-02 &                 4 &                \\
   77156829 &     696.02 &          1.33 &        14.78 &   6.6 &   0.26 &  2.42e-04 &                 4 &  Likely Planet \\
   77253676 &     697.01 &          2.29 &         8.61 &   9.8 &   0.04 &  0.00e+00 &                 0 &  Likely Planet \\
  141527579 &     698.01 &          2.02 &        15.09 &   7.7 &   0.11 &  6.14e-04 &                 2 &  Likely Planet \\
  149302744 &     699.01 &          3.26 &        14.80 &   3.1 &   0.23 &  3.65e-02 &                31 &                \\
  149302744 &     699.02 &          3.36 &        33.62 &   2.9 &   0.17 &  3.09e-02 &                35 &                \\
  150428135 &     700.02 &          3.03 &        37.42 &  -0.3 &   0.16 &  7.21e-03 &                 6 &                \\
  150428135 &     700.03 &          2.75 &         9.98 &   4.2 &   0.08 &  4.03e-03 &                 6 &                \\
  237914496 &     702.01 &          2.25 &         3.57 &   8.8 &   0.25 &  7.56e-04 &                 1 &  Likely Planet \\
  237928815 &     703.01 &          2.37 &         8.67 &   7.1 &   0.06 &  2.24e-03 &                 1 &                \\
  237928815 &     703.02 &          2.64 &        45.12 &   1.3 &   0.07 &  2.91e-03 &                 1 &                \\
  260004324 &     704.01 &          1.12 &         3.81 &   1.9 &   0.48 &  2.38e-01 &                17 &     Likely NFP \\
  391904697 &     705.01 &          2.74 &        47.02 &   1.8 &   0.16 &  2.51e-02 &                 4 &                \\
  396720998 &     709.01 &          3.99 &        32.38 &  25.9 &   1.00 &  0.00e+00 &                 0 &                \\
   38510224 &     711.01 &          2.32 &        18.38 &   6.7 &   0.30 &  5.44e-03 &                 2 &                \\
  150151262 &     712.01 &          2.92 &         9.53 &  10.3 &   0.13 &  1.28e-02 &                 3 &                \\
  167600516 &     713.01 &          2.23 &        36.00 &   6.4 &   0.07 &  7.54e-03 &                 3 &                \\
  167600516 &     713.02 &          1.39 &         1.87 &   3.9 &   0.50 &  2.91e-02 &                 7 &                \\
  219195044 &     714.01 &          3.46 &         4.32 &   8.4 &   0.19 &  8.73e-03 &                 3 &                \\
  219195044 &     714.02 &          3.56 &        10.18 &   5.7 &   0.16 &  5.15e-03 &                 3 &                \\
   38571020 &     721.01 &          2.72 &        12.29 &   1.9 &   0.17 &  1.27e-02 &                 5 &                \\
   38509907 &     722.01 &          3.36 &        15.30 &   4.5 &   0.20 &  4.97e-02 &                 5 &                \\
  177077336 &     723.01 &          1.30 &         1.42 &   4.0 &   0.59 &  3.96e-02 &                 5 &                \\
   34068865 &     731.01 &          0.72 &         0.32 &  10.3 &   0.72 &  1.26e-01 &                10 &     Likely NFP \\
   36724087 &     732.01 &          0.94 &         0.77 &  16.1 &   0.57 &  2.54e-01 &                 5 &     Likely NFP \\
   36724087 &     732.02 &          1.73 &        12.25 &  17.4 &   0.20 &  3.97e-02 &                 4 &                \\
  106402532 &     733.01 &          2.01 &         4.89 &   8.3 &   0.09 &  7.67e-03 &                 7 &                \\
  181804752 &     736.01 &          2.05 &         4.99 &  27.2 &$<$0.01 &  1.08e-10 &                 1 &      Validated \\
  181804752 &     736.02 &          0.97 &         0.95 &  13.3 &   0.71 &  1.47e-01 &                 3 &     Likely NFP \\
  219189765 &     737.01 &          6.78 &         1.73 &  20.9 &   1.00 &  0.00e+00 &                 0 &                \\
  310009611 &     740.01 &          3.59 &         2.13 &  13.3 &   0.78 &  7.24e-01 &               111 &     Likely NFP \\
  359271092 &     741.01 &          0.87 &         7.58 &   6.4 &   0.13 &  1.32e-02 &                14 &                \\
  444842193 &     745.01 &          2.47 &         1.08 &   9.1 &   0.53 &  1.89e-01 &                31 &     Likely NFP \\
   73228647 &     755.01 &          1.97 &         2.54 &  11.0 &   0.20 &  4.47e-02 &                 9 &                \\
   73649615 &     756.01 &          2.78 &         1.24 &  14.2 &   0.56 &  3.59e-01 &                 9 &     Likely NFP \\
  130924120 &     757.01 &          3.67 &        17.47 &  11.5 &   0.03 &  5.25e-05 &                 1 &  Likely Planet \\
  165317334 &     761.01 &          3.10 &        10.56 &  13.2 &   0.05 &  2.78e-02 &                 4 &                \\
  178709444 &     762.01 &          7.51 &         3.47 &  31.6 &   0.94 &  0.00e+00 &                 0 &                \\
  178819686 &     763.01 &          3.29 &         5.60 &   5.8 &   0.06 &  1.84e-02 &                 4 &                \\
  178819686 &     763.02 &          3.40 &        12.28 &   5.7 &   0.04 &  9.95e-03 &                 4 &                \\
  219401954 &     765.01 &          2.54 &         0.86 &   7.5 &   0.14 &  1.92e-03 &                 1 &                \\
  277634430 &     771.01 &          7.26 &         2.33 &   9.0 &   0.95 &  3.07e-01 &                 6 &     Likely NFP \\
  286864983 &     772.01 &          6.76 &        11.02 &  23.9 &   0.23 &  1.84e-34 &                 1 &  Likely Planet \\
  306996324 &     776.01 &          2.28 &        15.66 &  11.6 &   0.04 &  1.31e-04 &                 3 &  Likely Planet \\
  306996324 &     776.02 &          1.68 &         8.24 &   6.6 &   0.05 &  3.38e-03 &                 3 &                \\
  334305570 &     777.01 &          7.32 &        16.60 &  18.3 &   0.11 &  0.00e+00 &                 1 &  Likely Planet \\
  374095457 &     779.01 &          3.38 &         0.78 &  24.8 &   0.07 &  1.08e-02 &                 7 &                \\
  429358906 &     782.01 &          2.04 &        16.05 &   9.4 &   0.12 &  0.00e+00 &                 0 &  Likely Planet \\
  451645081 &     783.01 &          3.09 &        16.23 &   8.0 &   0.29 &  1.63e-05 &                15 &  Likely Planet \\
  460984940 &     784.01 &          2.14 &         2.80 &  13.4 &   0.22 &  5.37e-02 &                18 &                \\
  374829238 &     785.01 &          4.13 &        18.64 &  12.1 &   0.31 &  2.08e-02 &                21 &                \\
  375059587 &     786.01 &          2.45 &        12.67 &   3.9 &   0.15 &  2.11e-03 &                 4 &                \\
  350584963 &     787.01 &          1.62 &         2.13 &   2.9 &   0.69 &  2.63e-03 &                 5 &                \\
  349829627 &     788.01 &          4.46 &         6.49 &   4.1 &   0.08 &  5.32e-04 &                 5 &  Likely Planet \\
  300710077 &     789.01 &          1.10 &         5.45 &   6.7 &   0.44 &  9.17e-02 &                 6 &                \\
  278895705 &     795.01 &          3.49 &         8.76 &   3.5 &   0.20 &  5.92e-03 &                 8 &                \\
  277099925 &     796.01 &          4.03 &         0.81 &  11.5 &   0.81 &  4.29e-01 &                36 &     Likely NFP \\
  271596225 &     797.01 &          1.32 &         1.80 &   6.1 &   0.62 &  7.40e-02 &                 7 &                \\
  271596225 &     797.02 &          1.49 &         4.14 &   3.3 &   0.51 &  5.13e-02 &                 6 &                \\
  255685030 &     799.01 &          2.11 &         5.54 &   6.6 &   0.20 &  8.72e-03 &                 3 &                \\
  179308757 &     800.01 &          5.89 &         0.97 &   7.0 &   0.85 &  6.75e-01 &                35 &     Likely NFP \\
  177258735 &     801.01 &          1.34 &         0.78 &   3.0 &   0.61 &  6.36e-03 &                 8 &                \\
  167303382 &     802.01 &          0.98 &         3.69 &   3.2 &   0.24 &  1.43e-02 &                 9 &                \\
   41227743 &     804.01 &          1.77 &         1.42 &   4.4 &   0.49 &  1.97e-02 &                25 &                \\
   38460940 &     805.01 &          6.44 &         4.12 &   7.1 &   0.59 &  9.24e-03 &                 2 &                \\
   33831980 &     806.01 &          4.14 &        21.92 &   8.9 &   0.14 &  1.02e-03 &                 2 &                \\
   30853470 &     807.01 &          1.81 &         5.27 &   3.2 &   0.38 &  1.75e-03 &                 2 &                \\
   30122649 &     808.01 &          2.69 &         9.74 &  12.3 &   0.02 &  3.81e-03 &                 6 &                \\
  388106759 &     810.01 &          2.66 &        28.30 &   1.9 &   0.17 &  1.14e-02 &                 6 &                \\
  125405602 &     821.01 &          3.18 &        13.82 &  11.4 &   0.14 &  0.00e+00 &                 0 &  Likely Planet \\
  158978373 &     823.01 &          7.19 &        13.54 &   9.7 &   0.31 &  1.98e-02 &                 2 &                \\
  276128561 &     829.01 &          4.86 &         3.29 &  16.4 &   0.28 &  8.35e-02 &                 9 &                \\
  350332997 &     832.01 &          6.02 &         1.92 &  18.5 &   0.72 &  1.19e-01 &                 2 &     Likely NFP \\
  405700729 &     835.01 &          3.59 &         4.79 &  18.8 &   0.57 &  1.55e-02 &                 8 &                \\
  440887364 &     836.01 &          2.73 &         8.59 &  14.1 &   0.01 &  0.00e+00 &                 0 &      Validated \\
  440887364 &     836.02 &          1.84 &         3.82 &   5.8 &   0.07 &  8.50e-04 &                 1 &  Likely Planet \\
  238898571 &     863.01 &          1.27 &         0.53 &   2.2 &   0.74 &  5.61e-02 &                 7 &                \\
  231728511 &     864.01 &          3.04 &         0.52 &   5.7 &   0.81 &  7.85e-04 &                 1 &                \\
  358460246 &     867.01 &          2.49 &        15.40 &   2.5 &   0.16 &  2.83e-02 &                 7 &                \\
  200807066 &     869.01 &          2.62 &        26.48 &   5.9 &   0.08 &  2.39e-03 &                 3 &                \\
  219229644 &     870.01 &          2.30 &        22.04 &   2.8 &   0.09 &  0.00e+00 &                 0 &  Likely Planet \\
  219344917 &     871.01 &          2.27 &        28.69 &   5.3 &   0.10 &  3.87e-03 &                 2 &                \\
  220459826 &     872.01 &          2.65 &         2.24 &   8.3 &   0.29 &  5.15e-02 &                 4 &                \\
  237920046 &     873.01 &          1.73 &         5.93 &   3.6 &   0.32 &  2.22e-02 &                 4 &                \\
  232025086 &     874.01 &          2.45 &         5.90 &   1.3 &   0.12 &  2.60e-02 &                 3 &                \\
   14165625 &     875.01 &          2.47 &        11.02 &  11.5 &   0.04 &  1.07e-03 &                 2 &                \\
   32497972 &     876.01 &          3.09 &        38.70 &   0.1 &   0.04 &  0.00e+00 &                 0 &  Likely Planet \\
  210873792 &     900.01 &          2.50 &         4.84 &   6.1 &   0.26 &  5.28e-02 &                10 &                \\
  261257684 &     904.01 &          2.72 &        18.35 &   9.6 &   0.03 &  2.22e-06 &                 1 &  Likely Planet \\
  350153977 &     908.01 &          3.27 &         3.18 &  13.1 &   0.20 &  2.77e-02 &                 3 &                \\
  369327947 &     910.01 &          1.01 &         2.03 &   5.7 &   0.28 &  5.66e-04 &                 4 &  Likely Planet \\
  406941612 &     912.01 &          1.90 &         4.68 &  15.0 &   0.07 &  4.05e-05 &                 4 &  Likely Planet \\
  407126408 &     913.01 &          2.57 &        11.09 &   6.9 &   0.02 &  4.77e-07 &                 3 &  Likely Planet \\
  259863352 &    1051.01 &          3.30 &        21.70 &   2.0 &   0.09 &  4.21e-04 &                 1 &  Likely Planet \\
  317060587 &    1052.01 &          3.30 &         9.14 &   3.4 &   0.10 &  3.82e-03 &                 4 &                \\
  366989877 &    1054.01 &          3.28 &        15.51 &  11.1 &   0.01 &  6.64e-09 &                 1 &      Validated \\
  320004517 &    1055.01 &          4.05 &        17.47 &  12.2 &   0.02 &  0.00e+00 &                 0 &  Likely Planet \\
  421894914 &    1056.01 &          2.85 &         5.31 &   1.9 &   0.16 &  1.06e-03 &                 2 &                \\
   31553893 &    1058.01 &          3.36 &        11.11 &   4.8 &   0.02 &  1.53e-02 &                 5 &                \\
  299799658 &    1062.01 &          2.38 &         4.11 &   8.9 &   0.14 &  1.29e-03 &                 1 &                \\
  406976746 &    1063.01 &          2.35 &        10.07 &  10.5 &   0.03 &  0.00e+00 &                 0 &  Likely Planet \\
   79748331 &    1064.01 &          2.80 &         6.44 &   4.7 &   0.05 &  1.45e-02 &                 3 &                \\
   79748331 &    1064.02 &          2.95 &        12.23 &  13.9 &   0.04 &  1.29e-02 &                 3 &                \\
  327301957 &    1074.01 &          2.98 &        13.93 &  13.7 &   0.03 &  6.16e-03 &                 2 &                \\
  351601843 &    1075.01 &          2.00 &         0.60 &  11.4 &   0.41 &  5.32e-02 &                 3 &                \\
  370133522 &    1078.01 &          1.26 &         0.52 &  17.0 &   0.41 &  1.55e-01 &                 2 &     Likely NFP \\
  161032923 &    1080.01 &          2.26 &         3.97 &   8.3 &   0.17 &  1.11e-02 &                11 &                \\
  261108236 &    1082.01 &          3.62 &        16.35 &   7.1 &   0.06 &  2.19e-02 &                 4 &                \\
  322270620 &    1083.01 &          2.48 &        12.96 &   8.4 &   0.10 &  3.22e-03 &                 3 &                \\
  383390264 &    1098.01 &          3.38 &        10.18 &   4.0 &   0.06 &  2.99e-06 &                 3 &  Likely Planet \\
  290348383 &    1099.01 &          3.49 &         6.44 &  13.3 &   0.07 &  2.48e-02 &                 1 &                \\
  409934330 &    1114.01 &          6.06 &         2.49 &  37.4 &   0.32 &  0.00e+00 &                 0 &  Likely Planet \\
  304100538 &    1116.01 &          2.34 &         5.01 &   4.5 &   0.16 &  1.64e-02 &                 4 &                \\
   29960110 &    1201.01 &          3.05 &         2.49 &   5.3 &   0.29 &  2.82e-01 &                 1 &     Likely NFP \\
   23434737 &    1203.01 &          3.76 &        25.49 &  11.0 &   0.01 &  0.00e+00 &                 0 &      Validated \\
  467666275 &    1204.01 &          1.97 &         1.38 &   4.0 &   0.38 &  3.57e-02 &                55 &                \\
  287776397 &    1205.01 &          4.01 &         2.39 &   2.7 &   0.15 &  1.64e-02 &                11 &                \\
  364393429 &    1207.01 &          1.39 &         2.63 &   2.1 &   0.51 &  1.16e-01 &                37 &     Likely NFP \\
  273985865 &    1208.01 &          1.92 &         3.42 &   5.1 &   0.35 &  1.32e-01 &                 3 &     Likely NFP \\
   30037565 &    1209.01 &          4.19 &        40.72 &   1.8 &   0.19 &  8.24e-03 &                20 &                \\
   50312495 &    1211.01 &          3.74 &        14.71 &   6.4 &   0.06 &  6.88e-03 &                 1 &                \\
  451606970 &    1214.01 &          5.51 &        38.36 &   6.8 &   0.67 &  6.28e-01 &                34 &     Likely NFP \\
  453260209 &    1215.01 &          1.20 &         1.21 &   4.4 &   0.71 &  3.88e-01 &                 6 &     Likely NFP \\
  141527965 &    1216.01 &          1.75 &         4.55 &   3.5 &   0.30 &  4.21e-03 &                10 &                \\
  248092710 &    1217.01 &          4.59 &        41.46 &   6.0 &   0.17 &  1.08e-01 &                 4 &     Likely NFP \\
  294781547 &    1218.01 &          2.27 &        13.77 &   4.3 &   0.22 &  5.09e-03 &                11 &                \\
  294981566 &    1219.01 &          2.36 &         1.91 &   7.6 &   0.34 &  2.44e-02 &                48 &                \\
  374997123 &    1222.01 &          2.32 &        10.19 &   1.4 &   0.23 &  1.27e-03 &                 4 &                \\
  382437043 &    1223.01 &          4.08 &        14.64 &   1.3 &   0.20 &  2.64e-02 &                 9 &                \\
  299798795 &    1224.01 &          1.95 &         4.18 &   4.6 &   0.08 &  0.00e+00 &                 0 &  Likely Planet \\
  150428703 &    1225.01 &          2.33 &        13.90 &   4.0 &   0.20 &  5.32e-02 &                 8 &                \\
  177115354 &    1226.01 &          2.48 &         3.93 &   6.5 &   0.34 &  3.54e-02 &                 8 &                \\
  300038935 &    1228.01 &          4.43 &        29.05 &   5.5 &   0.05 &  2.00e-04 &                 4 &  Likely Planet \\
  287156968 &    1230.01 &          3.23 &        25.06 &  11.4 &   0.01 &  0.00e+00 &                 0 &      Validated \\
  447061717 &    1231.01 &          3.37 &        24.25 &  17.3 &   0.03 &  2.46e-32 &                 3 &  Likely Planet \\
  260647166 &    1233.01 &          3.19 &        14.18 &  13.3 &   0.01 &  1.63e-05 &                 4 &      Validated \\
  260647166 &    1233.02 &          3.19 &        19.59 &  10.6 &   0.03 &  6.08e-05 &                 4 &  Likely Planet \\
  260647166 &    1233.03 &          2.45 &         6.20 &   8.6 &   0.02 &  2.14e-03 &                 5 &                \\
  260647166 &    1233.04 &          2.02 &         3.80 &   7.3 &   0.07 &  8.97e-03 &                 6 &                \\
  153951307 &    1238.01 &          2.34 &         3.29 &  14.0 &   0.06 &  3.34e-03 &                 1 &                \\
  153951307 &    1238.02 &          1.57 &         0.76 &   2.3 &   0.50 &  2.00e-02 &                 2 &                \\
  154716798 &    1239.01 &          4.75 &        12.64 &  24.0 &   0.11 &  0.00e+00 &                 0 &  Likely Planet \\
  198212955 &    1242.01 &          1.81 &         0.38 &   7.6 &   0.64 &  9.25e-02 &                 3 &                \\
  219698776 &    1243.01 &          2.97 &         4.66 &  13.7 &   0.48 &  3.14e-03 &                 2 &                \\
  219850915 &    1244.01 &          2.55 &         6.40 &   8.7 &   0.07 &  1.65e-02 &                 2 &                \\
  229781583 &    1245.01 &          2.29 &         4.82 &  12.8 &   0.15 &  1.74e-02 &                 6 &                \\
  230127302 &    1246.01 &          3.88 &        18.65 &   9.7 &   0.06 &  3.00e-02 &                 5 &                \\
  230127302 &    1246.02 &          3.00 &         4.31 &   4.3 &   0.07 &  1.81e-02 &                 5 &                \\
  230127302 &    1246.03 &          2.73 &         5.90 &   3.6 &   0.08 &  1.34e-02 &                 5 &                \\
  232540264 &    1247.01 &          3.10 &        15.92 &  14.1 &   0.02 &  3.53e-04 &                 2 &  Likely Planet \\
  232976128 &    1249.01 &          3.43 &        13.08 &  12.0 &   0.06 &  3.99e-03 &                 1 &                \\
  237222864 &    1255.01 &          3.12 &        10.29 &  13.8 &   0.10 &  5.31e-06 &                 1 &  Likely Planet \\
  355867695 &    1260.01 &          2.46 &         3.13 &   6.9 &   0.05 &  4.05e-03 &                 3 &                \\
  355867695 &    1260.02 &          2.87 &         7.49 &   5.9 &   0.05 &  8.71e-04 &                 2 &  Likely Planet \\
  406672232 &    1263.01 &          1.84 &         1.02 &   2.6 &   0.34 &  1.88e-01 &                16 &     Likely NFP \\
  467179528 &    1266.01 &          2.60 &        10.90 &  12.3 &   0.02 &  1.68e-05 &                 1 &  Likely Planet \\
  467179528 &    1266.02 &          2.10 &        18.80 &   5.8 &   0.21 &  2.42e-05 &                 1 &  Likely Planet \\
  198241702 &    1269.01 &          2.31 &         4.25 &   6.4 &   0.10 &  1.02e-02 &                 4 &                \\
  417948359 &    1272.01 &          5.97 &         3.32 &  17.1 &   0.93 &  6.06e-01 &                 7 &     Likely NFP \\
   13499636 &    1275.01 &          2.85 &        11.32 &   6.7 &   0.11 &  4.91e-02 &                 3 &                \\
  153949511 &    1277.02 &          3.14 &        37.07 &   6.8 &   0.06 &  0.00e+00 &                 0 &  Likely Planet \\
  224297258 &    1279.01 &          2.81 &         9.61 &  14.7 &   0.03 &  0.00e+00 &                 0 &  Likely Planet \\
  232971294 &    1281.01 &          2.58 &         6.39 &   5.6 &   0.12 &  7.58e-04 &                 3 &  Likely Planet \\
  352764091 &    1287.01 &          3.30 &         9.60 &   8.4 &   0.09 &  6.06e-03 &                 5 &                \\
  269701147 &    1339.02 &          3.15 &        28.58 &   8.1 &   0.01 &  3.78e-05 &                 1 &      Validated \\
  229747848 &    1347.01 &          2.06 &         0.85 &   7.0 &   0.59 &  6.32e-02 &                 6 &                \\
  199444169 &    1410.01 &          3.38 &         1.22 &  14.1 &   0.13 &  2.91e-02 &                 5 &                \\
  116483514 &    1411.01 &          1.37 &         1.45 &   8.2 &   0.20 &  1.45e-02 &                 3 &                \\
  148782377 &    1415.01 &          4.99 &        14.42 &  11.8 &   0.02 &  0.00e+00 &                 0 &  Likely Planet \\
  158025009 &    1416.01 &          1.76 &         1.07 &   5.9 &   0.19 &  5.32e-02 &                 2 &                \\
  346418409 &    1423.01 &          3.70 &         2.76 &  25.0 &   0.63 &  5.48e-01 &                 2 &     Likely NFP \\
  418959198 &    1424.01 &          2.90 &         4.90 &  15.3 &   0.04 &  1.89e-02 &                 3 &                \\
  293954617 &    1430.01 &          2.15 &         7.43 &  10.1 &   0.03 &  5.12e-06 &                 1 &  Likely Planet \\
  138588540 &    1434.01 &          2.22 &        29.89 &   8.8 &   0.03 &  3.71e-08 &                 1 &  Likely Planet \\
  153976959 &    1435.01 &          1.31 &         0.69 &   6.4 &   0.55 &  1.69e-02 &                 2 &                \\
  154383539 &    1436.01 &          1.78 &         0.87 &  12.0 &   0.63 &  4.40e-02 &                 2 &                \\
  198356533 &    1437.01 &          2.72 &        18.84 &   4.2 &   0.02 &  1.79e-03 &                 3 &                \\
  233617847 &    1440.01 &          3.06 &        15.52 &   4.0 &   0.49 &  4.53e-01 &                 2 &     Likely NFP \\
  233951353 &    1441.01 &          2.73 &        22.10 &   4.4 &   0.06 &  2.67e-03 &                 5 &                \\
  235683377 &    1442.01 &          1.21 &         0.41 &   8.0 &   0.84 &  1.01e-01 &                 3 &     Likely NFP \\
  258514800 &    1444.01 &          1.46 &         0.47 &   3.5 &   0.65 &  9.12e-02 &                 6 &                \\
  259172391 &    1445.01 &          3.26 &         9.81 &  13.3 &   0.05 &  1.09e-02 &                 6 &                \\
  294471966 &    1446.01 &          2.45 &         6.32 &   5.2 &   0.12 &  2.13e-02 &                14 &                \\
  343628284 &    1448.01 &          3.28 &         8.11 &  11.4 &   0.12 &  3.43e-02 &                 8 &                \\
  356158613 &    1449.01 &          4.02 &        24.71 &  10.2 &   0.39 &  3.69e-02 &                16 &                \\
  356158613 &    1449.02 &          1.82 &         2.37 &   6.0 &   0.57 &  4.54e-01 &                31 &     Likely NFP \\
  377293776 &    1450.01 &          1.10 &         2.04 &   3.0 &   0.72 &  2.06e-02 &                 8 &                \\
  417931607 &    1451.01 &          3.08 &        33.07 &   5.2 &   0.09 &  2.87e-04 &                 1 &  Likely Planet \\
  420112589 &    1452.01 &          1.77 &        11.06 &   2.1 &   0.46 &  3.86e-01 &                 5 &     Likely NFP \\
  198390247 &    1453.02 &          1.21 &         4.31 &   1.8 &   0.34 &  2.10e-02 &                 4 &                \\
   16920150 &    1459.01 &          4.68 &         9.16 &  12.9 &   0.45 &  1.43e-01 &                 2 &     Likely NFP \\
  188768068 &    1462.01 &          1.83 &         2.18 &   7.8 &   0.09 &  9.03e-05 &                 1 &  Likely Planet \\
  229944666 &    1464.01 &          3.08 &        11.33 &  10.6 &   0.18 &  1.83e-02 &                 5 &                \\
  237086564 &    1466.01 &          2.79 &         1.87 &  12.6 &   0.11 &  3.91e-02 &                 7 &                \\
  240968774 &    1467.01 &          2.14 &         5.97 &   9.5 &   0.10 &  4.84e-03 &                 6 &                \\
  243185500 &    1468.01 &          2.65 &        15.53 &  11.4 &   0.09 &  3.78e-04 &                 1 &  Likely Planet \\
  243185500 &    1468.02 &          1.38 &         1.88 &   7.9 &   0.32 &  1.51e-01 &                 2 &     Likely NFP \\
  284441182 &    1470.01 &          2.42 &         2.53 &   1.9 &   0.13 &  3.87e-02 &                14 &                \\
  306263608 &    1471.01 &          4.26 &        20.77 &   6.6 &   0.02 &  3.23e-05 &                 1 &  Likely Planet \\
  306955329 &    1472.01 &          4.78 &         6.36 &  14.1 &   0.09 &  8.98e-03 &                 3 &                \\
  352413427 &    1473.01 &          2.90 &         5.26 &   5.4 &   0.02 &  2.43e-03 &                 3 &                \\
  428679607 &    1669.01 &          2.65 &         2.68 &   4.7 &   0.09 &  2.29e-03 &                 5 &                \\
  259168516 &    1680.01 &          1.55 &         4.80 &   4.8 &   0.30 &  1.08e-01 &                 2 &     Likely NFP \\
  321041369 &    1681.01 &          3.62 &         1.54 &  19.8 &   0.69 &  9.95e-02 &                 3 &                \\
   58542531 &    1683.01 &          2.72 &         3.06 &   8.1 &   0.06 &  0.00e+00 &                 0 &  Likely Planet \\
   28900646 &    1685.01 &          1.56 &         0.67 &  17.2 &   0.65 &  5.35e-02 &                 6 &                \\
  102672709 &    1686.01 &          4.24 &         6.70 &   5.7 &   0.04 &  3.33e-03 &                 5 &                \\
  103448870 &    1687.01 &          4.39 &        10.26 &   9.0 &   0.60 &  7.98e-07 &                 3 &                \\
  268334473 &    1691.01 &          3.68 &        16.73 &  12.2 &   0.02 &  0.00e+00 &                 0 &  Likely Planet \\
  288636342 &    1692.01 &          5.10 &        17.73 &   7.8 &   0.03 &  7.77e-03 &                 5 &                \\
  353475866 &    1693.01 &          1.46 &         1.77 &   4.9 &   0.31 &  4.59e-02 &                 5 &                \\
  396740648 &    1694.01 &          5.62 &         3.77 &  24.4 &   0.61 &  0.00e+00 &                 0 &                \\
  422756130 &    1695.01 &          2.22 &         3.13 &  12.1 &   0.23 &  1.40e-02 &                 6 &                \\
  470381900 &    1696.01 &          3.18 &         2.50 &   6.1 &   0.28 &  1.58e-01 &                 8 &     Likely NFP \\
 1884091865 &    1697.01 &          3.13 &        10.69 &   5.0 &   0.34 &  1.72e-01 &                 4 &     Likely NFP \\
   15863518 &    1713.01 &          4.17 &         0.56 &  22.7 &   0.81 &  5.98e-02 &                 5 &                \\
   14336130 &    1716.01 &          3.02 &         8.09 &  13.7 &   0.05 &  6.84e-03 &                 2 &                \\
  257241363 &    1718.01 &          3.96 &         5.59 &  22.6 &   0.02 &  3.10e-05 &                 1 &  Likely Planet \\
   85242435 &    1722.01 &          4.29 &         9.61 &  14.1 &   0.02 &  2.72e-04 &                 2 &  Likely Planet \\
   71431780 &    1723.01 &          3.16 &        13.72 &   9.0 &   0.02 &  7.34e-04 &                 3 &  Likely Planet \\
   81212286 &    1724.01 &          2.26 &         0.69 &  14.4 &   0.84 &  5.21e-01 &                 3 &     Likely NFP \\
  241225337 &    1727.01 &          2.58 &         1.83 &  10.7 &   0.33 &  3.13e-02 &                 5 &                \\
  285048486 &    1728.01 &          4.92 &         3.49 &  25.5 &   0.10 &  0.00e+00 &                 1 &  Likely Planet \\
  318022259 &    1730.01 &          2.58 &         6.22 &  22.7 &   0.03 &  1.69e-02 &                 3 &                \\
  318022259 &    1730.02 &          1.48 &         2.16 &   8.8 &   0.22 &  1.08e-01 &                 3 &     Likely NFP \\
  470987100 &    1732.01 &          2.66 &         4.12 &  15.0 &   0.06 &  1.59e-02 &                 2 &                \\
  159418353 &    1739.01 &          2.24 &         8.30 &   7.9 &   0.10 &  0.00e+00 &                 0 &  Likely Planet \\
  174041208 &    1740.01 &          2.51 &        19.43 &   7.6 &   0.05 &  4.23e-04 &                 3 &  Likely Planet \\
  232650365 &    1746.01 &          1.83 &         2.53 &   2.3 &   0.50 &  1.95e-01 &                 3 &     Likely NFP \\
  408636441 &    1759.01 &          3.50 &        37.70 &  10.6 &   0.03 &  2.36e-13 &                 5 &  Likely Planet \\
  420112587 &    1760.01 &          1.74 &        11.06 &   8.4 &   0.56 &  4.91e-01 &                 5 &     Likely NFP \\
    4897275 &    1774.01 &          2.90 &        16.71 &   8.0 &   0.01 &  1.13e-04 &                 1 &      Validated \\
   21535395 &    1776.01 &          1.58 &         2.80 &   7.6 &   0.29 &  9.57e-05 &                 1 &  Likely Planet \\
   29191624 &    1777.01 &          2.82 &        14.65 &   5.2 &   0.05 &  3.13e-04 &                 1 &  Likely Planet \\
   39699648 &    1778.01 &          3.43 &         6.52 &   9.9 &   0.07 &  3.54e-03 &                 1 &                \\
  160045097 &    1782.01 &          2.47 &         4.99 &   7.4 &   0.05 &  4.87e-03 &                 1 &                \\
  229938290 &    1783.01 &          0.73 &         1.42 &   0.3 &   0.74 &  3.19e-02 &                 3 &                \\
  286916251 &    1794.01 &          3.84 &         8.78 &   4.0 &   0.04 &  1.16e-03 &                 2 &                \\
  368435330 &    1797.01 &          3.30 &         3.65 &  17.4 &   0.03 &  4.46e-05 &                 1 &  Likely Planet \\
  198153540 &    1798.01 &          2.33 &         8.02 &  10.0 &   0.05 &  1.87e-04 &                 2 &  Likely Planet \\
    8967242 &    1799.01 &          1.93 &         7.09 &   8.3 &   0.09 &  2.94e-06 &                 1 &  Likely Planet \\
  119584412 &    1801.01 &          2.15 &        21.28 &   5.2 &   0.07 &  0.00e+00 &                 0 &  Likely Planet \\
  138762614 &    1802.01 &          2.46 &        16.80 &   8.4 &   0.08 &  2.67e-02 &                 2 &                \\
  144401492 &    1803.01 &          4.51 &        12.89 &  17.4 &   0.29 &  0.00e+00 &                 0 &  Likely Planet \\
  144401492 &    1803.02 &          3.08 &         6.29 &  13.9 &   0.02 &  0.00e+00 &                 1 &  Likely Planet \\
  148679712 &    1804.01 &          3.01 &         4.93 &  16.6 &   0.05 &  5.08e-03 &                 4 &                \\
  165763244 &    1805.01 &          3.23 &        24.07 &   8.8 &   0.03 &  1.40e-03 &                 3 &                \\
  166648874 &    1806.01 &          3.04 &        15.15 &  11.4 &   0.02 &  0.00e+00 &                 0 &  Likely Planet \\
  180695581 &    1807.01 &          1.48 &         0.55 &  15.3 &   0.28 &  8.66e-03 &                 1 &                \\
  390651552 &    1827.01 &          1.40 &         1.47 &  21.3 &   0.02 &  1.05e-02 &                 1 &                \\
   27194429 &    1831.01 &          7.18 &         0.56 &  12.2 &   0.87 &  2.98e-02 &                 2 &                \\
  307956397 &    1832.01 &          7.71 &         4.15 &  22.5 &   0.71 &  2.28e-01 &                 1 &     Likely NFP \\
  347332255 &    1835.01 &          2.15 &         5.64 &  11.2 &   0.57 &  0.00e+00 &                 1 &                \\
  381714186 &    1839.01 &          2.49 &         1.42 &  11.7 &   0.27 &  0.00e+00 &                 0 &  Likely Planet \\
  202426247 &    1860.01 &          1.61 &         1.07 &   4.3 &   0.54 &  3.39e-01 &                 4 &     Likely NFP \\
  390651552 &    1827.01 &          1.40 &         1.47 &  21.3 &   0.02 &  1.05e-02 &                 1 &                \\
   27194429 &    1831.01 &          7.18 &         0.56 &  12.2 &   0.87 &  2.98e-02 &                 2 &                \\
  307956397 &    1832.01 &          7.71 &         4.15 &  22.5 &   0.71 &  2.28e-01 &                 1 &     Likely NFP \\
  347332255 &    1835.01 &          2.15 &         5.64 &  11.2 &   0.57 &  0.00e+00 &                 1 &                \\
  381714186 &    1839.01 &          2.49 &         1.42 &  11.7 &   0.27 &  0.00e+00 &                 0 &  Likely Planet \\
  202426247 &    1860.01 &          1.61 &         1.07 &   4.3 &   0.54 &  3.39e-01 &                 4 &     Likely NFP \\
  188589164 &    2013.01 &          1.03 &         2.61 &  12.8 &   0.20 &  0.00e+00 &                 0 &  Likely Planet \\
  368287008 &    2015.01 &          3.38 &         3.35 &  16.4 &   0.26 &  1.23e-01 &                 1 &     Likely NFP \\
  219508169 &    2016.02 &          2.80 &         2.46 &  14.3 &   0.04 &  4.13e-03 &                 2 &                \\
  357501308 &    2018.01 &          2.22 &         7.44 &  12.2 &   0.02 &  1.86e-04 &                 2 &  Likely Planet \\
  159781361 &    2019.01 &          5.78 &        15.35 &  11.9 &   0.02 &  0.00e+00 &                 0 &  Likely Planet \\
   11996814 &    2022.01 &          4.37 &         0.45 &  25.0 &   0.99 &  4.35e-04 &                 7 &                \\
   16884216 &    2023.01 &          2.47 &        11.19 &  17.4 &   0.03 &  0.00e+00 &                 0 &  Likely Planet \\
 \enddata
 \tablecomments{This table is published in its entirety in machine-readable format.}
\end{deluxetable*}

.

\end{document}